\documentclass[useAMS,usenatbib]{mnras}
\usepackage[fleqn]{amsmath}

\usepackage{newtxtext,newtxmath}

\usepackage[T1]{fontenc}
\usepackage{ae,aecompl}

\usepackage{amsfonts,bm}
\usepackage{graphicx}
\usepackage[utf8]{inputenc}
\usepackage{adjustbox}   %Centering issues
\usepackage{subfig}

\newcommand{\tild}{~}
\newcommand{\Rho}{\mathcal{R}}

\newcommand  *{\diff}   {\mathop{}\!\mathrm{d}}
\renewcommand*{\vec}[1] {\boldsymbol{#1}}
\newcommand  *{\uvec}[1]{\hat{\vec{#1}}}
\newcommand  *{\s}[1]   {\mathsf{#1}}
\newcommand  *{\mat}[1] {\vec{\s{#1}}}
\newcommand  *{\Ups}    {\Upsilon}
\newcommand  *{\The}    {\Theta}
\newcommand  *{\I}      {\mathrm{i}}
\DeclareMathOperator{\sinc}{sinc}
\newcommand  *{\Exp}[1] {\mathrm{e}^{\textstyle #1}}
\newcommand  *{\phn}    {\phantom{-}}
\newcommand  *{\pfrac}[2] {\left(\frac{#1}{#2}\right)}

\title[Triaxial Deprojection]{Non-parametric Triaxial Deprojection of Elliptical Galaxies}

\author[S.~de Nicola et al.]{%
Stefano de Nicola,$^{\!\!1}$\thanks{E-mail: denicola@mpe.mpg.de}
Roberto P. Saglia,$^{\!\!1,2}$
Jens Thomas,$^{\!\!1}$ 
Walter Dehnen,$^{\!\!2,3}$ and 
\newauthor
Ralf Bender $^{\!\!2,1}$
\\
$^{1}$ Max-Planck Institute for Extraterrestrial Physics, Giessenbachstrasse 1, D-85748, Garching (Germany) \\
$^{2}$ Universit{\"a}ts-Sternwarte Muenchen, Scheinerstrasse 1, D-81679, Munich, Germany\\
$^{3}$ University of Leicester, Dept.~for Astronomy \& Physics, University Rd, LE1 7RH, UK
}

\date{Accepted XXX. Received YYY; in original form ZZZ}

\pubyear{2020}

\begin{document}
\label{firstpage}
\pagerange{\pageref{firstpage}--\pageref{lastpage}}
\maketitle

\begin{abstract}
We present a grid-based non-parametric approach to obtain a triaxial three-dimensional luminosity density from its surface brightness distribution. Triaxial deprojection is highly degenerate and our approach illustrates the resulting difficulties. Fortunately, for massive elliptical galaxies, many deprojections for a particular line of sight can be discarded, because their projection along other lines of sight does not resemble elliptical galaxies. The near-elliptical isophotes of these objects imply near ellipsoidal intrinsic shapes. In fact, deprojection is unique for densities distributed on ellipsoidal shells. The constrained non-parametric deprojection method we present here relaxes this constraint and assumes that the contours of the luminosity density are boxy/discy ellipsoids with radially varying axis ratios. With this approach we are able to reconstruct the intrinsic triaxial densities of our test models, including one drawn from an $N$-body simulation. The method also allows to compare the relative likelihood of deprojections at different viewing angles. We show that the viewing orientations of individual galaxies with nearly ellipsoidal isophotal shapes can be constrained from photometric data alone.
\end{abstract}


%%%%%%%%%%%%%%%%%%%%%%%%%%%%%%%%%%%%%%%%%%%%%%%%%%

%%%%%%%%%%%%%%%%% BODY OF PAPER %%%%%%%%%%%%%%%%%%
%INTRODUZIONE:

\section{Introduction}
\label{sec_introduction}

Photometric observations of galaxies provide us with the surface brightness (SB) profile and the shapes of the isophotes, possibly as a function of wavelength.  From here, we need to constrain the intrinsic, three-dimensional luminosity and, possibly, stellar mass density $\rho$, as a starting point to study the dynamics of galaxies. This step can be performed fitting the galaxy kinematics through the powerful \cite{Sch79} method, where the stellar mass density, along with a dark matter (DM) profile and a black hole (BH) mass, is used to build a potential by integrating the Poisson's equation through which the orbits are computed \citep{Gebhardt03, Jens04, Jens05}. Another potent dynamical modelling approach is the made-to-measure $N$-body technique \citep{SyerTremaine1996,NMAGIC2007,Dehnen2009}, where an $N$-body model is adapted to fit the data subject to any constraints. A less general but popular alternative is to solve the Jeans equation, typically assuming  cylindrical symmetry \citep{Cappellari08}.

Computer tomography solves the problem of reconstructing the three dimensional structure of a body by combining a number of two-dimensional projections taken at different angles covering a semi-circle. Astronomers have access to only one line of sight (LOS).  For an axisymmetric system this means that the true density can be reconstructed only when the object can be assumed to be seen edge-on \citep{Rybicki87, GB96}. In general, at any assumed inclination angle $i$ (defined as the angle between the LOS and the equatorial plane) a 'cone of ignorance' of opening $90\degr-i$ is generated in Fourier space, such that any density inside this cone will project to nothing along the assumed LOS. Such \emph{conus densities} are unphysical on their own, since they are necessarily negative somewhere. However, to some extent, they make the deprojection at any assumed $i<90\degr$ non-unique. \citet{GB96}, \citet{VDB97} and \citet{Kochanek96} discuss extensively conus densities.  \citet{Gerhard96} considers the extension of the Fourier slice theorem \citep{Rybicki87} to the triaxial case, where the degeneracy of the problem is increased further, since only 4 planes in Fourier space are constrained by the measured surface brightness.

Although deprojecting SBs is a mathematically ill-posed question, a number of parametric and non-parametric approaches have been implemented to sample the space of possible three dimensional density distributions of galaxies. Parametric algorithms have the natural advantage of yielding smooth solutions and being fast, while non-parametric methods trade off naturally smooth solutions and very short computational time for an approach that can find a much broader family of solutions. In both cases, exploiting additional \emph{statistical} information, like the ellipticity distribution determined from the observations of millions of galaxies on the sky, can help reducing the ambiguity of the deprojection of individual objects.

The most widely used parametric method is the Multi-Gaussian Expansion (MGE, \citealt{Bendinelli91, Emsellem94, Cappellari02}). This routine can be directly applied to images and fits a SB distribution with a sum of Gaussians. This Multi-Gaussian \emph{model} can be deprojected analytically for a given set of viewing angles (see the next section for their geometric definition), under the assumption that each 2D Gaussian component of the SB deprojects into a 3D Gaussian component of the density (see Section~\ref{sec:ellipsoidal}).  This approach has several benefits, however, it just yields one deprojection per set of viewing angles, and there is no guarantee that this intrinsic deprojected density is correct. By construction, all these deprojections project \emph{exactly} to the same Multi-Gaussian model, for the respective viewing angles. Thus, this approach does not allow to rank different deprojections based on their different relative likelihoods. Although the range of possible viewing angles is limited by the requirement that the axis ratios $p$ and $q$ should always follow the constraint $0\leq q \leq p\leq 1$ and can be further constrained by assumptions about the minimal physically plausible intrinsic flattening etc.

Non-parametric algorithms are the best choice with respect to the intrinsic degeneracy issue. However, the ability to find any mathematically possible solution comes to the expense that these algorithms need some kind of penalized approach in order to keep the deprojection under control and filter out non-physical solutions. One well-working non-parametric axisymmetric algorithm is the one presented in \citet{Magorrian99}, hereafter M99. This algorithm implements a penalized Metropolis Monte-Carlo algorithm that starting from an isophotal table deprojects the SB under the assumption of axisymmetry, also allowing for penalty functions that make the solution smoother or bias it towards a more boxy/discy shape. Being axisymmetric, the code has the limitation of not allowing for isophotal twist, a typical indication of triaxiality.

In this work, we implement a triaxial version of this fully non-parametric approach.  Unlike for the axisymmetric case, finding suitable smoothing or penalty functions turns out difficult in the extension to triaxiality and we here follow a different approach to ``penalize'' the deprojection: we take advantage of the empirical fact that iso-density contours of massive ellipticals do not deviate strongly from ellipsoidal shapes. This suggests a smoothing towards ellipsoidal intrinsic shapes and we develop a constrained non-parametric tool, where the density is stratified onto deformed (discy-boxy) ellipsoids. We show that exactly ellipsoidal deprojections are unique when the line-of-sight (LOS) is known (and different from one of the principal axes). We use our ellipsoidal code to explore, for the first time, how tightly the viewing angles of triaxial objects can be constrained from surface photometry only.

The paper is organized as follows. Section~\ref{Sec.2} provides the mathematical background of the triaxial deprojection problem and introduces the concept of {\bf cloaked densities}, the triaxial analogous of the konus densities encountered in the axisymmetric case. In Section~\ref{sec_triaxial_toymodel} we show the details of the transformation of the M99's code from the axisymmetric to the triaxial case and discuss the degeneracies connected to non-parametric deprojections. Section~\ref{sec_deprojectellipsoids} demonstrates that ellipsoidal deprojections of density distributions stratified on ellipsoidals are unique if the viewing angles are known and different from the principal axes. Section~\ref{sec_semiparameteric} presents the modifications implemented to constrain the solution on deformed ellipsoids.  Section~\ref{sec_range_tests} explores the range of observables (ellipticities, position angle twists, $a_4$ coefficients) generated by projection effects along with illustrating the reliability of the algorithm, while Section~\ref{sec_viewingangles} explores how tightly the viewing angles can be constrained. Section~\ref{par.MGE} compares the performances of our approach with the MGE strategy. Section~\ref{sec_conclusions} summarizes our findings and conclusions. The Appendices present a number of analytic \emph{cloacked densities} (see Section~\ref{Sec.2}) and discuss how to deal with the presence of discs.

%%%%%%%%%%%%%%%%%%%%%
\begin{figure*}
    \resizebox{.3\linewidth}{!}{\includegraphics{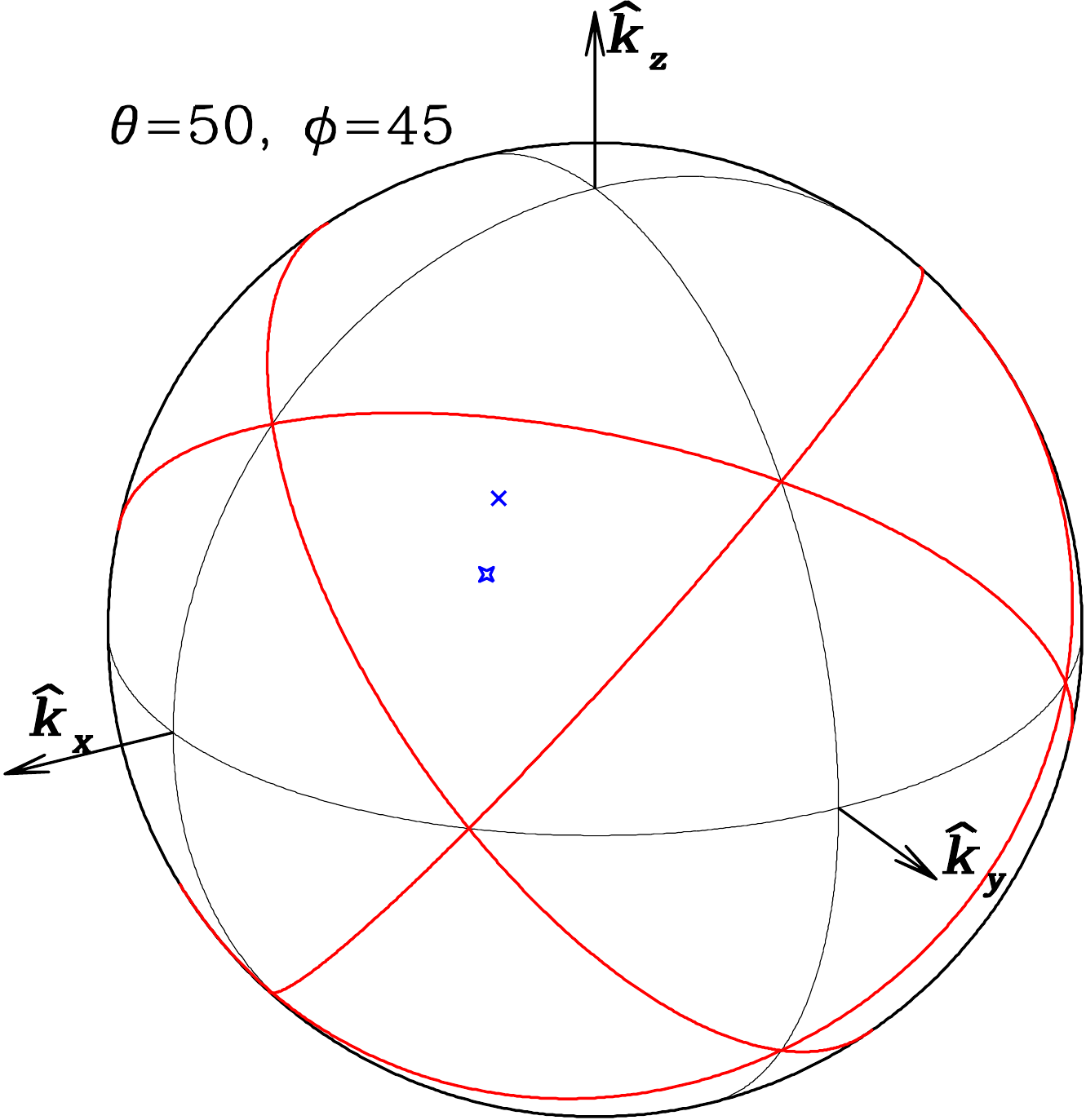}}
    \hfil
    \resizebox{.3\linewidth}{!}{\includegraphics{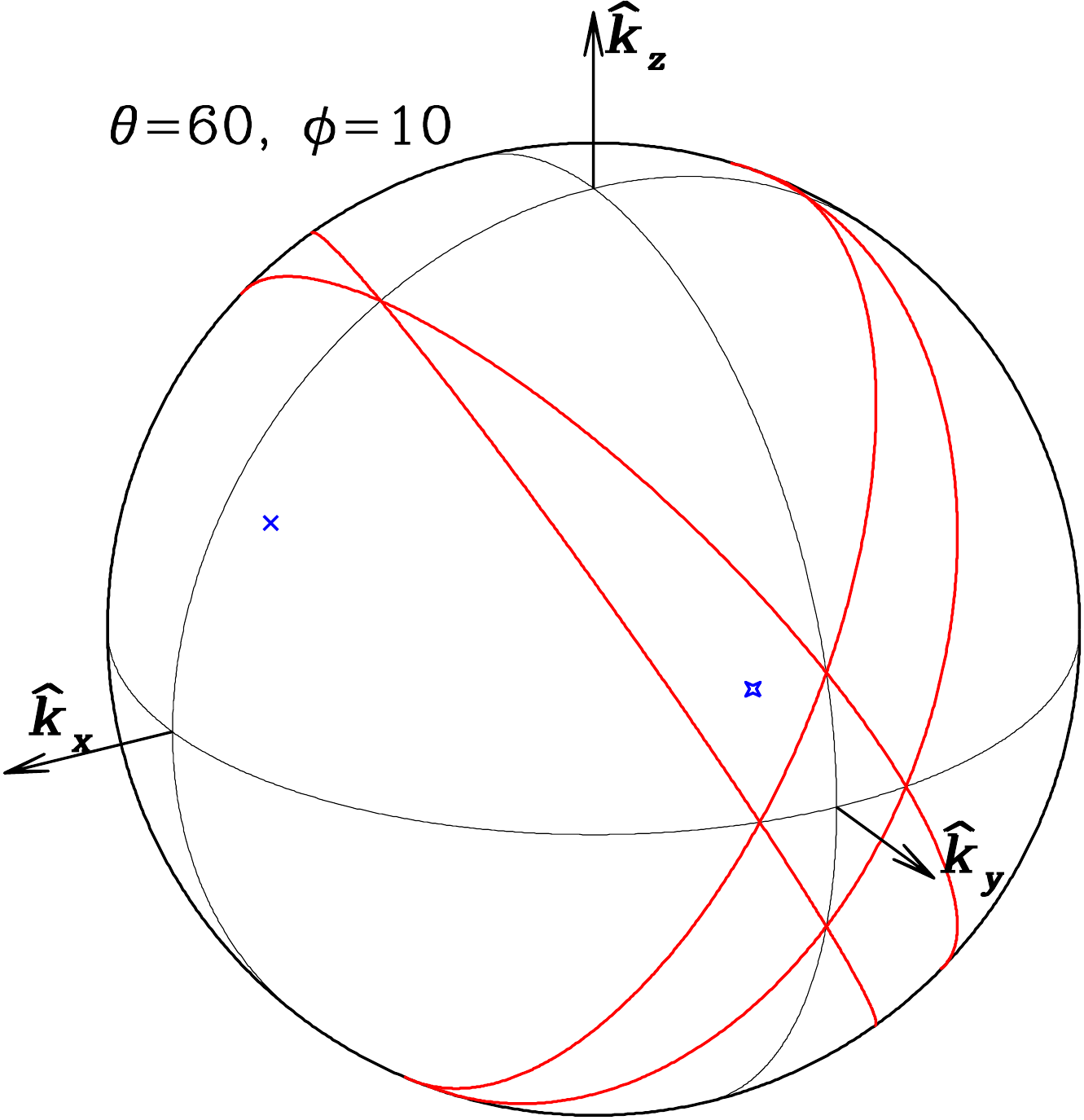}}
    \hfil
    \resizebox{.3\linewidth}{!}{\includegraphics{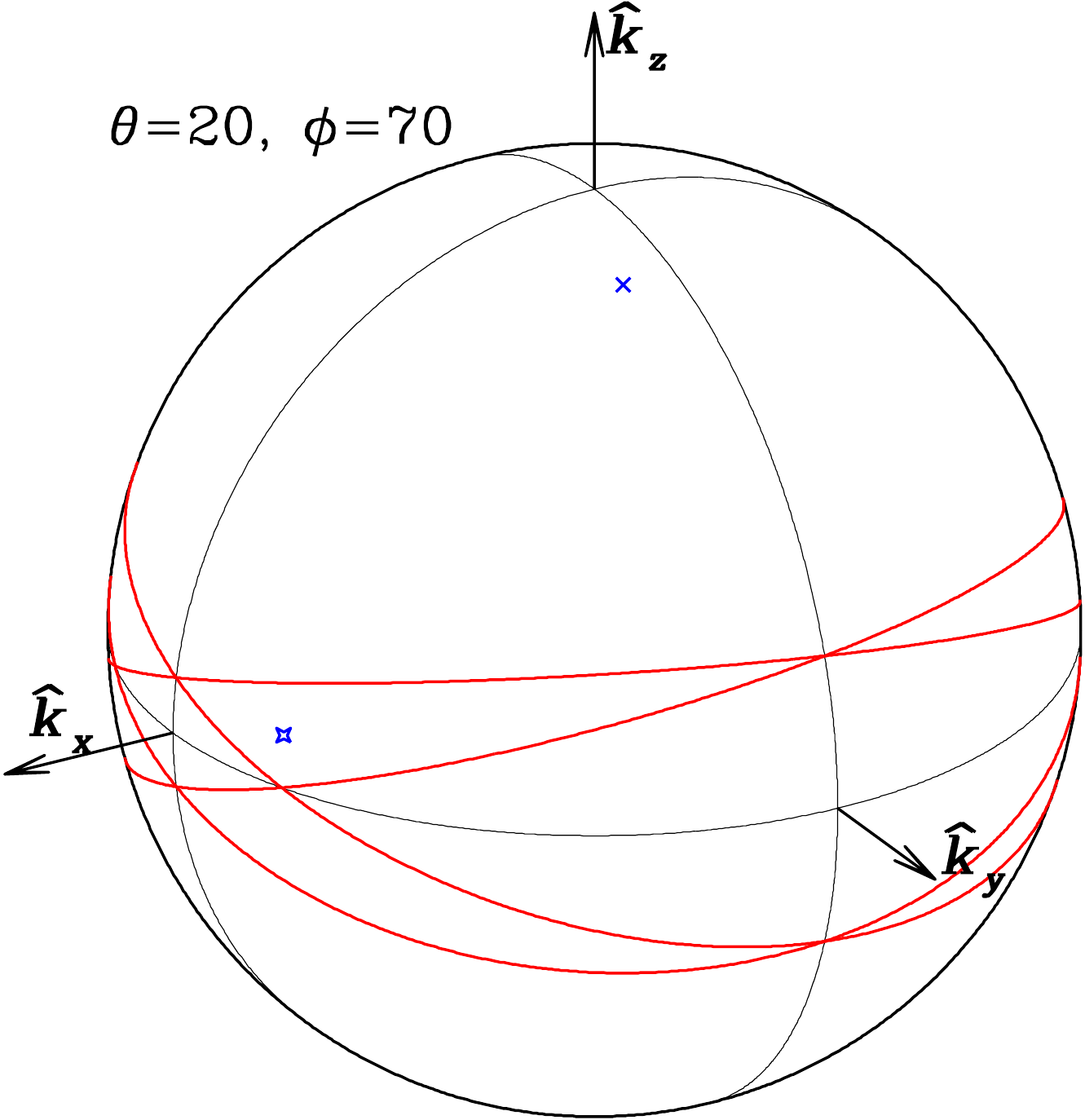}}
    \caption{The projection along various LOSs (as indicated, also by the cross) constrains the 3D density only on four planes in Fourier space, shown as great circles (red) on a unit sphere. If the LOS is near a fundamental plane or axis, large parts of Fourier space are completely unconstrained. The planes dissect Fourier space into funnel-shaped regions: a four-sided funnel around each fundamental axis and a three-sided funnel between them in each octant around the reciprocal LOS (star).}
    \label{fig:proj}
\end{figure*}
%%%%%%%%%%%%%%%%%%%%

%%%%%%%%%%%%%%%%%%%%%%%%%%%%%%%%%%%%%%%%%%%%%
%Qua vanno ricontrollate le definizioni delle formule...
%%%%%%%%%%%%%%%%%%%%%%%%%%%%%%%%%%%%%%%%%%%%%%%
\section{The Fourier-slice theorem \& cloaked densities}  
\label{Sec.2}

It is well known that the deprojection of an axisymmetric density is not unique, unless the object is viewed precisely edge-on \citep{GB96}. The deprojection of a triaxial galaxy is even less constrained: let $\rho(\vec{r})$ be a 3D density distribution of a transparent galaxy and
\begin{align}
	\label{eq:Sigma}
	\Sigma(x,y)=\int\rho(\vec{r})\diff z
\end{align}
its surface density when projected along the $z$ axis. Now consider the Fourier transforms of both $\rho$ and $\Sigma$:
\begin{align}
	\hat{\rho}(\vec{k}) &=
		\int\rho(\vec{r})\Exp{-i\vec{k}\cdot\vec{r}}\diff\vec{r},\\
	\hat{\Sigma}(k_x,k_y) &= 
		\int\Sigma\,\Exp{-i(k_xx+k_yy)}\diff x\diff y
	\nonumber \\ &= 
		\int\rho(\vec{r})\Exp{-i(k_xx+k_yy)}\diff\vec{r},
\end{align}
where equation~\eqref{eq:Sigma} has been used. Thus (Fourier-slice theorem, \citealt{Rybicki87}),
\begin{equation}
	\hat{\Sigma}(k_x,k_y) = \hat{\rho}(\vec{k})\big|_{k_z=0}.
\end{equation}
Of course, nothing is special about the choice of the $z$-axis and a more general form of the theorem states that the Fourier transform $\hat{\Sigma}$ of the projection of $\rho$ along the LOS direction
\begin{equation}
	\label{eq:LOS}
	\vec{\ell} = (\sin\theta\cos\phi,\,\sin\theta\sin\phi,\,\cos\theta)^t
\end{equation}
equals $\hat{\rho}$ in the plane $\vec{\ell}\cdot\vec{k}=0$. If $\rho$ is assumed/known to be triaxial, then so is $\hat{\rho}$ and knowledge of $\Sigma$ constrains $\hat{\rho}$ on the four planes $\vec{\ell}_i\cdot\vec{k}=0$ with
\begin{equation}
  \vec{\ell}_1 \equiv \vec{\ell},
  \quad
  \vec{\ell}_2 \equiv %(-\ell_x, \ell_y, \ell_z)^t
	\begin{pmatrix} -\ell_x \\ \phn \ell_y \\ \phn \ell_z \end{pmatrix}
	,\quad
	\vec{\ell}_3 \equiv %(\ell_x,-\ell_y,\ell_z)^t
	\begin{pmatrix} \phn \ell_x \\ - \ell_y \\ \phn \ell_z \end{pmatrix}
	,\quad
	\vec{\ell}_4 \equiv %(-\ell_x,-\ell_y,\ell_z)^t
	\begin{pmatrix} - \ell_x \\ - \ell_y \\ \phn \ell_z \end{pmatrix}
	,
\end{equation}
which are the reflections of $\vec{\ell}$ off the symmetry axes. These four planes dissect the space into distinct regions as depicted in Fig.~\ref{fig:proj}.  Of course, if $\rho(\vec{r})$ is triaxial, the conditions that $\hat{\rho}(\vec{k})=0$ on any one of these planes are mutually identical.

Hereafter, we denote a density distribution that projects to $\Sigma=0$ at all sky positions a \textbf{cloaked density}. We also denote as \textbf{cloak} the set of LOS $\vec{\ell}$ such that the projection of $\rho(\vec{r})$ is invisible. For every non-trivial density, the cloak can only cover a small but possibly continuous set of directions.

A projection along of one of the principal axes provides the least amount of information and does not constrain the density distribution along those axes. On the other hand, a projection along a line of sight far from any principal axis does not constraint the Fourier transform $\hat{\rho}$ near these axes, such that distributions $\rho$ with $\hat{\rho}\neq0$ only around one principal axis are cloaked. Adding or subtracting such cloaked densities typically adds or subtracts a disc perpendicular to the respective principal axis. In Appendix~\ref{sec_cloaked} we study cloaked densities in more detail and present several straightforward and elegant schemes for constructing them as well as near-invisible densities.\\
Some mathematical properties of these cloaked densities are as follows.
Let $\rho_1(\vec{r})$ and $\rho_2(\vec{r})$ be two cloaked densities and $f(\vec{r})$ an arbitrary function, then the following are also invisible when projected along $\vec{\ell}$.
\begin{enumerate}
\item \label{item:cloak:superimpose} Linear combinations of $\rho_1(\vec{r})$ and $\rho_2(\vec{r})$, whereby the cloak shrinks to the intersection of the cloaks of $\rho_1$ and $\rho_2$;
\item \label{item:cloak:differential} any linear differential of $\rho_1(\vec{r})$ with respect to either $\vec{r}$ or any parameters (or both);
\item \label{item:cloak:convolve} a convolution of $\rho_1(\vec{r})$ with $f(\vec{r})$;
\item \label{item:cloak:convolve:rho} a convolution of $\rho_1(\vec{r})$ with $\rho_2(\vec{r})$, whereby the cloak extends to the union of the cloaks of $\rho_1$ and $\rho_2$.
\end{enumerate}

The high degree of degeneracy in the triaxial deprojection problem suggests to approach the problem in a non-parametric fashion.  For the axisymmetric case, such code already exists \citep{Magorrian99} and has been successfully applied to many galaxies.  In the next section we present our triaxial extension of the axisymmetric code of M99.

\section{Non-parametric triaxial deprojection} \label{sec_triaxial_toymodel}

\subsection{Extension to the triaxial case} \label{sec_triaxial}
We start with a short overview of J.Magorrian's algorithm. The most significant points are:
\begin{itemize}
\item Both the SB and the intrinsic density are placed onto elliptical polar grids. In the axisymmetric case, a natural choice for the flattenings of the two grids is given by the inclination angle $i$ and by the relation\footnote{We use primes to denote coordinates and quantities defined in projection, i.e.\ on the sky.}
  \begin{equation}
    q' = \sqrt{q^2 \sin^2 i + \cos^2 i}, \,\, q = b/a
    \label{qp}
  \end{equation}
  where q' is the mean value of $1- \varepsilon$, $\varepsilon$ being the measured ellipticity.

\item The program minimizes a likelihood function
  \begin{equation}
    \mathcal{L} = -\frac{1}{2} \chi^2 + P
  \end{equation}
  where $\chi^{2}$ is given by
  \begin{equation} \chi^2 =
    \sum_{i=0}^{n_{m'}}\sum_{j=0}^{n_{\theta'}} \left( \frac{S_{ij} -
        \hat{S}_{ij}}{\Delta S_{ij}} \right)^2,
    \label{chisq}
  \end{equation}
  and $P$ is a penalty term used to penalize against unsmooth solutions or to drift the solution towards a certain shape. In equation~\eqref{chisq}, $S_{ij}$ and $\hat{S}_{ij}$ refer to the observed and the model SB, respectively, while $\Delta S_{ij}$ is the error coming from the observations. The grid has dimensions $n_{m'} \times n_{\theta'}$.
\item Using a Metropolis algorithm \citep{Metropolis53}, the program starts from an initial guess given by a double-power-law profile~\eqref{eq:double-power-law} to seek an intrinsic density projecting to a good fit to the observed SB profile.
\end{itemize}

In the triaxial case, some modifications are needed. First, we choose to represent the SB onto a grid of the form
\hspace{-2mm} % this is to prevent LaTeX from producing a VERTICAL gap to the equation below.
\begin{equation} \label{grid_sky}
    x'_{ij} = m'_i \cos\theta'_j,\quad
    y'_{ij} = \eta m'_i \sin\theta'_j
\end{equation}
where $\eta$ is used to flatten the grid along $y$ ($\eta < 1$), along $x$ ($\eta > 1$) or to keep it circular ($\eta = 1$). Typically we sample $m'_i$ with 50 points and $\theta'_j$ with 11 points from 0 to $\pi$.

The triaxial intrinsic density $\rho_{ijk} \equiv \rho(x_{ijk},y_{ijk},z_{ijk})$ is sampled onto an ellipsoidal grid of the form:
\begin{equation} \label{grid_intr}
  \begin{array}{l}
    x_{ijk} = m_i \sin\theta_j \cos\phi_k,\\
    y_{ijk} = P m_i \sin\theta_j \sin\phi_k, \\
    z_{ijk} = Q m_i \cos \theta_j.
  \end{array}
\end{equation}
Hereafter we define $\Rho \equiv \log\rho$.  The radial variable $m_i$ ranges the semi-minor axis of the innermost isophote to a few ($\sim$4) times the semi-major axis of the outermost isophote with typically 50 logarithmic steps, $\theta$ and $\phi$ go from $0$ to $\pi/2$ with 11 linearly spaced steps, and $P$, $Q$ are the two flattenings of the grid. $P$ and $Q$ can be chosen freely, their values do not influence the solutions discussed in Section~\ref{sec_semiparameteric}, but have an impact on the computing time needed to achieve the final solution.

\begin{figure}
    \centering
    \includegraphics[width=40mm]{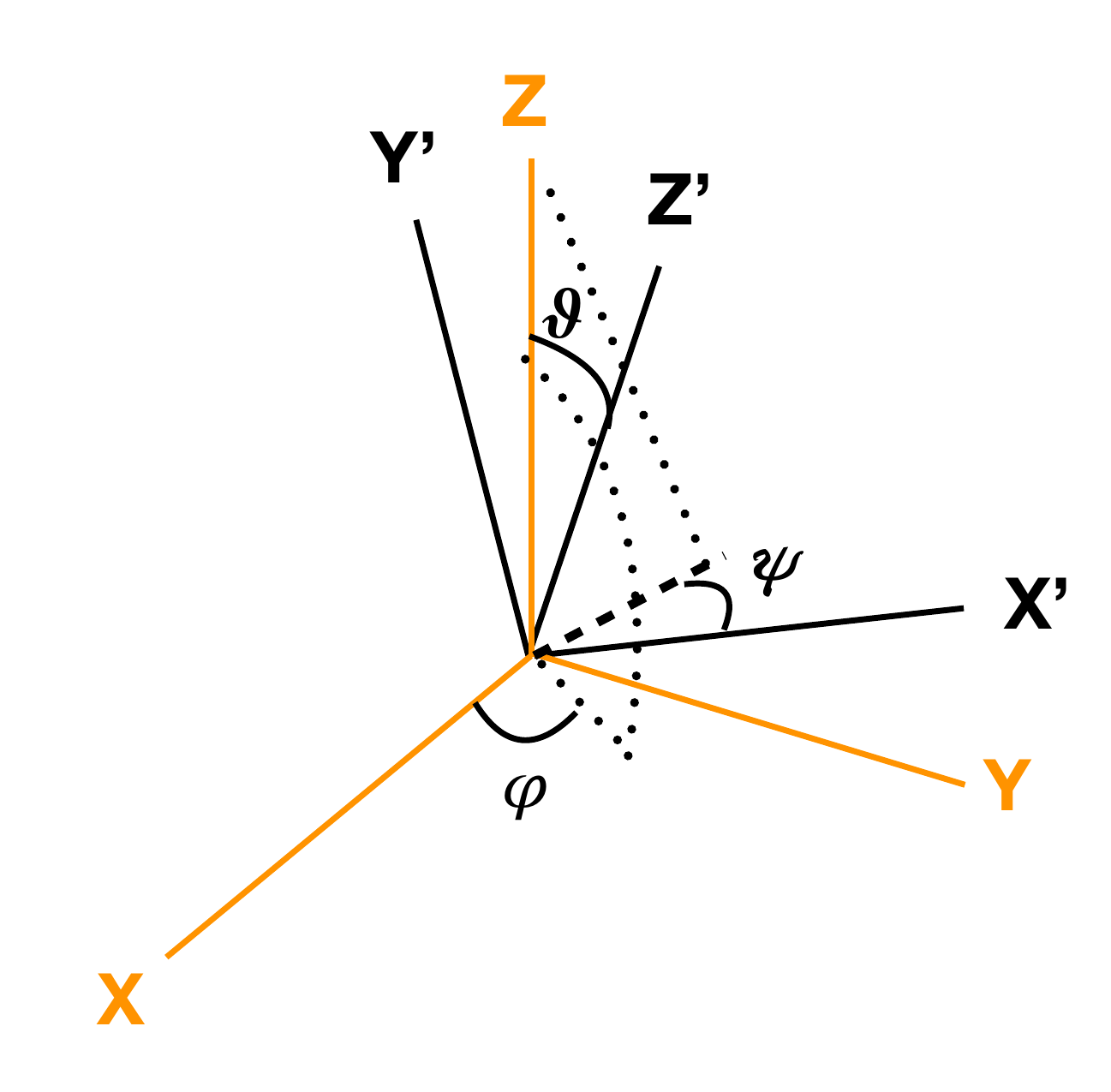}
    \caption{Geometric meaning of the viewing angles $\theta$ and $\phi$, which determine the LOS direction $\uvec{z}'$, and $\psi$, which is a rotation around the LOS itself.}
    \label{Fig.FOR}
\end{figure}

The two coordinate systems $(x',y',z')$ and $(x,y,z)$ are related by a rotation. Instead of using Euler angles, we follow the convention of \cite{Binney1985} and \cite{deZeeuwFranx1989} and use the polar coordinates $(\theta,\,\phi)$ of the LOS \eqref{eq:LOS} plus a rotation in the plane of the sky to parameterise this coordinate transform. Then
\begin{equation}
	\label{eq:projection}
  	\begin{pmatrix} x' \\ y' \\ z' \end{pmatrix}
		= \mat{R}_\psi \cdot \mat{P}
			\cdot \begin{pmatrix} x \\ y \\ z \end{pmatrix},
	\quad\text{with}\quad
		\mat{R}_\psi = 
		\begin{pmatrix} \sin\psi & -\cos\psi & 0 \\
						  \cos\psi & \phantom{-}\sin\psi & 0 \\
						 	0 & 0 & 1 \end{pmatrix}
\end{equation}
and the projection matrix (\citealt{deZeeuwFranx1989}, equation~3.2)
\begin{equation}
	\label{eq:matrix:P}
	\mat{P} = 
  	\begin{pmatrix}
		\phantom{\cos\theta}-\sin\phi & 
		\phantom{-\cos\theta}\cos\phi & 0 \\
		-\cos\phi\cos\theta & -\sin\phi\cos\theta & \sin\theta \\ 
		\phantom{-}\cos\phi\sin\theta &
		\phantom{-}\sin\phi\sin\theta & \cos\theta
	\end{pmatrix}.
\end{equation}
The inverse transform is simply $\vec{r}=\mat{P}^t\cdot\mat{R}_\psi^t\cdot\vec{r}'$. $\psi$ is the angle between the projection of the $z$-axis onto the sky and the $x'$-axis, measured counterclockwise, see also Fig.~\ref{Fig.FOR}.

In the axisymmetric case, the orientation of the SB major axis determines the appropriate value for $\psi$, while owing to axial symmetry the angle $\phi$ has no effect, so that only the inclination $i=\theta$ is of importance. Conversely, for the triaxial case all three viewing angles must be considered.

%%%%%%%%%%%%
%%%%%%%%%%%%%

\subsubsection{The penalty function}
\label{sec_penaltyfunction}
The penalty function consists of two terms. The first one 
\begin{align}
    P_{\mathrm{sm}} = & \phantom{+}\, \frac{\mathcal{C}}{\lambda_m} \sum_{i,j,k}
    \left[\frac{\Rho_{i+1,j,k} - 2 \Rho_{i,j,k} + \Rho_{i-1,j,k} } {\Delta \log m}\right]^2
    \nonumber \\ &+
    \frac{\mathcal{C}}{\lambda_\theta} \sum_{i,j,k} \left[\frac{\Rho_{i,j+1,k} - 2 \Rho_{i,j,k} + \Rho_{i,j-1,k} } {\Delta\theta}\right]^2
    \nonumber \\ &+
    \frac{\mathcal{C}}{\lambda_\phi} \sum_{i,j,k} \left[\frac{\Rho_{i,j,k+1} - 2 \Rho_{i,j,k} + \Rho_{i,j,k-1} } {\Delta\phi}\right]^2,
\label{smoothing_rho}
\end{align}
(with $\mathcal{C} = -\sqrt{2 n_{m'}n_{\theta'} }$) penalizes un-smooth solutions and extends equation~(9) of M99 to the triaxial case.  We use typically values for $\lambda_m$ between 0.5 and 1.2, which is up to an order of magnitude smaller than the default value $\lambda_m = 6$ for the axisymmetric code, while for $\lambda_{\theta}$ and $\lambda_{\phi}$ we usually adopt a value of 0.5 \citep{Magorrian99}.

The second term of the penalty function 
\begin{align}
    P_{\mathrm{nn}} = &\phantom{+}\, \frac{\mathcal{C}_{nn}}{n_m} \sum_{i,k} \max \left\{0,\left(\Rho_{i,2,k} - \Rho_{i,1,k} \right)\right\}^2
    	\nonumber \\ &+
    \frac{\mathcal{C}_{nn}}{n_m} \sum_{i,k} \max \left\{0,\left(\Rho_{i,n_{\theta},k} - \Rho_{i,n_{\theta} - 1,k} \right)\right\}^2
    	\nonumber \\ &+
    \frac{\mathcal{C}_{nn}}{n_m} \sum_{i,j} \max \left\{0,\left(\Rho_{i,j,2} - \Rho_{i,j,1} \right)\right\}^2
    	\nonumber \\ &+
    \frac{\mathcal{C}_{nn}}{n_m} \sum_{i,k} \max \left\{0,\left(\Rho_{i,j,n_{\phi}} - \Rho_{i,j,n_{\phi} - 1} \right)\right\}^2,
\label{Pnn}
\end{align}
(with $\mathcal{C}_{nn} = -4\mathcal{C}$) generalizes equation~(6) of M99 and penalizes models whose isocontours have negative $\partial \Rho / \partial \theta$ at $\theta = 0$ and $\pi/2$ (and the same for $\phi$).

The Metropolis algorithm works in the same way as in Magorrian's code. The problem here is that since we go up one dimension, there will be a significant larger amount of points that shall be modified by the code. For instance, in the axisymmetric case we sample $\log \rho$ on a $50 \times 11$ grid, while in the triaxial case we take a $50 \times 11 \times 11$ grid. Since in the Magorrian's code \emph{all} points of the SB grid are recomputed after each iteration, even a modest increase in the grid dimension leads to a significant increase in computational time. To speed up things, after the initial guess for $\Rho$ has been computed, we vary each $\Rho_{ijk}$ by a large factor, say 100, project the intrinsic density along the LOS and verify which points of the SB grid are actually varied by a factor larger than 0.1\%. We tested that by using this mapping on the axisymmetric code we can get a triaxial \texttt{Python} code nearly as fast as the axisymmetric \texttt{C} code of M99.

\subsubsection{Seeing Convolution}
\label{sec_psf}
When the distance from the galaxy centre is significantly larger than the resolution of the observations, one can neglect point-spread-function (PSF) effects; however, when studying the central regions of a galaxy, correct BH masses can be derived only when the BH sphere of influence is well resolved and the PSF effects are taken into account \citep{Rusli2013}.  In our code, we added the option to perform the PSF convolution at every step of the Metropolis before comparing the projection to the observations. Typical dimensions of the (non-parametric) PSF matrix we use are about $100\times100$, but can be adapted to the specific photometric data; the PSF is supposed to be sampled from $-3\sigma_{\mathrm{obs}}$ to $3\sigma_{\mathrm{obs}}$, where $\sigma_{\mathrm{obs}}$ is the seeing of the observations. The PSF convolution is by far is the most time-consuming step and is the only step that has been parallelized. We postpone a detailed discussion of this part to the code to upcoming first
applications to real galaxies.

%%%%%%%%%%%%%%
\begin{figure}
    \includegraphics[width=\columnwidth]{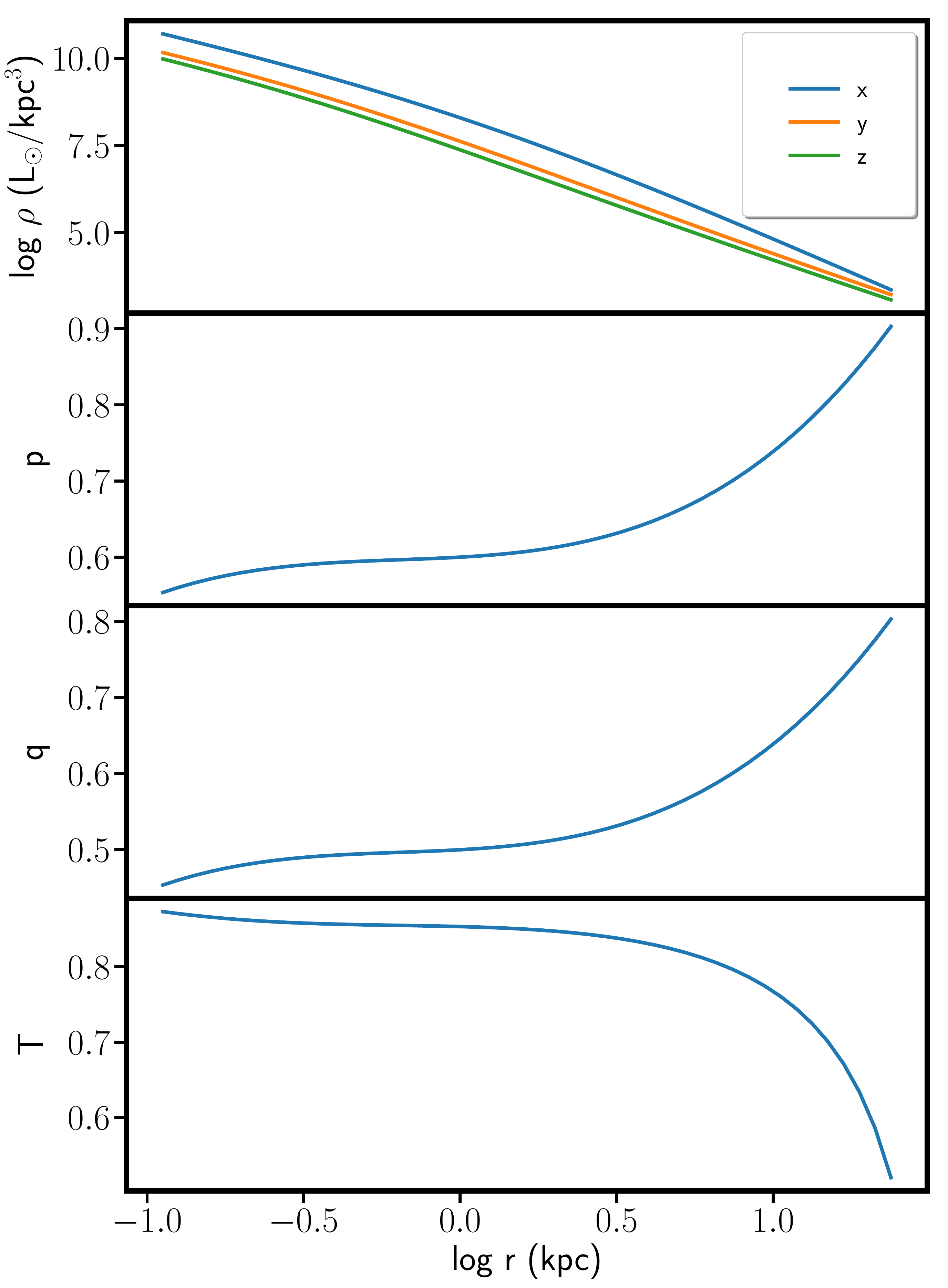}
    \vspace{-4mm}
    \caption{The \emph{ELLIP} model of Section~\ref{Sec.proj}, see also Table \ref{tab_models}. \emph{From top to bottom:} radial profiles of the density along the principal axes, $p$, $q$, and the triaxiality parameter $T$ (equation~\ref{eq:T}).}
    \label{Fig.toy}
\end{figure}
%%%%%%%%%%%%%%

\subsection{Exploring non-parametric triaxial deprojections}
\label{sec_toymodel}

\subsubsection{A benchmark model} 
\label{Sec.proj}
As a first step towards the testing of our deprojection algorithm, we consider as a benchmark a \citet{Jaffe83} model, which corresponds to the case $\alpha=2$ and $\beta=4$ of the double-power-law models \citep[][equation~2.64]{BinneyTremaine2008},
\begin{align}
	\label{eq:double-power-law}
	\rho(r) = \frac{\rho_0}{(r/s)^\alpha(1+r/s)^{\beta-\alpha}},
\end{align}
stratified on coaxial ellipsoids with specified radial profiles of the axis ratios. The values we chose for the total luminosity and the scale radius are $10^{10}\,L_{\odot}$ and $s=1\,$kpc, whereas the grids \eqref{grid_sky} and \eqref{grid_intr} have dimensions $30\times7$ and $50\times11\times11$, respectively. The SB grid extends from 0.1 to 10\,kpc with bin size $\sim0.16\,$dex, while the $\rho$ grid reaches out to 30 kpc and has bin size $\sim0.115\,$dex. We modelled the $p,\,q$ profiles to be cubic polynomials with coefficients such that $p$ increases from $\sim0.55$ to $\sim1$ from the innermost to the outermost density contour, while $q$ increases from $\sim0.45$ to $\sim0.8$, see Fig.~\ref{Fig.toy}. We also show the triaxiality parameter
\begin{equation}
	\label{eq:T}
	T = \frac{1 - p(r)^2}{1 - q(r)^2}.
\end{equation}
\citep{BinneyTremaine2008}. This model, hereafter referred to as \emph{ELLIP}, appears in several figures, listed in Table~\ref{tab_models} (on page~\pageref{tab_models}), which summarizes all models considered in this study. Although the $p,\,q,\,T$ profiles of this model are probably not representative of bright ellipticals, their mean values $\langle q \rangle \approx 0.6$ and $\langle T \rangle \approx 0.7$ are in line with the observed ranges $0.6\lesssim q\lesssim0.8$ and $0.4\lesssim T\lesssim0.8$ \citep{Tremblay1996, Vincent2005,Weijmans2014,Foster2017,Ene2018}.

The SB of \emph{ELLIP} is placed onto a grid with $\eta = 0.8$ (equation~\ref{grid_sky}) while the flattenings of the $\rho$ grid are $P = 0.7$ and $Q = 0.6$ (equation~\ref{grid_intr}). These values for $\eta$, $P$, and $Q$ will be used throughout the paper for all tests using a Jaffe density profile. As M99, we first compare the analytic expression of the Jaffe profile with our numerical projection, getting an RMS of $\sim$0.03\%, good enough for our purposes since it is smaller than typical uncertainties. The RMS can be simply obtained by multiplying equation~\eqref{chisq} by $\Delta S_{ij}^2$ and dividing the result by the number of grid points before taking the square root.

Due to the existence of cloaked densities, it is not a surprise that our non-parametric deprojection algorithm reconstructs a variety of densities, depending on many factors, such as the random seeds values and the shape of grid imposed by the choice of $P$ and $Q$.

\begin{figure}
\includegraphics[width=\columnwidth]{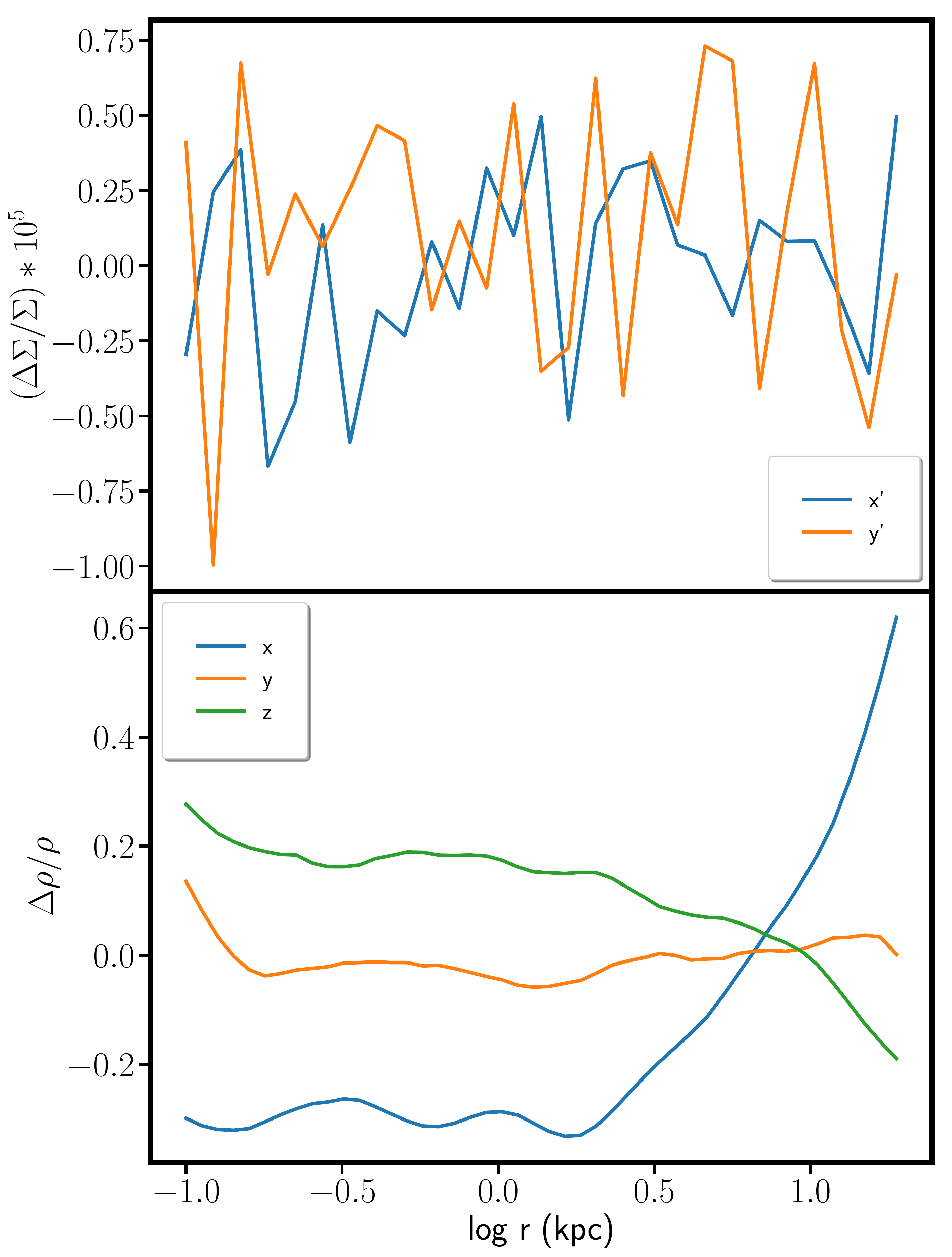}
    \vspace{-4mm}
    \caption{Relative differences between the true density along the principal axes (bottom) as well as SB (top) along the apparent major (blue) and minor (orange) axes of model \emph{ELLIP} projected for $\theta=\phi=\psi=45\degr$ and those obtained by our non-parametric deprojection. Although the fit to the observed SB is very good, the intrinsic density is far off the true value, a consequence of the non-uniqueness of triaxial deprojection.}
    \label{Fig.2}
\end{figure}

\subsubsection{An example of a cloaked density}
\label{sec_cloakdensities}

\begin{figure*}
    \includegraphics[height=68mm]{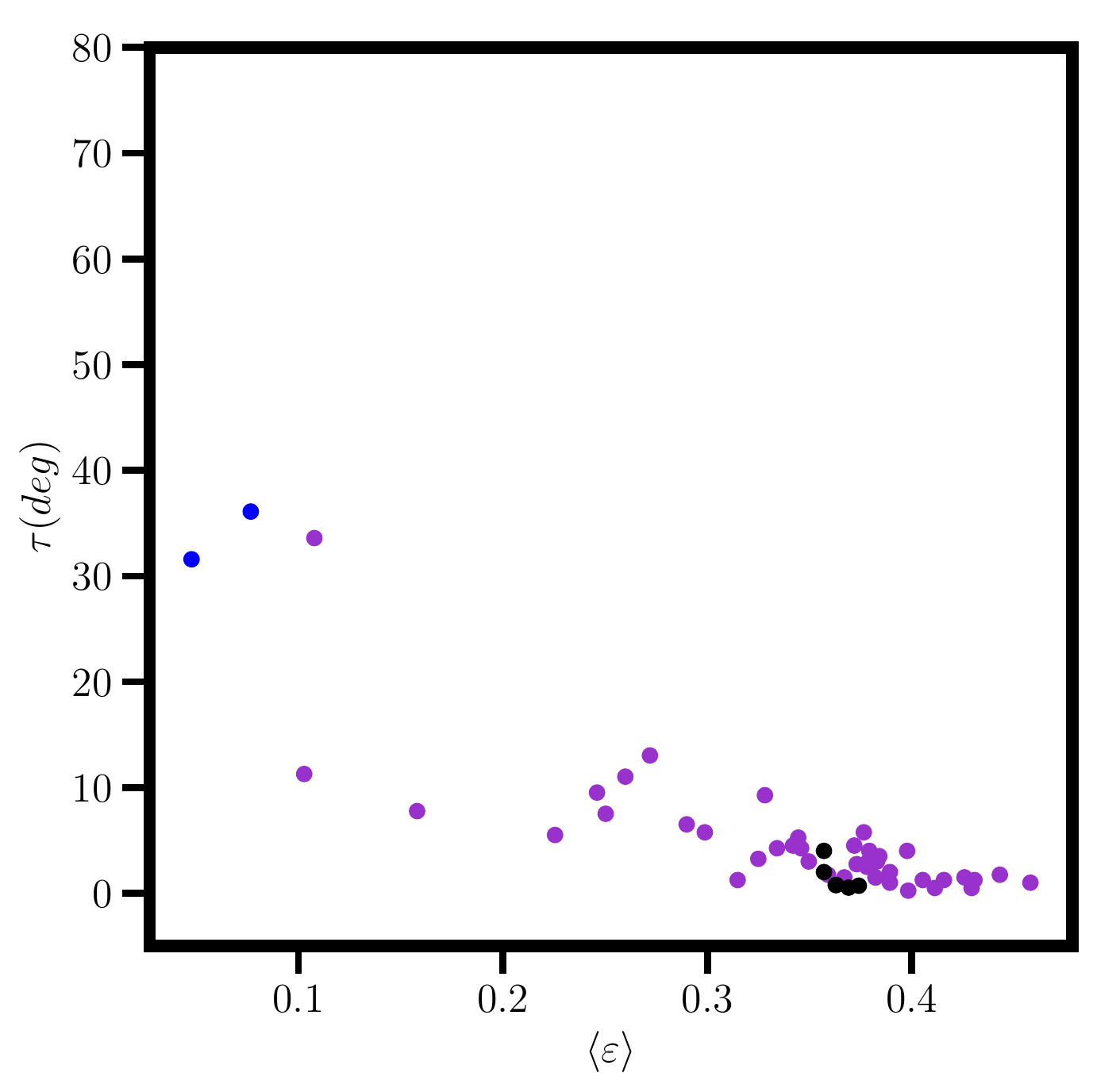}
    \hfil
    \includegraphics[height=68mm]{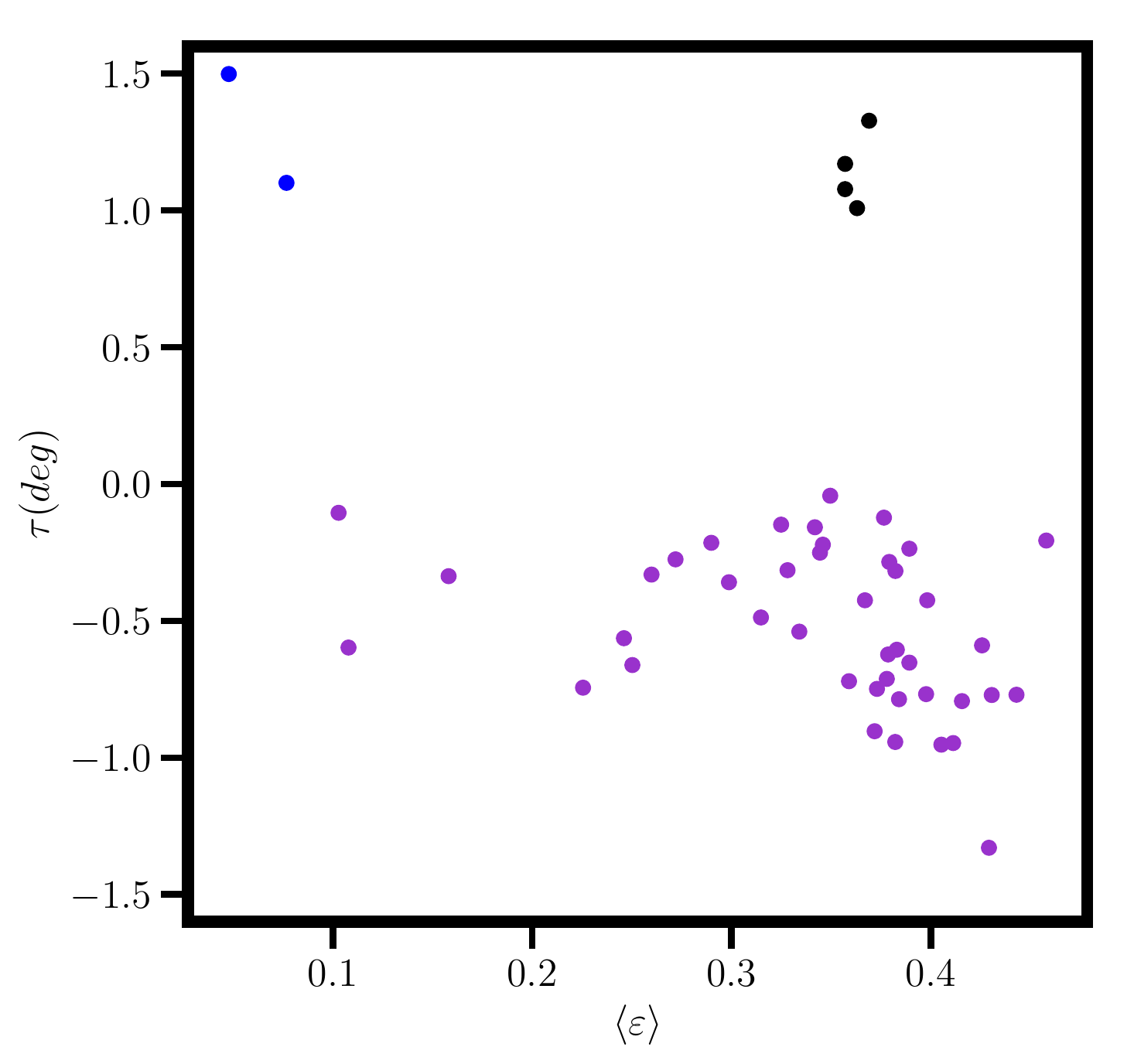}
    \caption{Correlations between mean ellipticity $\langle \varepsilon \rangle$ and twist angle $\tau$ (\emph{left}) and $\langle \varepsilon \rangle$ and mean $a_4$ (\emph{right}), when re-projecting an intrinsic density recovered by non-parametric deprojection for model \emph{ELLIP}. The discy isophotes ($\langle a_4 \rangle>0$) obtained for some viewing directions are very unusual for massive ellipticals. In the right panel, black points have $(\theta,\, \phi) \le (24,22)\degr$ (near the z-axis), while blue points have  $\theta\approx 70\degr$ and  $\phi\approx 6\degr$ (near the x-axis).}
      \label{Fig.isophote_correlations_np}
\end{figure*}

\begin{figure}
    \centering
    \includegraphics[width=\columnwidth]{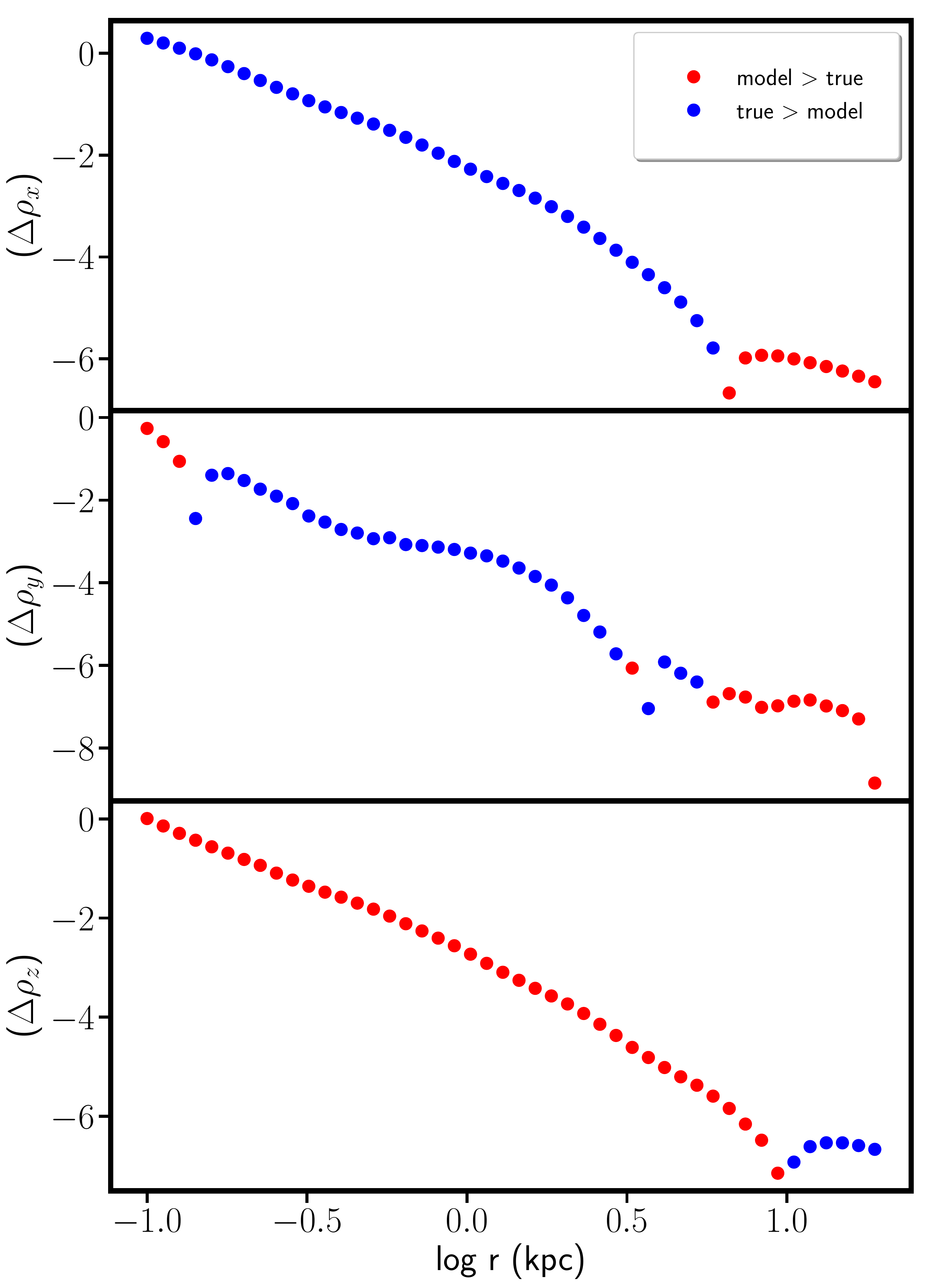}
    \vspace{-2mm}
    \caption{Radial profiles of the cloaked density hidden in the non-parametric deprojection of the \emph{ELLIP} model (see Fig. \ref{Fig.2}) along the three
      principal axes (\emph{top: major; middle:
      intermediate; bottom: minor}). Red/blue colours indicate that the recovered density is larger/smaller than
      the true density.}
    \label{Fig.konus}
\end{figure}

We illustrate this effect by deprojecting the projection of model \emph{ELLIP} for $\theta=\phi=\psi=45\degr$. In Section~\ref{sec_deprojectellipsoids} we show that a density that is stratified on ellipsoids, such as \emph{ELLIP}, admits a unique deprojection onto ellipsoids (unless it is viewed along one of the principal axes). But our non-parametric algorithm can find many more solutions (which are necessarily not stratified on ellipsoids), for whatever choice of viewing angles. Fig.~\ref{Fig.2}, bottom, shows the percentage differences between one of these solutions obtained using the true viewing angles $(\theta=\phi=\psi=45\degr)$ and the true density along the three principal axes. In projection this model agrees with the true SB to a striking 0.0007\% (Fig.\tild\ref{Fig.2}, top), but differs from the true space density by up to 60\% (Fig.\tild\ref{Fig.2}, bottom). Moreover, it is physically plausible, when compared to the properties of low and high luminosity ellipticals. Re-projecting it along a variety~ of viewing angles generates SBs with ellipticity lower than 0.5, twists\footnote{We define the \emph{twist angle} $\tau$ as the maximal variation across the position angles of the isophote major axis.}  $\tau$ exceeding 10$\degr$ only for low ellipticities (Fig.~\ref{Fig.isophote_correlations_np}, left) and higher-order shape coefficients $a_4$ (Fig.~\ref{Fig.isophote_correlations_np}, right) and $a_6$\footnote{Throughout this study, we adopt the definition of \citet{Bender87} for the isophote shape coefficients, normalizing them to the major axis value $a$ as $a_n \equiv a_n/a \times 100$.} spanning the range observed in discy or boxy ellipticals \citet{Bender88,Bender89}.  However, for a few viewing directions (close to the $z$- or $x$ axes: black and blue points in Fig.~\ref{Fig.isophote_correlations_np}, right) the isophotes are discy, which would rule out such a model for massive ellipticals, which are only observed with elliptical and boxy isophotes (see
discussion in Section~\ref{sec_range}).

\begin{figure*}
    \subfloat[($x,y$) plane. \label{Fig.konus_contour_xy}]
    	{\includegraphics[width=.33\linewidth,trim=1mm 1mm 1mm 1mm]{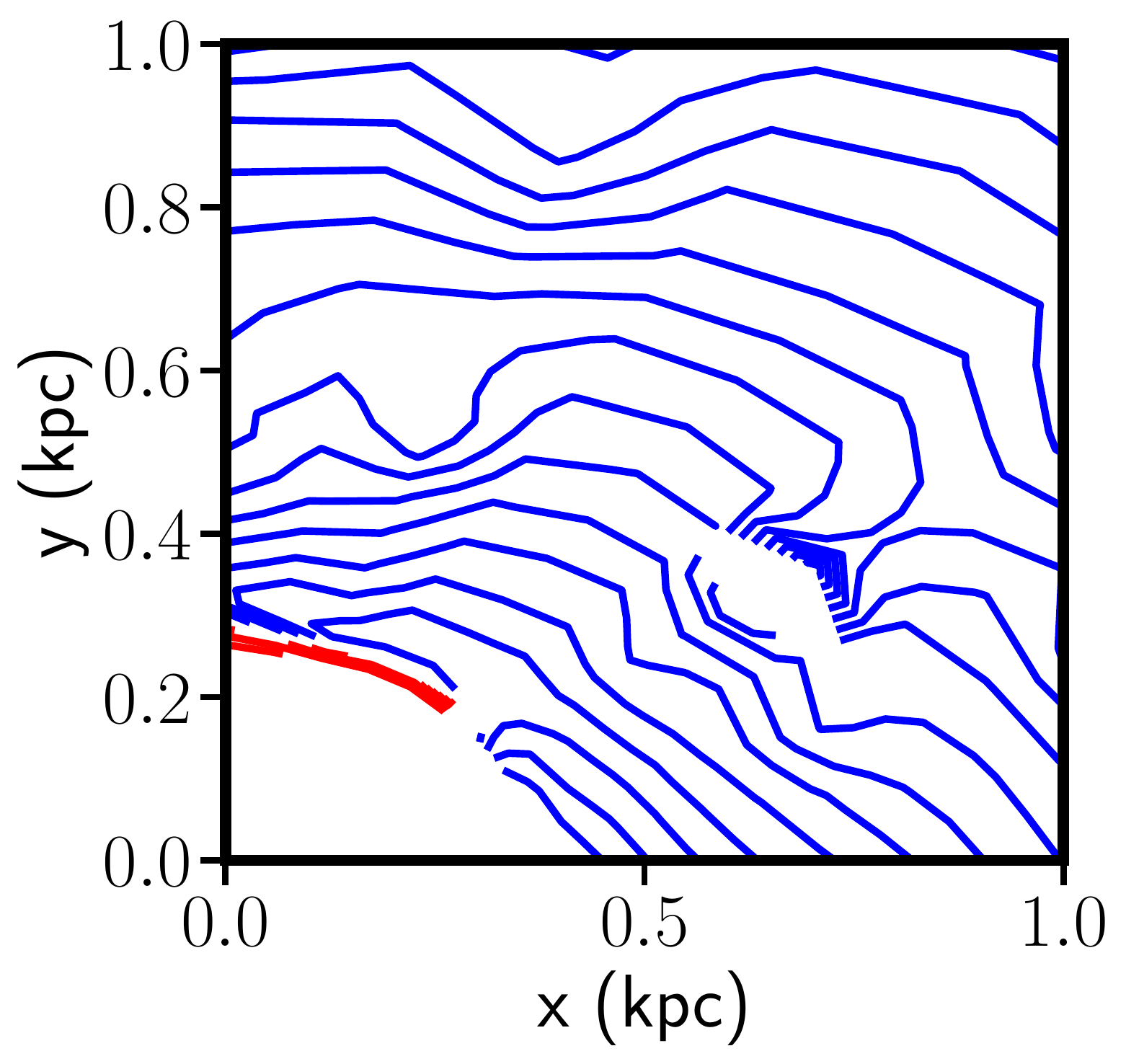}} \hfil
    \subfloat[($x,z$) plane. \label{Fig.konus_contour_xz}]
    	{\includegraphics[width=.33\linewidth,trim=1mm 1mm 1mm 1mm]{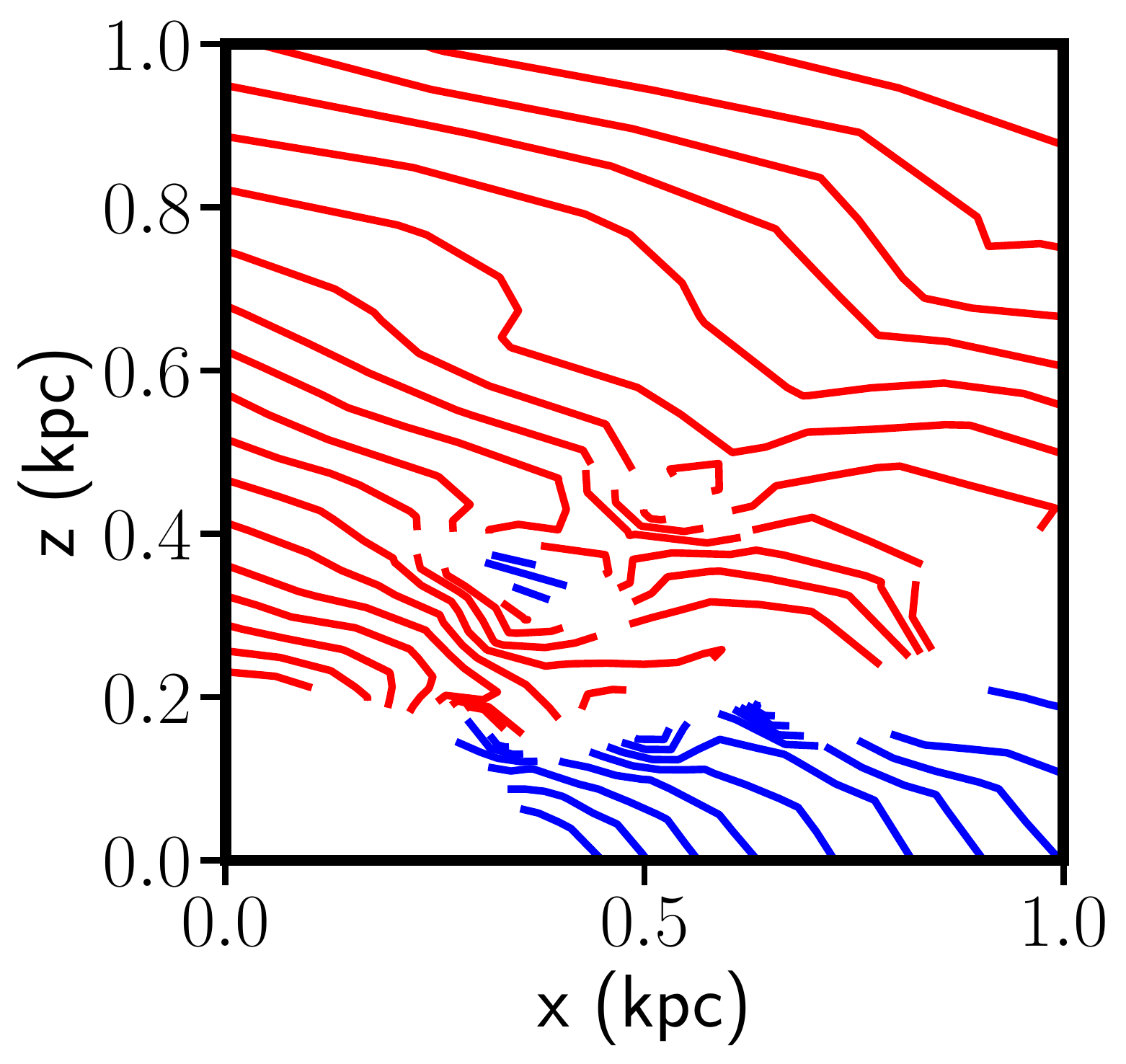}} \hfil
	\subfloat[($y,z$) plane. \label{Fig.konus_contour_yz}]
    	{\includegraphics[width=.33\linewidth,trim=1mm 1mm 1mm 1mm]{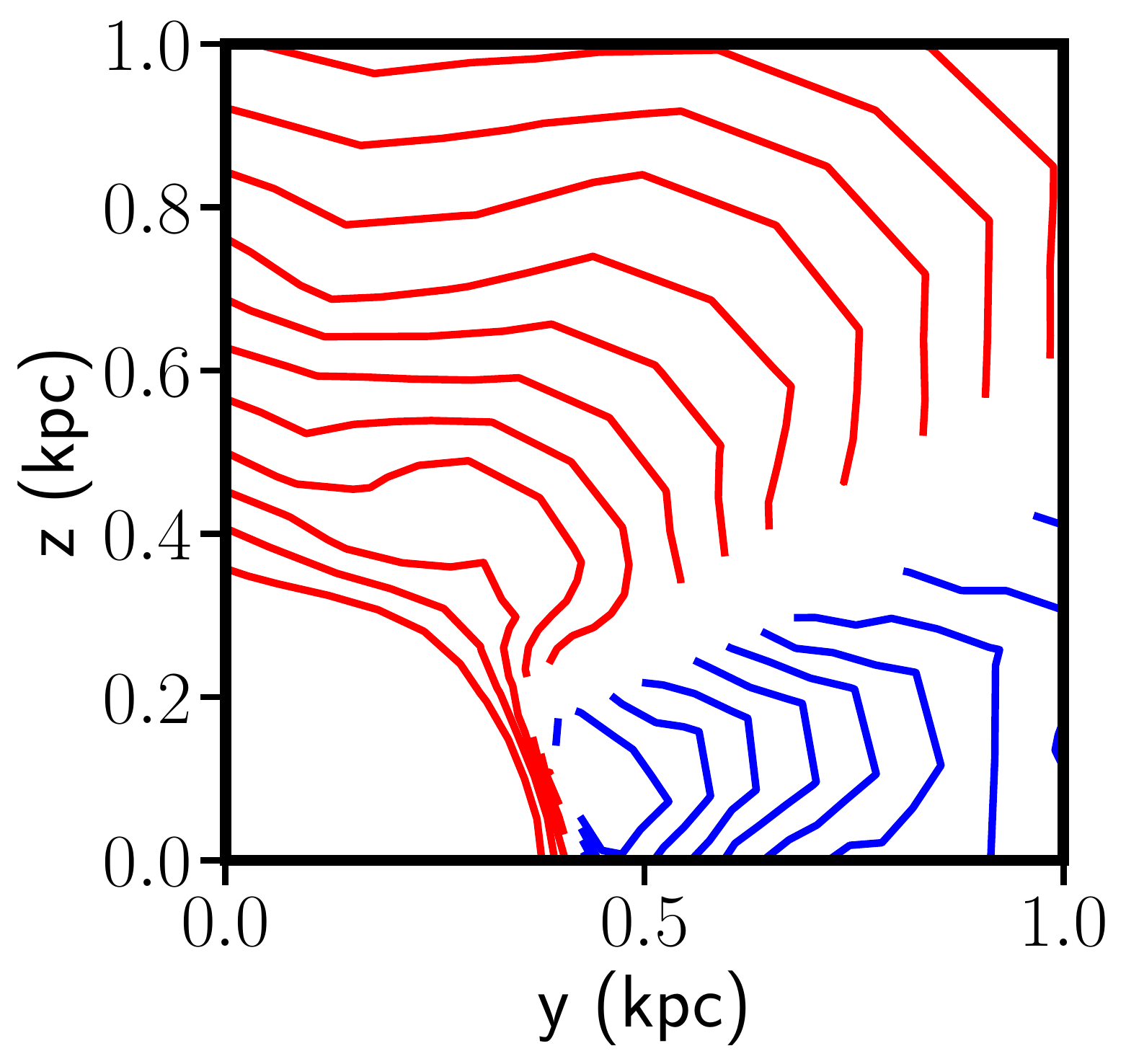}}
    \caption{Contours of the cloaked density shown in Fig. \ref{Fig.konus} on the (x,y) plane (\emph{left}), (x,z) plane (\emph{middle}), and (y,z) plane (\emph{right}). The contours line are red (blue) where the recovered density is larger (smaller) than the true.}
    \label{Fig.konus_contour}
\end{figure*}

Figs.~\ref{Fig.konus} \& \ref{Fig.konus_contour} show the actual cloaked density, i.e. the difference between the true density and the deprojection solution. It is a flattened structure almost orthogonal to the $z$-axis with negative density at low $z$, reminiscent of a (reversed) disc, causing discy isophotes in projection when seen near the $z-$axis.  A qualitative equivalent of the bottom plot of Fig. \ref{Fig.konus} (left) and the middle plot of Fig. \ref{Fig.konus_contour} is given in Fig. \ref{fig:cloackexample} for one of the possible analytical descriptions of cloacked densities discussed in Appendix \ref{sec_cloaked}

This example shows that, although the code does its job very well in producing a good fit to the observations, an efficient mechanisms is needed to suppress solutions that are unrealistic for massive ellipticals.

\section{Deprojection assuming approximately ellipsoidal isodensity contours}
\label{sec_deprojectellipsoids}

Although a non-parametric approach is desirable for exploring the broadest possible range of densities, it suffers from the large ambiguity in triaxial deprojections. Observationally, however, we know that the isophotes of massive elliptical galaxies do not deviate strongly from ellipses. This suggests, that the intrinsic density distributions of these galaxies are approximately ellipsoidal.  As we will show in this Section, the assumption of ellipsoidal density distributions makes the deprojection problem a lot more tractable.

\begin{figure*}
    \subfloat[$\xi = 0$, exact ellipsoids. \label{Fig.ell}]
        {\includegraphics[width=.33\linewidth,trim=1mm 1mm 1mm 1mm]{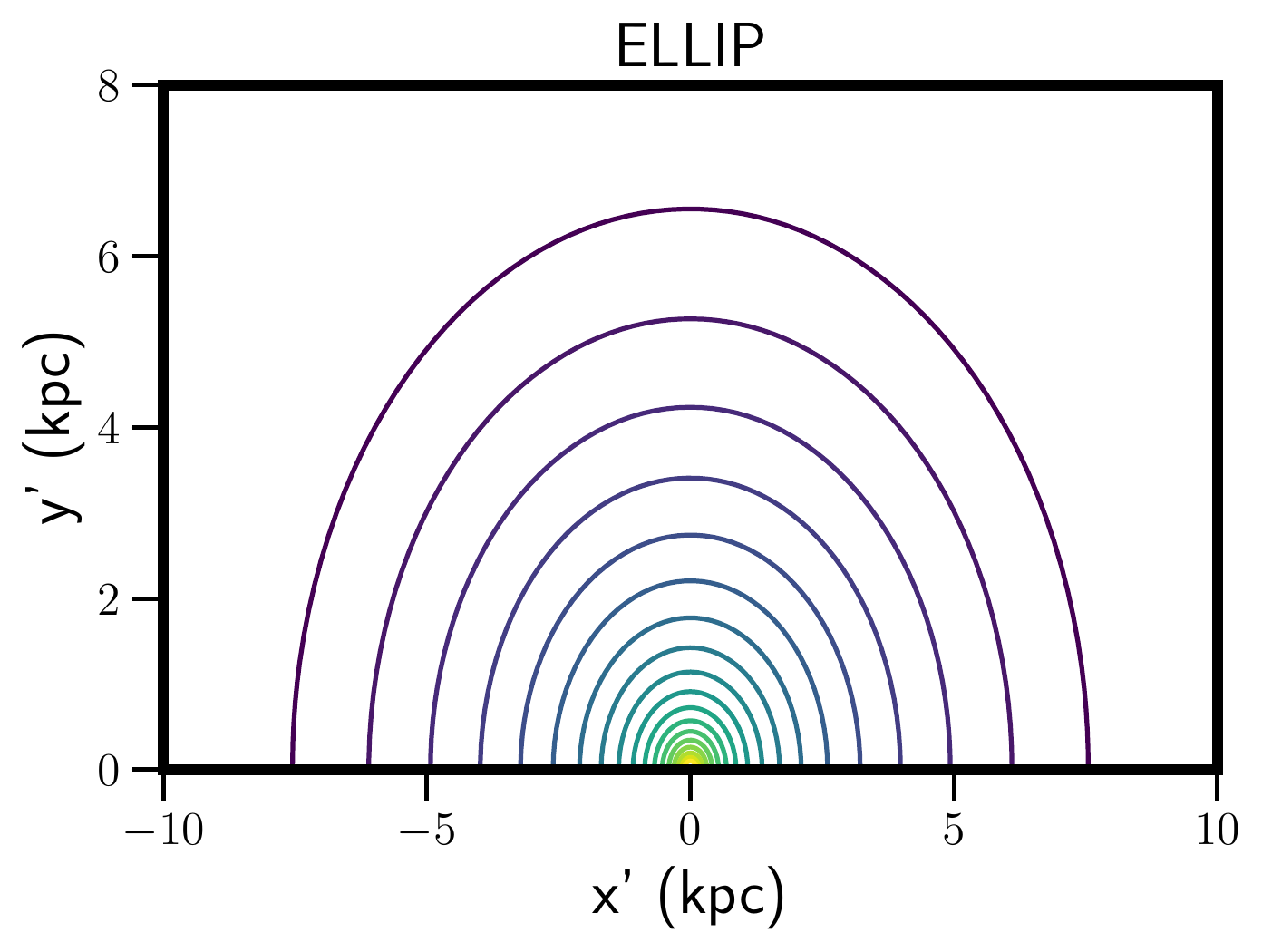}}
    \hfil
    \subfloat[$\xi = -0.5$, boxy bias. \label{Fig.boxy}]
	{\includegraphics[width=.33\linewidth,trim=1mm 1mm 1mm 1mm]{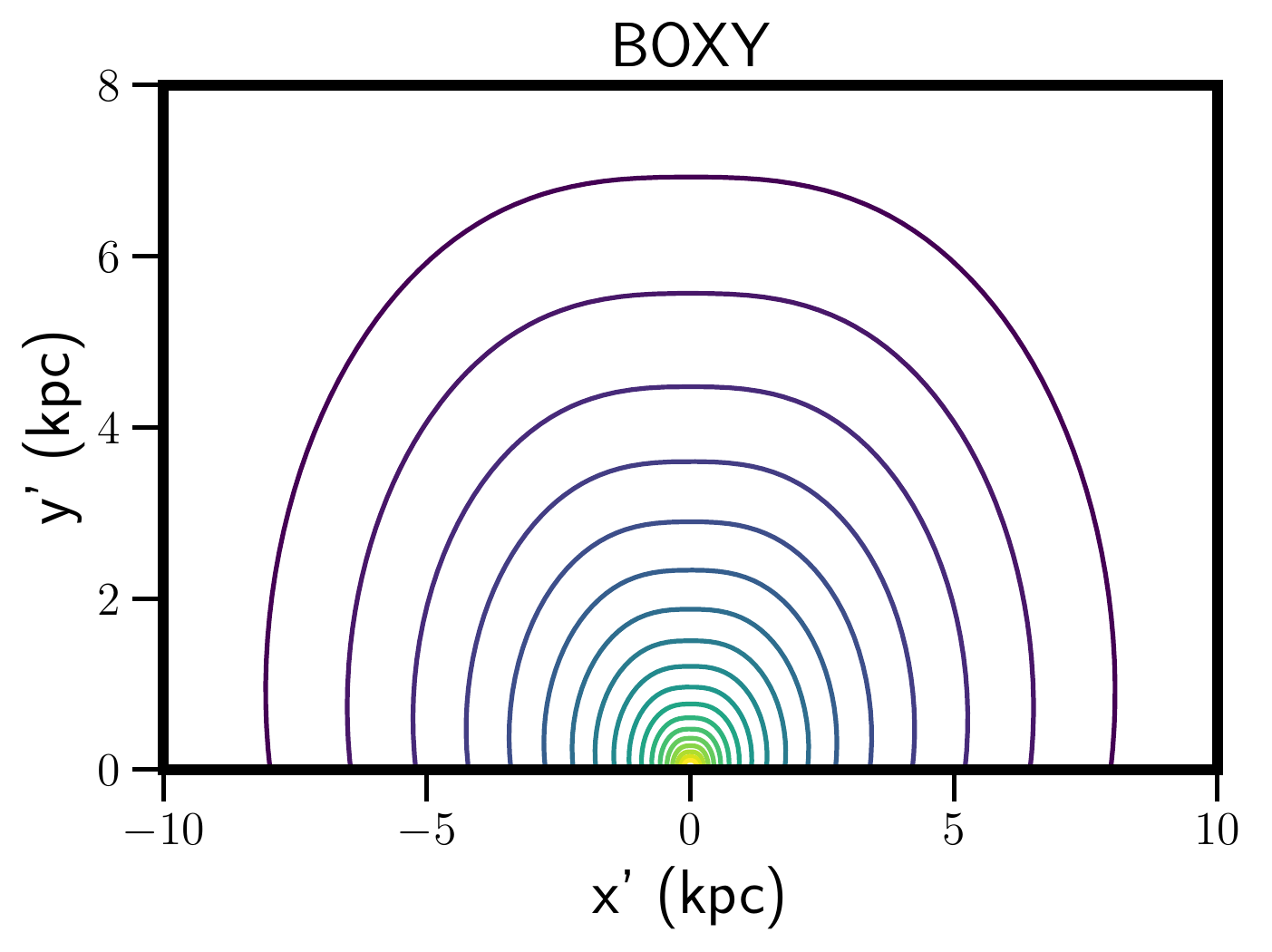}}
    \hfil
    \subfloat[$\xi = 0.3$, discy bias. \label{Fig.discy}]
	{\includegraphics[width=.33\linewidth,trim=1mm 1mm 1mm 1mm]{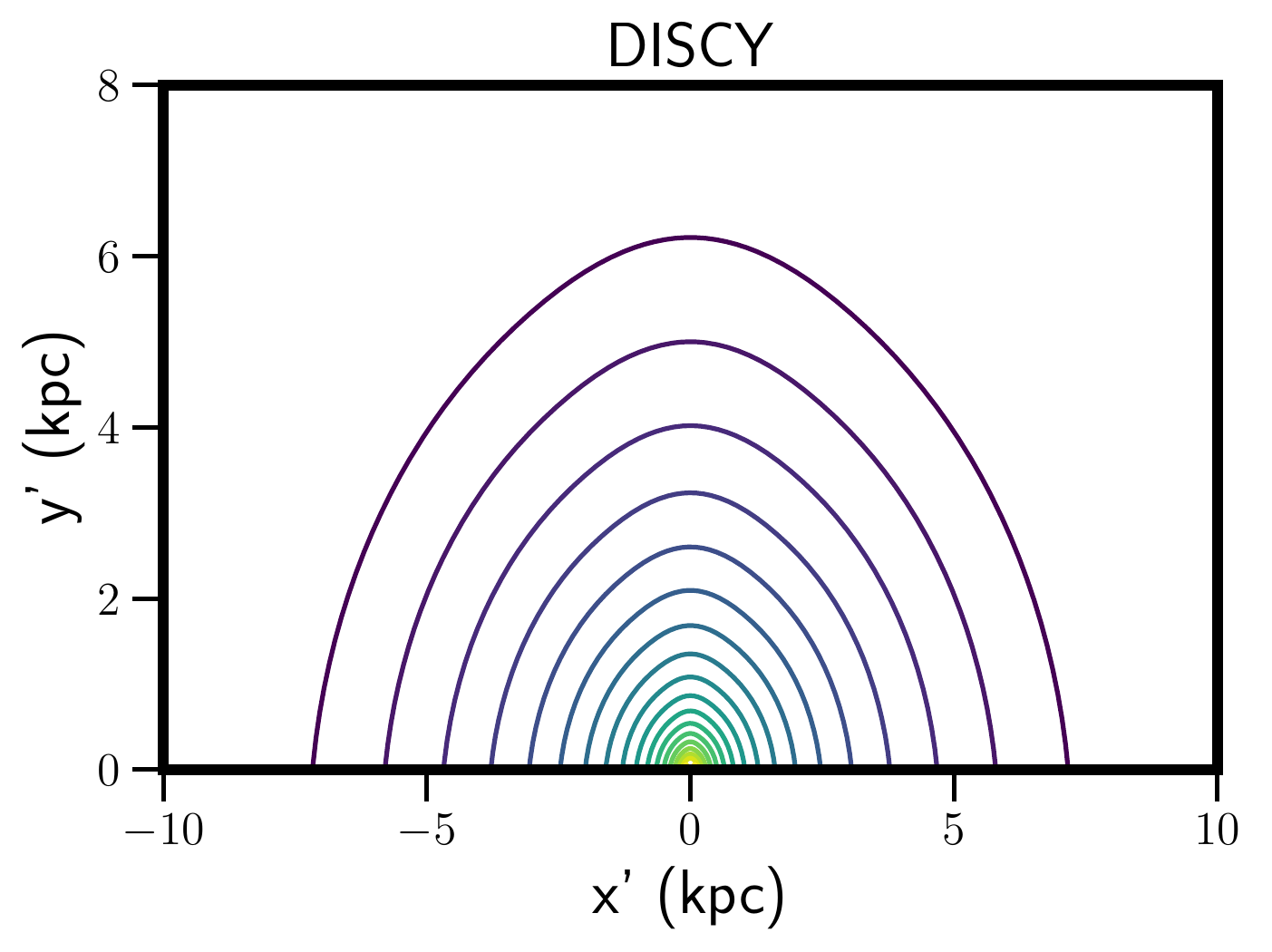}} 
    \caption{Jaffe models stratified on deformed ellipsoids \eqref{m} projected along the minor axis: model \emph{ELLIP} with $\xi=0$ (Fig.~\ref{Fig.ell}), model \emph{BOXY} with $\xi=-0.5$ (Fig.~\ref{Fig.boxy}) and model \emph{DISCY} with $\xi=0.3$ (Fig.~\ref{Fig.discy}).}
    \label{Fig.gamma_bias}
\end{figure*}

\subsection{Ellipsoidal projection and deprojection}
\label{sec:ellipsoidal}
First note that the projection
\begin{equation}
	\label{eq:sph:Sigma}
	\Sigma_s(R) = \int_{-\infty}^\infty\rho_s\Big(\!\sqrt{R^2+z^2}\Big)\diff z
	= 2\int_R^\infty\frac{ \rho_s(r)\,r\,\diff r}{\sqrt{r^2-R^2}}
\end{equation}
of a spherical galaxy with density $\rho_s(r)$ can (in principle) always be de-projected to obtain $\rho_s(r)$ via \citep[e.g.][problem 1.2]{BinneyTremaine2008}
\begin{equation}
    \label{eq:sph:Abel}
    \rho_s(r) = -\frac{1}{\pi}\int_r^\infty\frac{\diff R}{\sqrt{R^2-r^2}}
	\frac{\diff\Sigma_s}{\diff R}.
\end{equation}
Now, given parameters $a,\,b,\,c>0$, an ellipsoidal version of the galaxy has density
\begin{equation}
    \rho(\vec{r}) = (abc)^{-1} \rho_s(m)
\end{equation}
with
\begin{equation}
    m^2 = \frac{x^2}{a^2} + \frac{y^2}{b^2} + \frac{z^2}{c^2} = 
    \vec{r}^t\cdot\mat{C}^{-1}\cdot\vec{r},
\end{equation}
where $\mat{C}\equiv\mathrm{diag}(a^2,b^2,c^2)$. Complement the LOS direction $\vec{\ell}=\uvec{z}'$ with the two perpendicular unit vectors $\vec{\xi}$ and $\vec{\eta}$ spanning the sky, which we take to be the first and second rows of matrix $\mat{P}$~\eqref{eq:matrix:P}. The rotated coordinates are then
\begin{align}
    \tilde{\vec{r}} \equiv (\xi,\eta,\ell)^t = \mat{P}\cdot\vec{r}
\end{align}
and the projection integral becomes
\begin{align}
    \label{eq:ell:Sigma}
    \Sigma(\xi,\eta) = \int\diff\ell\,\rho\left(\sqrt{\tilde{\vec{r}}^t
        \cdot\tilde{\mat{C}}^{-1}\cdot\tilde{\vec{r}}}\right),
\end{align}
where $\tilde{\mat{C}}\equiv\mat{P}\cdot\mat{C}\cdot\mat{P}^t$. Using the Fourier slice theorem\footnote{The Fourier transform of $\rho$ is
\begin{align}
    \textstyle
    \hat\rho(\vec{k}) = &
    \int\rho(\vec{r})\,\Exp{-i\vec{k}\cdot\vec{r}}\,\diff\vec{r} 
    = \int\rho_s(m)\,\Exp{-i\vec{\kappa}\cdot\vec{m}}\,\diff\vec{m} = \hat{\rho}_s(\kappa)
    \nonumber \\
    & \qquad\text{with}\qquad
    \vec{m}\equiv\mat{C}^{-1/2}\cdot\vec{r}
	%=\left(\frac{x}{a},\frac{y}{b},\frac{z}{c}\right),
	,\quad
	\vec{\kappa}\equiv\mat{C}^{1/2}\cdot\vec{k}.
	%=\left(ak_x,bk_y,ck_z\right)
\end{align}
 Thus, $\hat{\rho}(\vec{k})$ is an ellipsoidal function, but with axis ratios inverted from those in real-space. According to the Fourier slice theorem
\begin{equation}
     \hat{\Sigma}(k_\xi,k_\eta) = \hat{\rho}(\vec{k})\big|_{k_\ell=0}
	=  \hat{\rho}_s\left(\sqrt{(k_\xi,k_\eta)^t\cdot\bar{\mat{C}}\cdot(k_\xi,k_\eta)}\right),
\end{equation}
where the $2\times2$ matrix $\bar{\mat{C}}$ is the $\xi$-$\eta$ part of $\tilde{\mat{C}}$, and equation~\eqref{eq:ellipsoidal:proj} follows.}  $\Sigma$ can be expressed in terms of the spherical projection as
\begin{equation}
    \label{eq:ellipsoidal:proj}
    \Sigma(\xi,\eta) = \left|\bar{\mat{C}}\right|^{-1/2}
    \Sigma_s\Big((\xi,\eta)^t\cdot\bar{\mat{C}}^{-1}\cdot(\xi,\eta)\Big),\end{equation}
where the $2\times2$ matrix $\bar{\mat{C}}$ is the $\xi$-$\eta$ part of $\tilde{\mat{C}}$ with components
\begin{subequations}
	\vspace{-3ex}
	\label{eq:bar:C:comp}
\begin{align}
  \bar{\s{C}}_{\xi\xi}   &= a^2\sin^2\phi+b^2\cos^2\phi, \\
  \bar{\s{C}}_{\xi\eta}  &= (a^2-b^2)\sin\phi\cos\phi\cos\theta, \\
  \bar{\s{C}}_{\eta\eta} &= (a^2\cos^2\phi+b^2\sin^2\phi)\cos^2\theta+c^2\sin^2\theta
\end{align}
\end{subequations}
and
\begin{align}
	\label{eq:det:Cbar}
	|\bar{\mat{C}}|
	&= |\mat{C}|\,(\vec{\ell}^t\cdot\mat{C}^{-1}\cdot\vec{\ell})
	\nonumber \\
	&= a^2b^2\cos^2\theta + (a^2\sin^2\phi+b^2\cos^2\phi)c^2\sin^2\theta.
\end{align}
Thus, \textbf{the projection of an ellipsoidal density is elliptic}. Moreover if the LOS is not within any of the fundamental planes, then the parameters of the matrix $\mat{C}$ can be recovered from $\bar{\mat{C}}$: an elliptic surface density has a unique ellipsoidal density\footnote{This is the basis of the \textbf{multi-Gaussian deprojection} method, when $\Sigma(\xi,\eta)$ is decomposed into a superposition of elliptical Gaussian surface densities, each of which is then deprojected into the corresponding unique ellipsoidal Gaussian density.}. In contrast, for $\theta=0$ (the projection along the minor axis) $\bar{\mat{C}}$ does not depend on $c$; for $\theta=\pi/2$ and $\phi=0$ (the projection along the major axis) $\bar{\mat{C}}$ does not depend on $a$; for $\theta=\pi/2$ and $\phi=\pi/2$ (the projection along the intermediate axis) $\bar{\mat{C}}$ does not depend on $b$.  Therefore, in these cases the deprojection of an elliptical SB onto an ellipsoidal density is not unique.

\subsection{Developing the algorithm}
\label{sec_semiparameteric}

Exploiting this result, we modified the fully non-parametric code in order to find the best solution on shells of a given shape (ellipsoids with possible boxy or discy deformations). Instead of searching for the density values on the three dimensional grid, we assume that four one-dimensional functions $\rho(x)$, $p(x)$, $q(x)$, and $\xi(x)$ of the distance $x$ along the major axis describe at every point the density $\rho(x,y,z)$ to be stratified on shells of the form:
\begin{equation}
    m^{2-\xi(x)} = x^{2-\xi(x)} + \left[\frac{y}{p(x)}\right]^{2-\xi(x)}
    							+ \left[\frac{z}{q(x)}\right]^{2-\xi(x)}.
    \label{m}
\end{equation}
$\xi=0$ obtains perfect ellipsoids, while $\xi>0$ and $\xi<0$ give discy and boxy deformations, respectively. Fig.~\ref{Fig.gamma_bias} shows examples of Jaffe models with different values of $\xi$: the $\xi=0$ \emph{ELLIP} case (left) described in the last section, the model \emph{BOXY} with $\xi=-0.5$ (middle), and the model \emph{discy} (right) with $\xi=0.3$.

The algorithm makes random changes to the function values $\rho(m_i)$, $p(m_i)$, $q(m_i)$, and $\xi(m_i)$ on the radial grid $m_i$ introduced in Section~\ref{sec_triaxial}, and then uses linear interpolation (and extrapolation) in $\log r$ along each of the grid directions ($j,\,k$) to update the density values on the three-dimensional grid. Changes that result in an intersection of density shells \eqref{m} for any of the grid directions are rejected.

\begin{table*}
    \centering
	\caption{The models considered in this study and the figures where they feature.}
	\begin{tabular}{ccc}
Name & Property & Figures \\
\hline
\emph{ELLIP} & Jaffe with $\xi=0$   & \ref{Fig.toy}, \ref{Fig.gamma_bias} (left), \ref{Fig.contours}, \ref{Fig.isoshapes}, \ref{Fig.Jaffe_ell_45}\\
\emph{DISCY} & Jaffe with $\xi=0.3$ & \ref{Fig.gamma_bias} (middle)\\
\emph{BOXY}  & Jaffe with $\xi=-0.5$ & \ref{Fig.gamma_bias} (right), \ref{Fig.maps},\ref{Fig.isophote_correlations}\\
\emph{DISCYBOXY} & Jaffe with $\xi$ from 0.3 to -0.5 & \ref{Fig.contours}, \ref{Fig.isoshapes}, \ref{Fig.recovery}, \ref{Fig.variable_grid_gammapq}, \ref{Fig.Jaffe_gamma_45}, \ref{Fig.mine_mge}\\
\emph{PQCROSS} & Jaffe with $\xi=0$ and $p$ \& $q$ profiles crossing & \ref{Fig.twist_strong}\\
\emph{NBODY} & $N$-body model & \ref{Fig.bad_depros}, \ref{Fig.Nbody_45}\\
\emph{LARGEDISC} & 50\% Jaffe with $\xi=0$ plus 50\% disc (equation~\ref{eq.rho_disc}) & \ref{Fig.isophotes_disc}, \ref{Fig.radprofs_disc50}\\ 
\emph{SMALLDISC} & 85\% Jaffe with $\xi=0$ plus 15\% disc (equation~\ref{eq.rho_disc}) & \ref{Fig.isophotes_disc}, \ref{Fig.radprofs_disc15}, \ref{Fig.disc_maps}\\ 
\hline
	\end{tabular}
	\label{tab_models}
\end{table*}

\begin{table}
    \centering
	\caption{The correspondence between $p$ and $q$ values and position of the principal axes of a galaxy according to the convention used in our model.}
	\begin{tabular}{cccc}
$p$ and $q$ values& Major& Intermediate& Minor\\
\hline
$q < p < 1$& $x$& $y$& $z$\\
$p < q < 1$& $x$& $z$& $y$\\
$q < 1 < p$& $y$& $x$& $z$\\
$1 < q < p$& $y$& $z$& $x$\\
$p < 1 < q$& $z$& $x$& $y$\\
$1 < p < q$& $z$& $y$& $x$\\
\hline
	\end{tabular}
	\label{tab_axis}
\end{table}

The initialization follows closely Section~2.2 of M99: the SB is placed onto an elliptical polar grid and the initial guess for the intrinsic density is found by fitting an ellipsoidal double-power-law model~\eqref{eq:double-power-law} via the Levenberg-Marquardt algorithm with the axis ratios fixed to those of the grid ($p=P,\, q=Q$).

Then, for each Metropolis iteration the code randomly chooses one of $\Rho=\log \rho$, $p$, $q$ or $\xi$ as the variable $X$ to change and applies a random change to one of its elements picked at random. As in M99, the change is made by setting $X_l\leftarrow X_l+r \Delta X_l$, where $r\in[-1,1]$ is a uniformly distributed random number, while the $\Delta$ arrays determine the maximum possible change. Initially, we set $\Delta\Rho=0.5$, $\Delta\xi=0.1$ and $\Delta p=\Delta q=0.05$ for all $l$, but multiply (divide) an element by 1.5 after a change that was rejected (accepted).

Since the density contours are constrained to be deformed ellipsoids, the smoothness penalty function \eqref{smoothing_rho} is better expressed directly in terms of the four functions actually fitted. Therefore, we replace \eqref{smoothing_rho} with
\begin{align}
	\label{eq:P:SM:deformed:ellipsoids}
  P_{\mathrm{sm}} &=\frac{\mathcal{C}}{\lambda_{\!\rho}} \sum_{l} \left[\frac{\Rho_{l+1} - 2 \Rho_{l} + \Rho_{l-1}}{\Delta \log m}\right]^2 \!
    %\nonumber \\ &
    + \frac{\mathcal{C}}{\lambda_{\!p}} \sum_{l} \left[\frac{p_{l+1} - 2 p_{l} + p_{l-1}}{\Delta \log m}\right]^2
     \nonumber \\ &
    + \frac{\mathcal{C}}{\lambda_q} \sum_{l} \left[\frac{q_{l+1} - 2 q_{l} + q_{l-1}}{\Delta \log m}\right]^2 \!
    %\nonumber \\ &
    + \frac{\mathcal{C}}{\lambda_\xi} \sum_{l} \left[\frac{\xi_{l+1} - 2 \xi_{l} + \xi_{l-1}}{\Delta \log m}\right]^2,
\end{align}
where typically $\lambda_{\rho} \sim 10 \lambda_{\xi} \sim 100 \lambda_{p,q}$, when the four terms in~\eqref{eq:P:SM:deformed:ellipsoids} are of comparable magnitude.

\begin{figure*}
	\subfloat[Maximum twist $\tau$. \label{Fig.twist_weak}]
		{\includegraphics[width=59mm,trim=1mm 2mm 7mm 1mm]{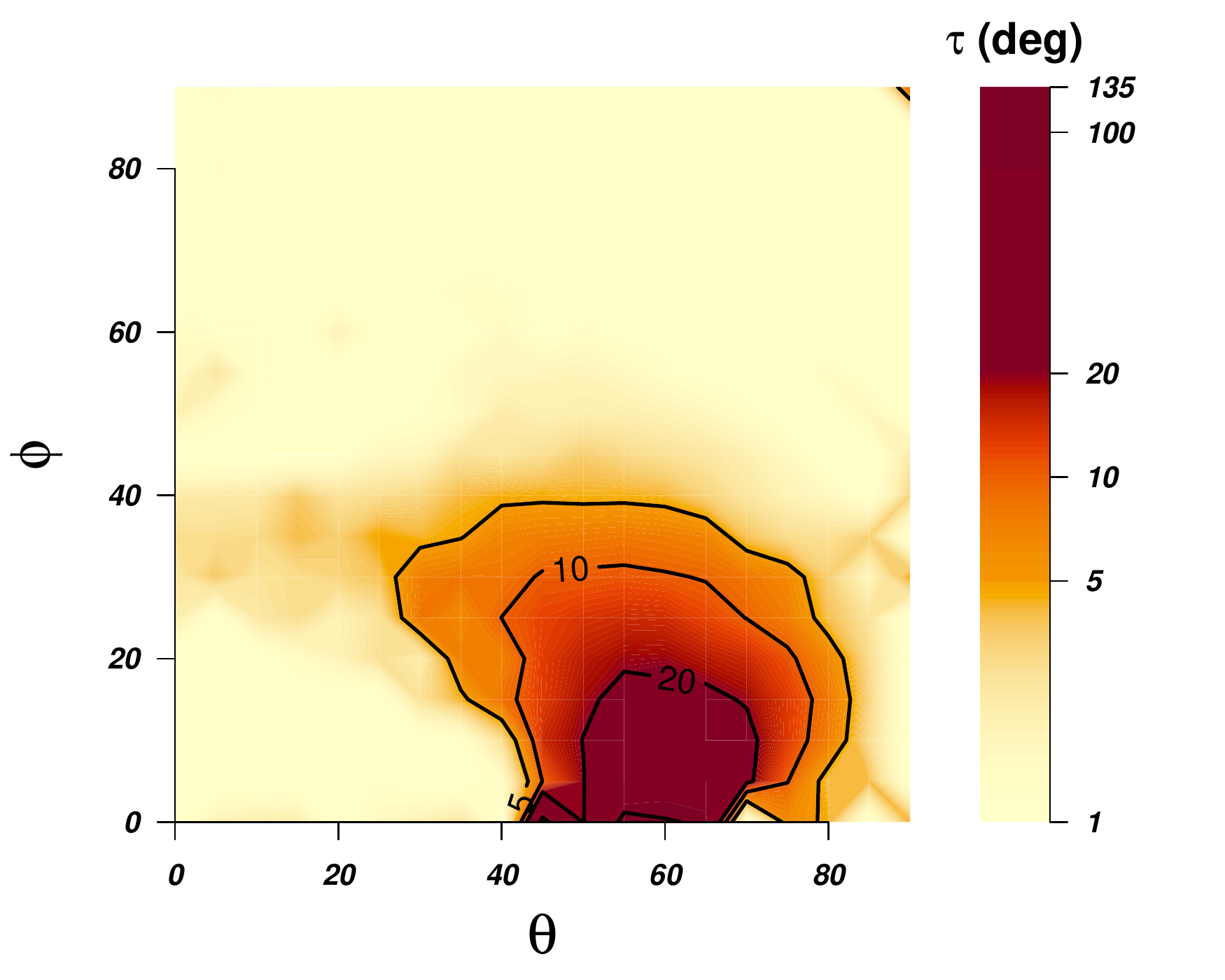}}
	\hfil
	\subfloat[Mean ellipticity $\langle\varepsilon\rangle$.
			\label{Fig.mean_ellipticity}]
		{\includegraphics[width=59mm,trim=1mm 2mm 7mm 1mm]{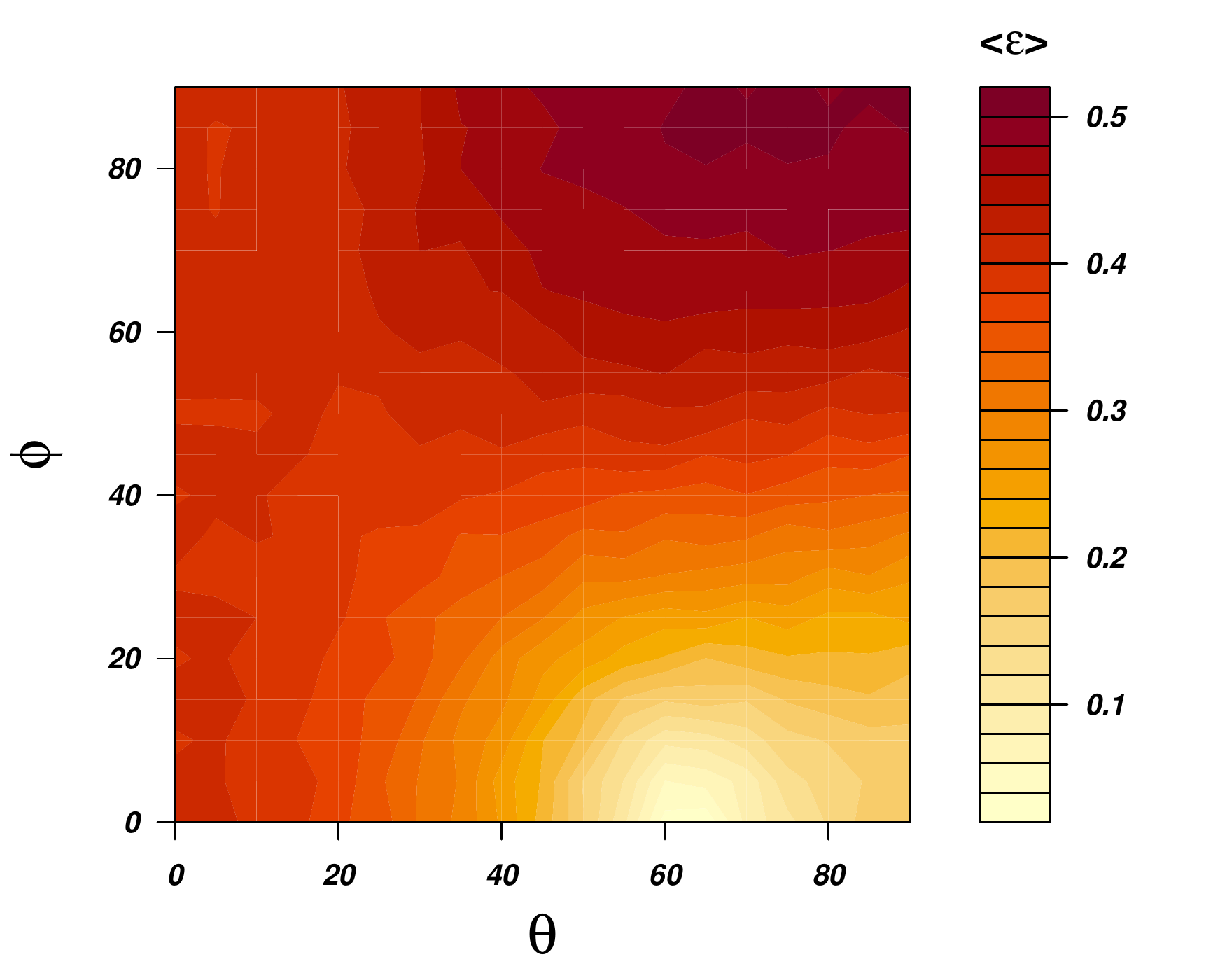}}
	\hfil
	\subfloat[Mean $a_4$ coefficient. \label{Fig.mean_a4}]
		{\includegraphics[width=59mm,trim=1mm 2mm 7mm 1mm]{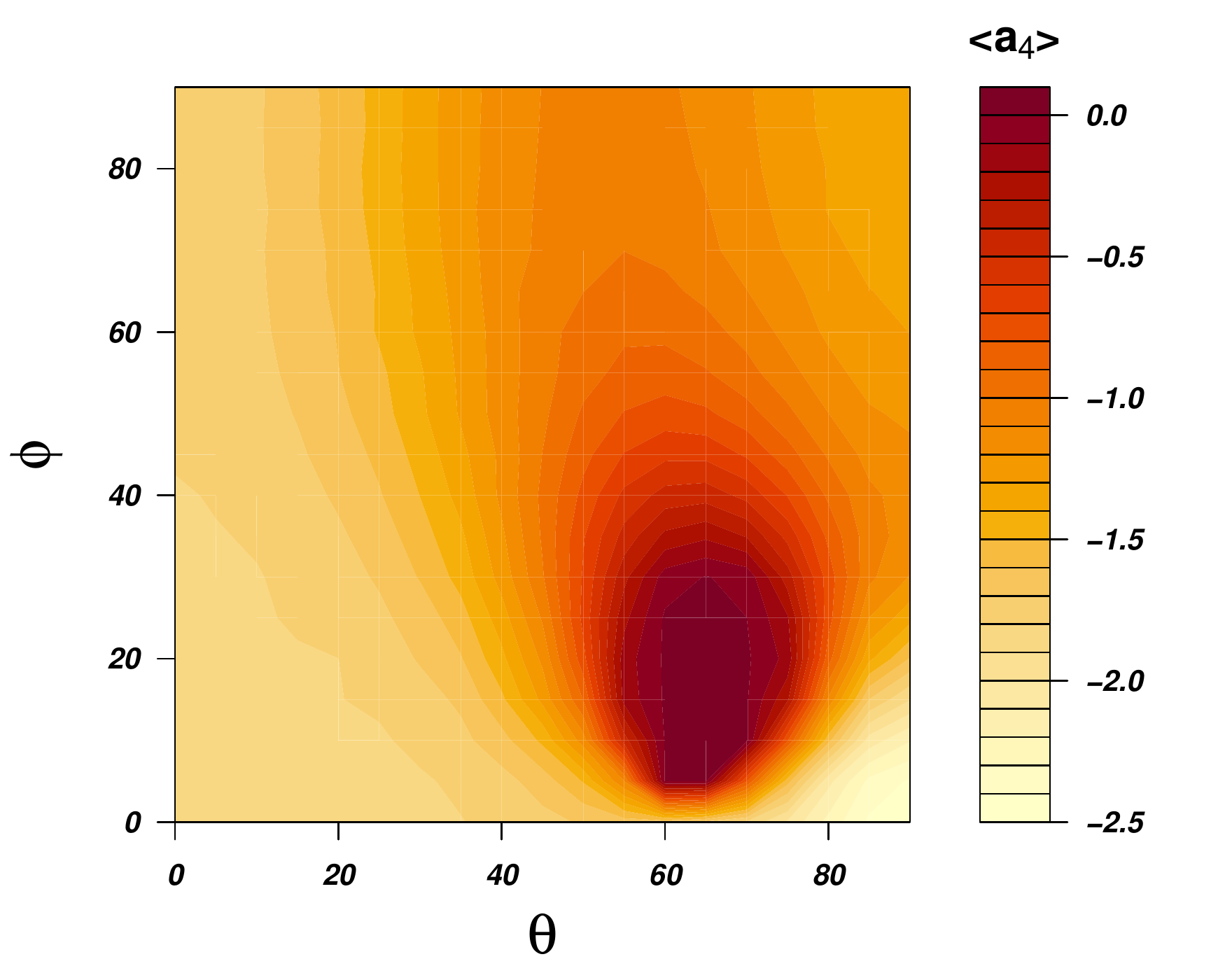}}
	\caption{Maps of the twist angle $\tau$, of the mean ellipticity $\langle\varepsilon\rangle$ and of the Fourier coefficient $a_4$ as a function of the projection angles $\theta$ and $\phi$ for model \emph{BOXY} described in Section~\ref{Sec.proj}, which has $\xi=-0.5$ (equation~\ref{m}). In (a) contours are drawn for twists of $5\degr$, $10\degr$, and $20\degr$. For octants not shown ($90\degr<\theta<180\degr$ and $90\degr<\phi<360\degr$) the results are identical by triaxial symmetry.} 
	\label{Fig.maps}
\end{figure*}

Since we cannot know which of the principal axes is major, minor or intermediate at any distance from the centre, we do not constrain the values of $p$ and $q$ a priori. Crossing $p$ and $q$ profiles result in changes of the relative axis ranking, summarized in Table~\ref{tab_axis}. As discussed in Section~\ref{Sec.proj}, some $p$ and $q$ profiles generate strong twists when re-projected and can be discarded a posteriori as unlikely.

\section{Tests of the ellipsoidal isodensity assumption for massive ellipticals}  \label{sec_range_tests}

In this section we argue that the deformed-ellipsoidal model described in the previous section when projected on the sky is able to match reasonably well the general properties of massive ellipticals (see e.g. \citealt{Foster2017, Goullaud18, KlugeEtAl2020}). Such galaxies have ellipticity distributed in the range [0,\,0.5] (unless sub-structures are present), tend to have slight boxy biases ($-1.5\lesssim a_4\lesssim0$) and not very large twists ($\tau\lesssim10\degr$). All these observables depend on the viewing angles $\theta$ and $\phi$ \emph{and} on the $p$, $q$ profiles (see e.g.\ equation~7-8 of \citealt{Cappellari02}), but not on $\psi$, since a rotation around the LOS rotates all isophotes by the same amount. Then, building on the result discussed in Section~\ref{sec:ellipsoidal} about the uniqueness of density deprojections stratified on ellipsoids, we will verify the performances of our numerical algorithm.

\subsection{The range of isophote shapes of triaxial elliptical galaxies}
\label{sec_range}

\subsubsection{Reproducing real massive ellipticals}
We consider the model \emph{BOXY}, similar to \emph{ELLIP} described in Fig.~\ref{Fig.toy}, but with $\xi=-0.5$, in order to reproduce the boxy bias observed in most massive ellipticals. We map the twist $\tau$, the mean ellipticity $\langle\varepsilon\rangle$ and the mean $a_4$ as functions of $\theta$ and $\phi$ (and assuming $\psi=45\degr$). The results are shown in Fig.~\ref{Fig.maps}. For most of the angles $\theta$ and $\phi$, the twist $\tau$ (Fig.~\ref{Fig.twist_weak}) is smaller than $5\degr$; larger values are obtained when observing the model between the intrinsic long and the intrinsic short axes. At these viewing directions, the compression of the short axis relative to the intermediate axes near the centre tends to elongate the isophotes along the projected direction of the (``longer'') intermediate axis. The compression of the short axis relative to the long axis in the outer parts likewise tends to elongate the isophotes, again along the direction of the ``longer'' axis. However, this time, the ``longer'' axis is the intrinsic long axis and therefore points to different projected direction. This gives rise to isophote twists, which depend on the exact profiles of $p(r)$ and $q(r)$. As pointed out in Section~\ref{Sec.proj}, the ones of the \emph{ELLIP} model are not necessarily representative of bright ellipticals, but illustrative of how twists can be generated. Moreover, their mean values match the observed ones reasonably well: the mean ellipticity (Fig.~\ref{Fig.mean_ellipticity}) is indeed in the range 0-0.5 and is roughly \emph{anti-correlated} with the twist, being lower when the twist is higher. As expected, it reaches the highest values for high $\theta$ and $\phi$, thus close to projections along the intermediate axis. Finally, the mean $a_4$ (Fig.~\ref{Fig.mean_a4}) spans the range $-2.5$-0.1, which is what we expected given the $\xi$ profile we have chosen.

\begin{figure*}
	\includegraphics[height=68mm,trim=1mm 1mm 2mm 2mm]{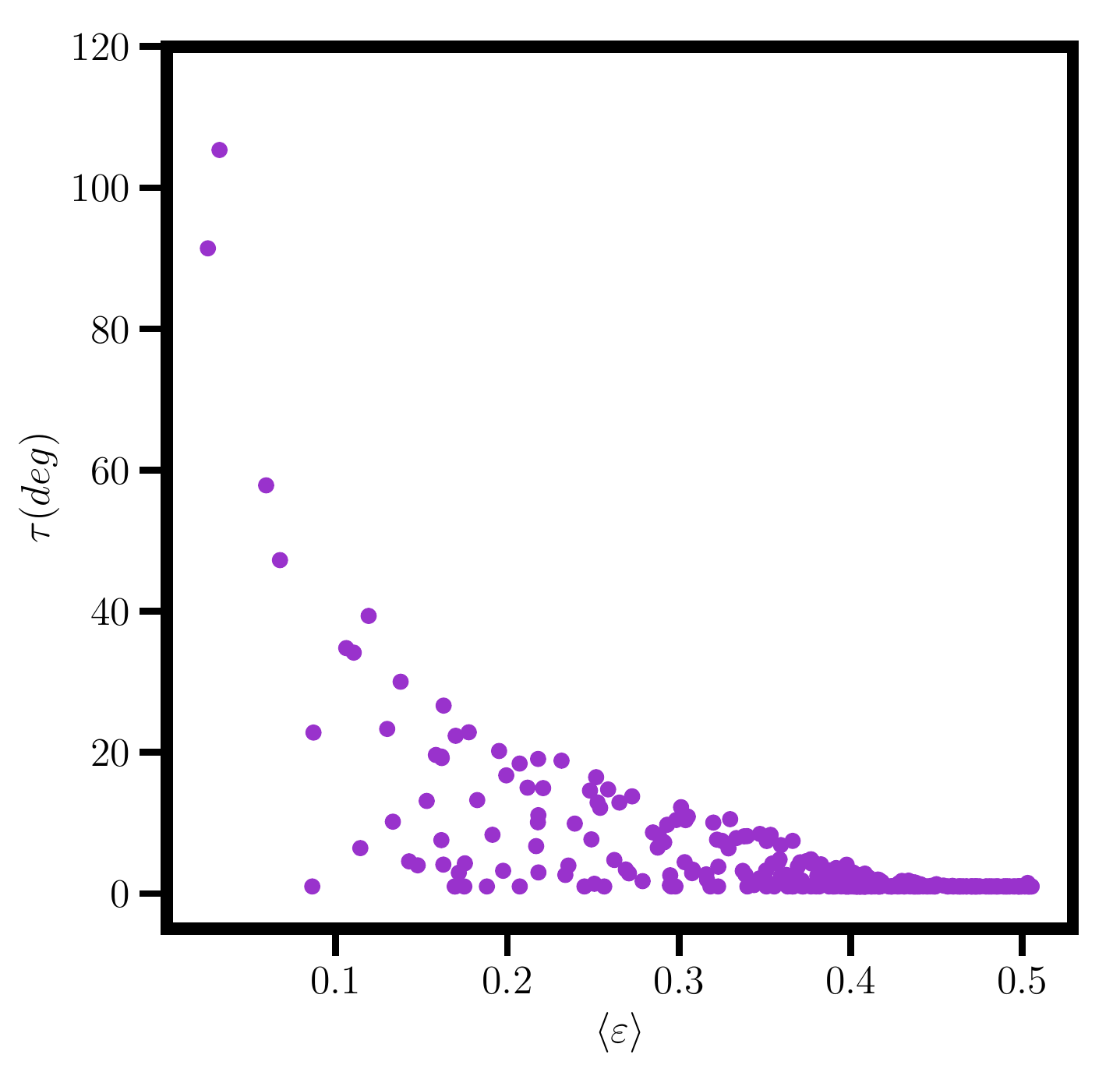}
	\hfil
	\includegraphics[height=68mm,trim=1mm 1mm 2mm 2mm]{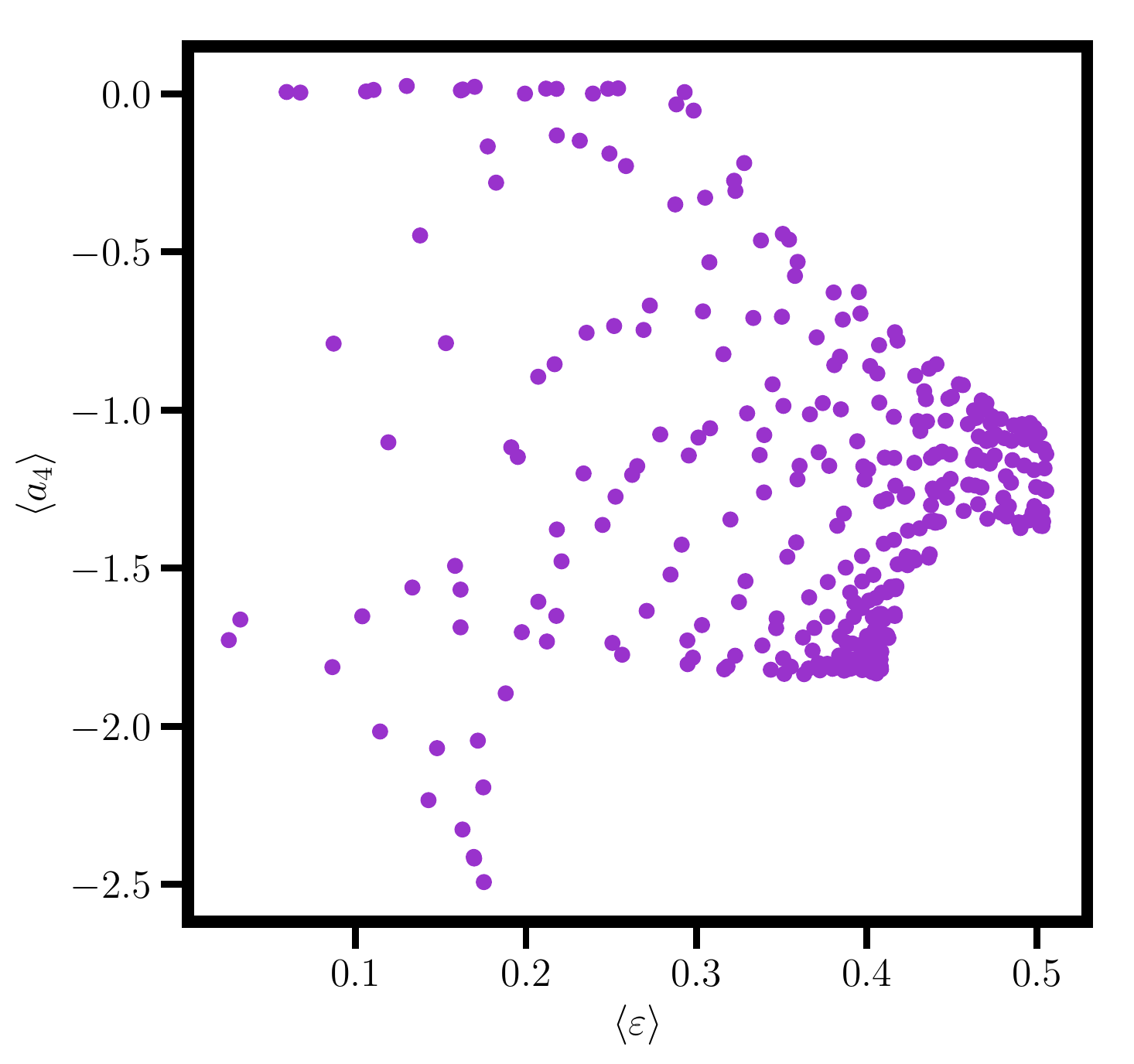}
	\caption{Same as Fig.~\ref{Fig.isophote_correlations_np} for model \emph{BOXY}. No particularly discy projections are present and the distributions are qualitative consistent with what observed for massive ellipticals \citet{Bender88, Bender89}.}
	\label{Fig.isophote_correlations}
\end{figure*}

As a final check, in Fig.~\ref{Fig.isophote_correlations} we show the analogous of Fig.~\ref{Fig.isophote_correlations_np} for the model considered here: similarly to what observed by e.g. \citet{Bender88, Bender89,Foster2017} for real bright ellipticals, Fig.~\ref{Fig.isophote_correlations} shows that the strongest twist happens where the isophotes are rounder and that the mean $a_4$ becomes more negative (i.e., the isophothes are boxier) as the isophotes become more flattened. Differently from what discussed in Fig.~\ref{Fig.isophote_correlations_np}, here we do not find any viewing directions yielding strongly discy isophotes.

\subsubsection{A case with strong twist}
\label{Par.twist_strong}

\begin{figure}
	\hfil
	\includegraphics[width=\columnwidth,trim=1mm 2mm 10mm 2mm]{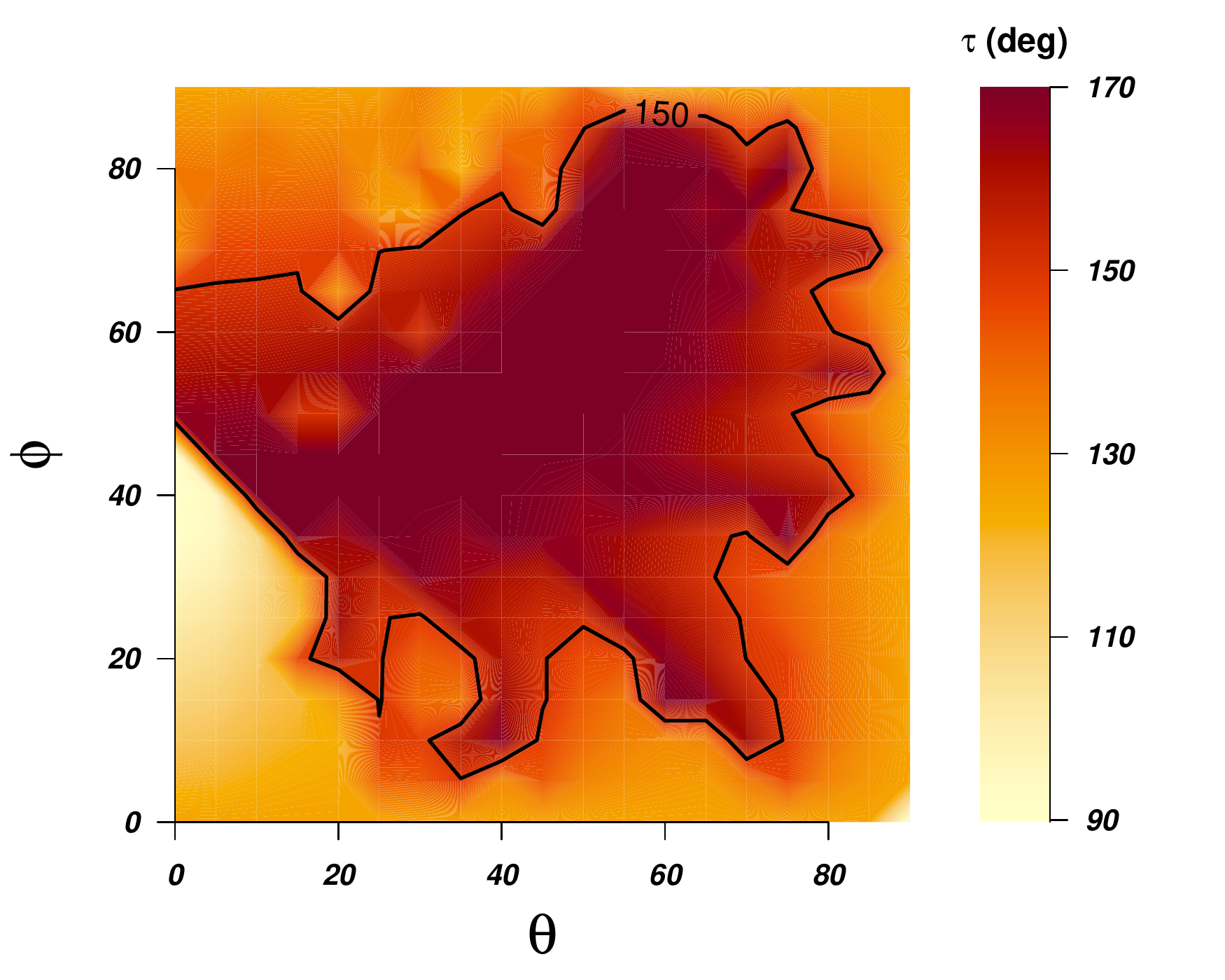}
	\hfil
 	\caption{Map of the twist angle $\tau$ for model \emph{PQCROSS} (see Section~\ref{Par.twist_strong}).}
	\label{Fig.twist_strong}
\end{figure}

Strong twists can be obtained when the orientation of the intrinsic long axis changes with radius. Fig.~\ref{Fig.twist_strong} shows the mapping of $\tau$ for model \emph{PQCROSS}: this is similar to model \emph{ELLIP}, but with $p$ decreasing linearly with $\log(r)$ from 1.3 to 0.6 and $q$ increasing linearly in $\log(r)$ from 0.6 to 1.3. In this case the intrinsic long axis of the model is along the $y$-direction in the inner regions, but along the $z$-direction at large radii; the model is near-spherical in the transition region. Since the orientation of the long and short axes changes with radius, there is now more than one region in the viewing-angle plane, where twists can occur.

The comparison of Figs.~\ref{Fig.twist_weak} and \ref{Fig.twist_strong} illustrates how the expected occurance rate of isophote twists is closely related to the direction stability of the long and short axes in triaxial galaxies. The above examples suggest that if the orientation of the intrinsic long axes would change with radius in many real massive galaxies, then we should observe strong isophote twists very often. While such strong twists indeed exist in individual galaxies (e.g. \citealt{Ximena16}), they are not characteristic for massive elliptical galaxies as a class \citep{KlugeEtAl2020,Ma14,Goullaud18}. Thus, in the following we will often restrict the analysis to the case of $p(r) > q(r)$. However, our code can also deproject without this condition, to cover individual galaxies where strong twists may be real.

In summary, an ellipsoidal density distribution with generically $q(r) < p(r)$ as just described qualitatively reproduces the observed properties of massive elliptical galaxies for any random projection angles.

\subsubsection{Hidden discs}

\begin{figure*}
	\subfloat[\emph{ELLIP}. \label{Fig.ell_contours}]{\includegraphics[height=52mm]{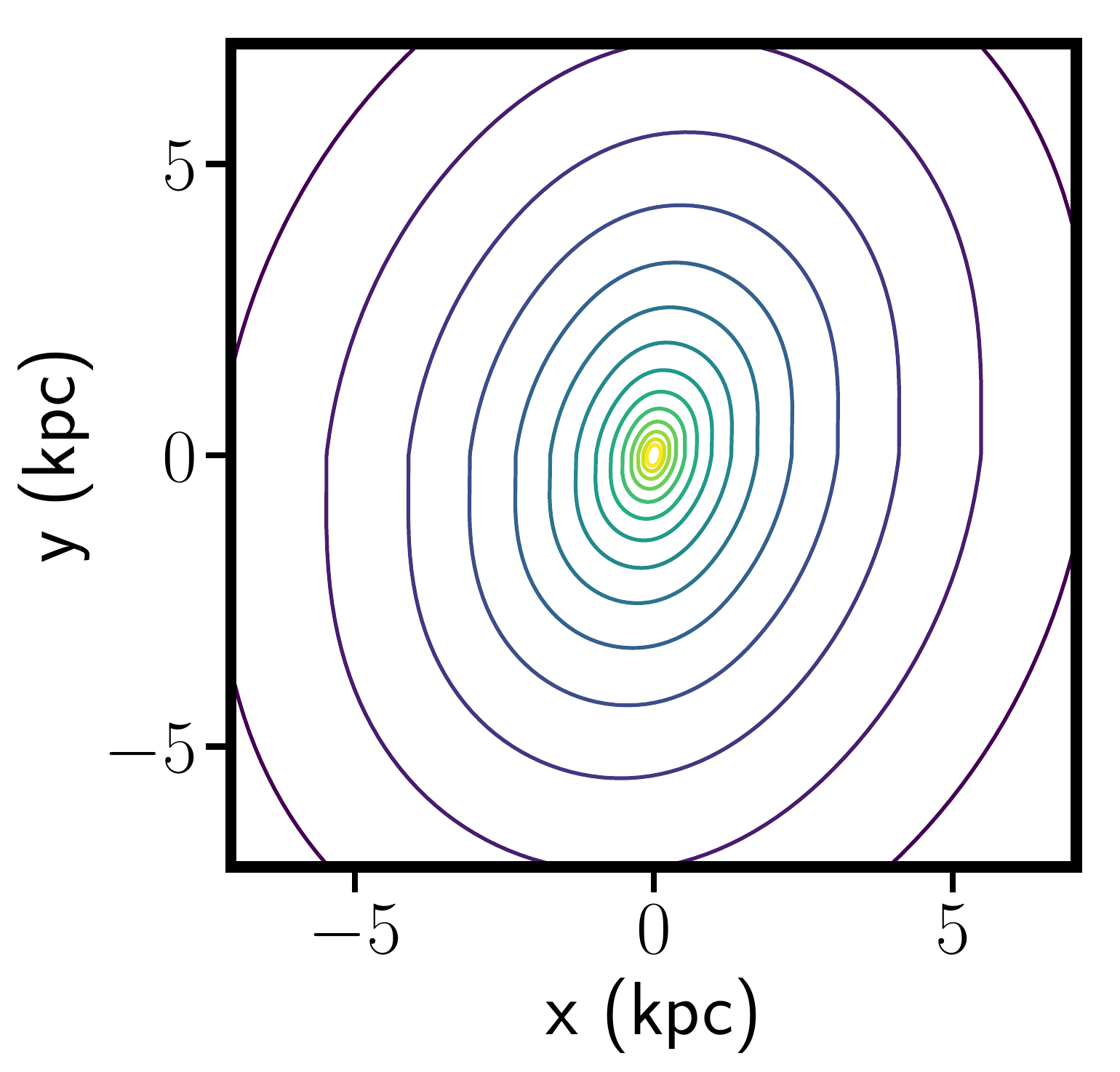}}
	\hfil
	\subfloat[\emph{DISCYBOXY}, inner regions. \label{Fig.discy_contours}]{\includegraphics[height=52mm]{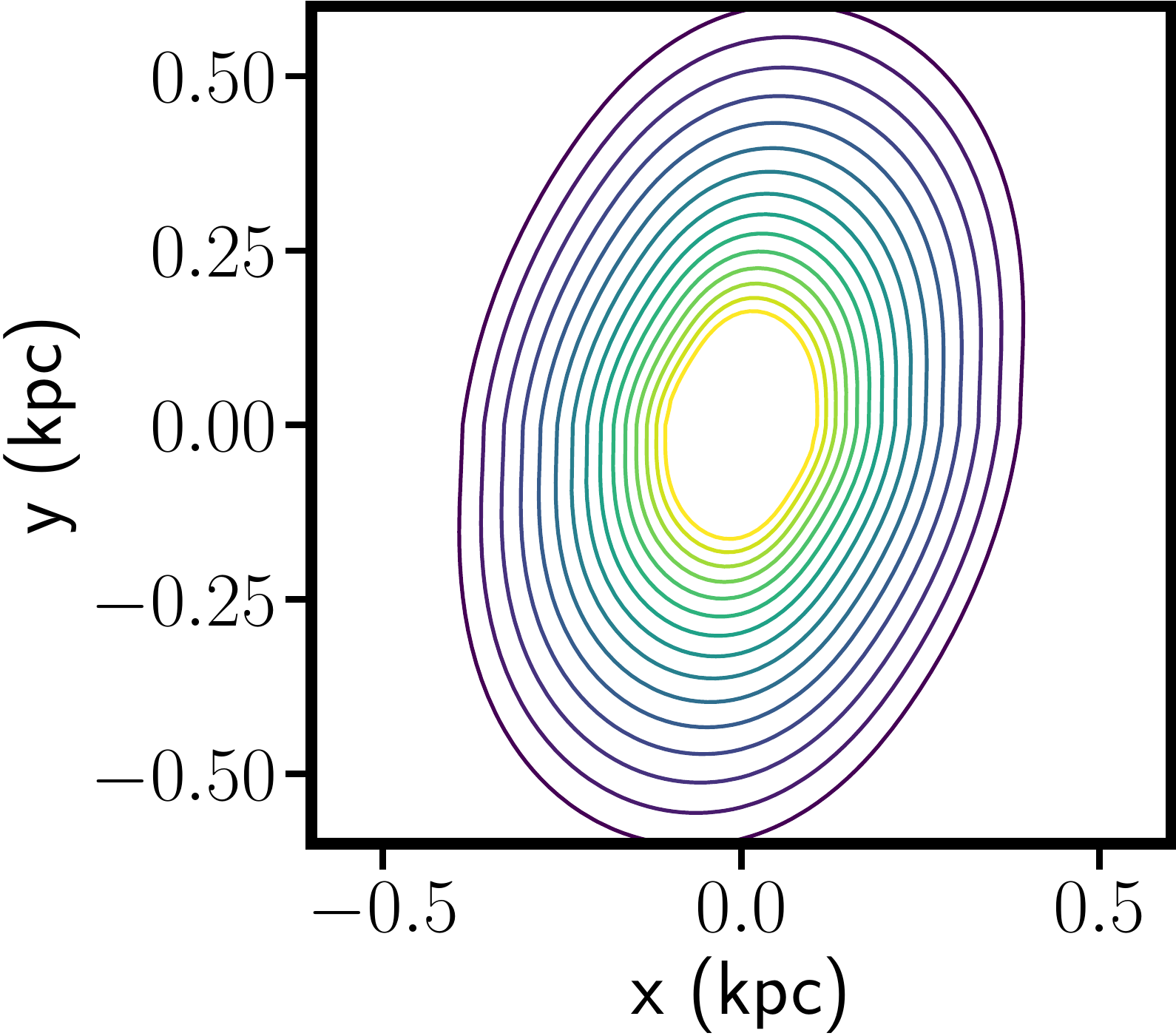}}
	\hfil
	\subfloat[\emph{DISCYBOXY}, outer regions.  \label{Fig.boxy_contours}]{\includegraphics[height=52mm]{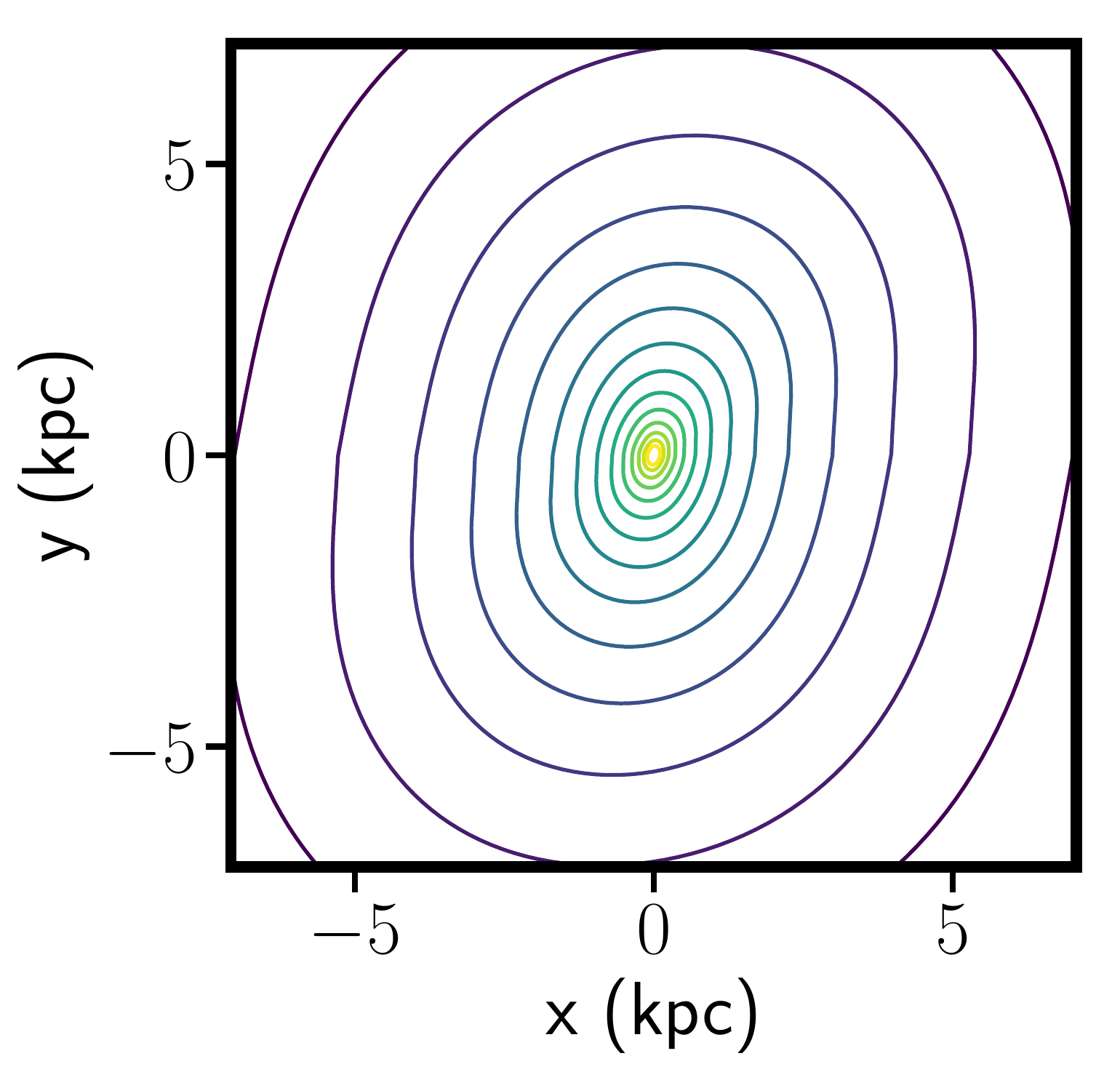}}

	\caption{Contours of the projected SB (using $\theta=\phi=\psi=45\degr$) of \emph{ELLIP} (left panel) and \emph{DISCYBOXY} (middle and right panels). \emph{DISCYBOXY} has discy isophotes near the centre and boxy ones in the outer regions. The contour are colour reflects the SB value.}
	\label{Fig.contours}
\end{figure*}

In the axisymmetric case, much of the deprojection degeneracy can be traced back to disc-like conus densities, which become quickly unidentifiable when the inclination is far enough from edge-on.  Lower-luminosity elliptical galaxies often show discy isophotal distortions and are intrinsically flattened strong rotators; for this class of objects consideration of embedded axisymmetric discs is indeed important. But for massive ellipticals, less flattened and mainly boxy objects, discs are either not present or contribute very little to the total density and therefore in the following we will ignore them. We discuss the effects of the superposition of an axisymmetric disc and a triaxial spheroidal body in Appendix~\ref{sec_discs}. There we also show that the presence of important discs also tends to produce strong isophote twists that are not observed in most massive ellipticals, as discussed above.

%and for the same model 

\begin{figure}
    \hfill\includegraphics[width=0.97\columnwidth]{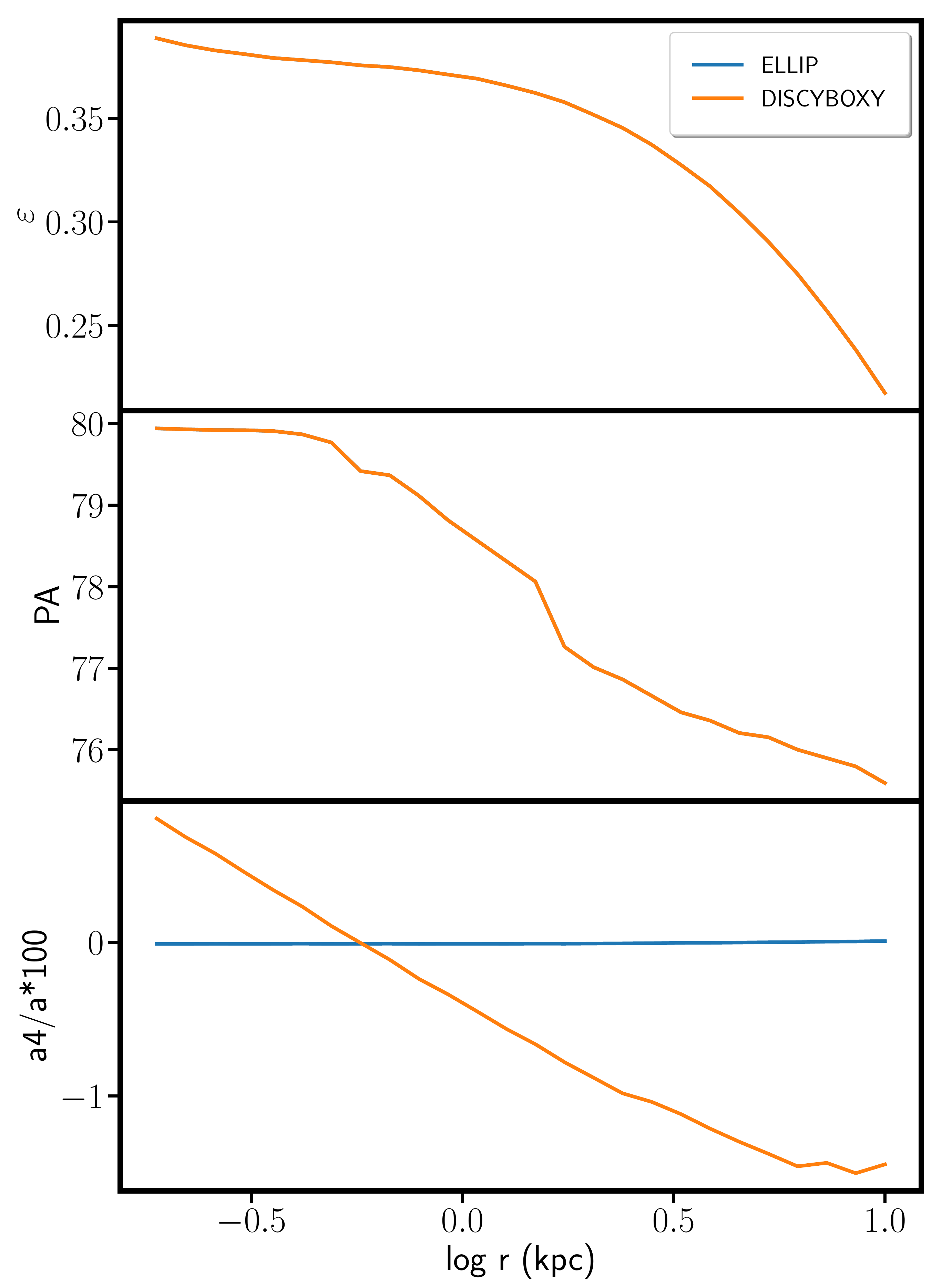}\hfill
    \vspace{-1mm}
    \caption{Radial profiles of the ellipticity $\varepsilon$ (upper panel), PA (middle panel) and $a_4$ coefficient (lower panel) for \emph{ELLIP} (blue line) and \emph{DISCYBOXY} (orange line) projected using $\theta=\phi=\psi=45\degr$. As expected, the ellipticity and PA profiles are identical while the \emph{DISCYBOXY} model has discy isophotes ($a_4>0$) in the central regions and boxy ones ($a_4<0$) in the outer parts.}
    \label{Fig.isoshapes}
\end{figure}

\subsection{Testing deprojections with constrained shapes} 
\label{sec_tests}

We consider the \emph{ELLIP} Jaffe model described in Section~\ref{Sec.proj} with $\xi=0$ and the \emph{DISCYBOXY} model, where $\xi$ decreases linearly $\log(r)$ from 0.3 to $-0.5$, i.e. discy in the innermost regions and boxy outside, and project them along the direction $\theta=\phi=\psi=45\degr$.  Fig.~\ref{Fig.contours} shows their surface brightness contours, and Fig.~\ref{Fig.isoshapes} their ellipticity, PA and $a_4$ profiles. First, we provide the code with the correct viewing angles, $p$, $q$ and $\xi$ profiles and let it search for the density profile.  Secondly, we let the code also search for $p$, $q$ non-parametrically, starting from an initial guess of $p = q = 1$ across the whole grid.  Finally, we let the code recover also the $\xi$ profile for \emph{DISCYBOXY}

%Qui plot Rho, pq, pq gamma
\begin{figure}
	\includegraphics[width=\columnwidth]{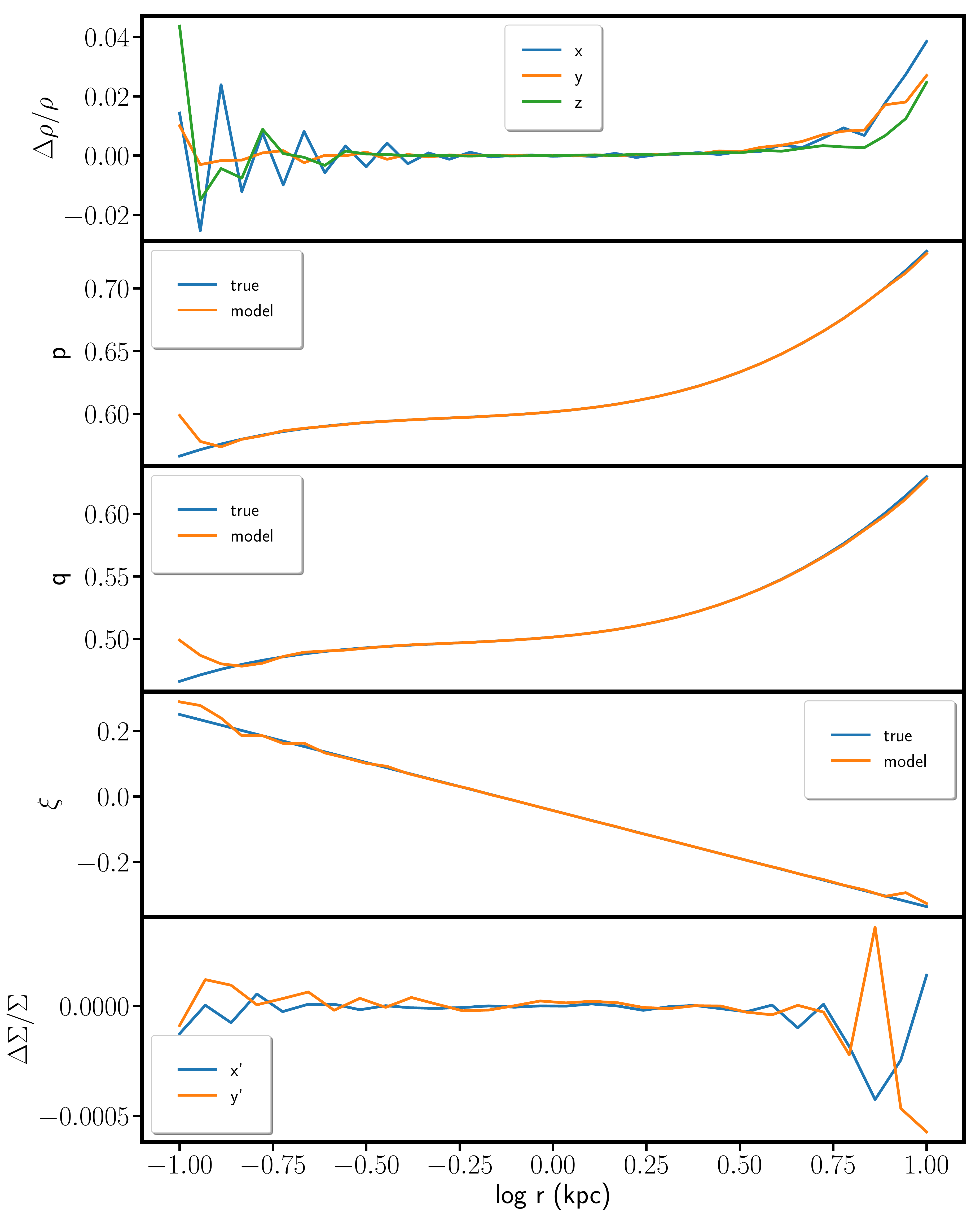}
	\vspace{-3mm}
	\caption{\emph{From top to bottom:} relative difference between the true (blue) and the recovered (orange) intrinsic density along the principal axes $x$, $y$, $z$ of model \emph{DISCYBOXY} using the constrained shape deprojection algorithm; recovered $p$, $q$, and $\xi$ profiles superimposed to the true ones ; percentage difference between the true and the recovered SB along the principal axes.}
	\label{Fig.recovery} 
\end{figure}

\begin{figure}
    \includegraphics[width=\columnwidth]{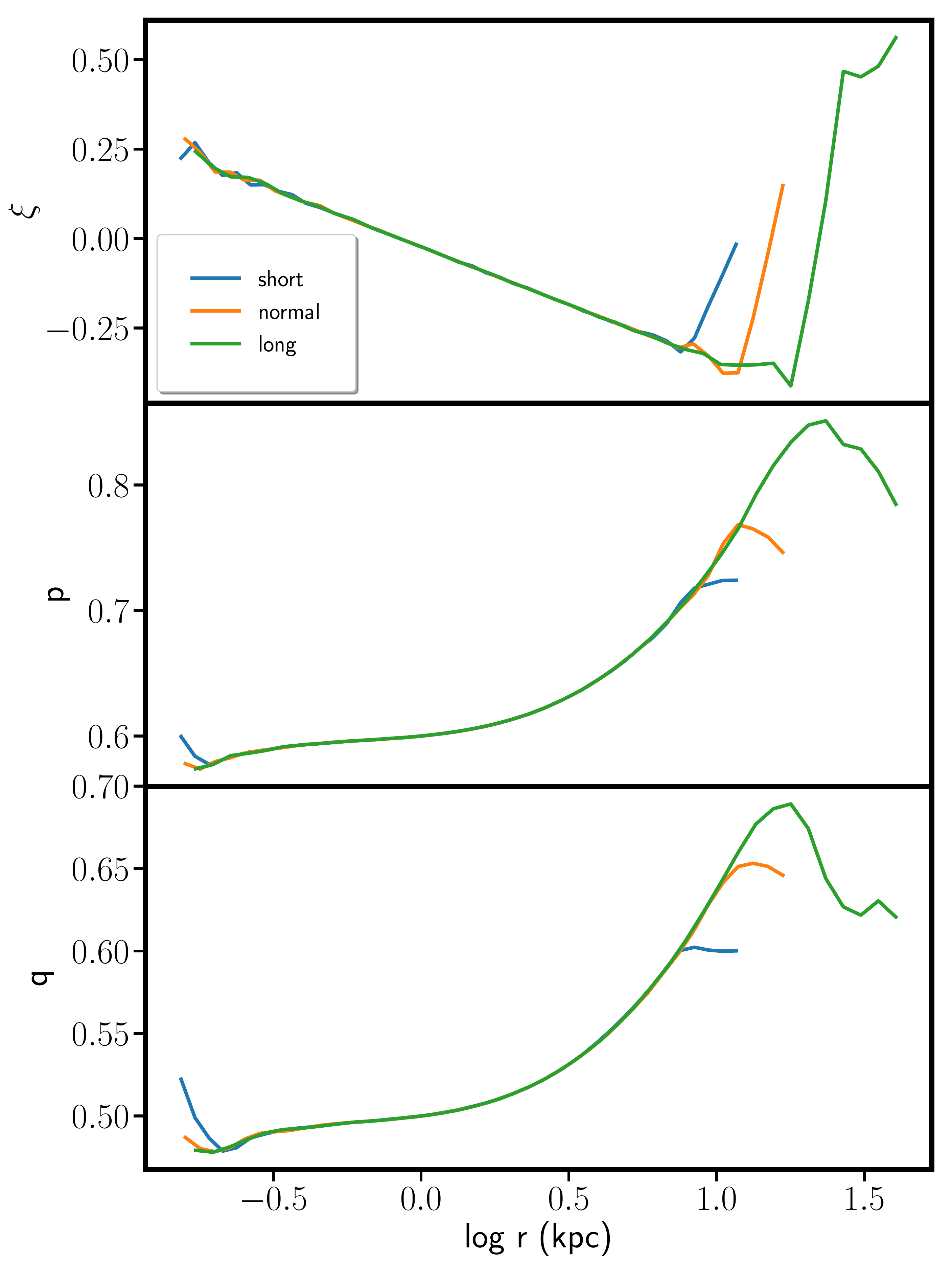}
    \vspace{-3mm}
    \caption{Recovery of $p$, $q$, and $\xi$ profiles for model \emph{DISCYBOXY} by assuming three $\rho$ grids of different extension using the constrained shape deprojection algorithm. The more extended the grid, the farther out our code is able to recover the true profiles.}
    \label{Fig.variable_grid_gammapq}
\end{figure}

\begin{figure}
	\includegraphics[width=\columnwidth]{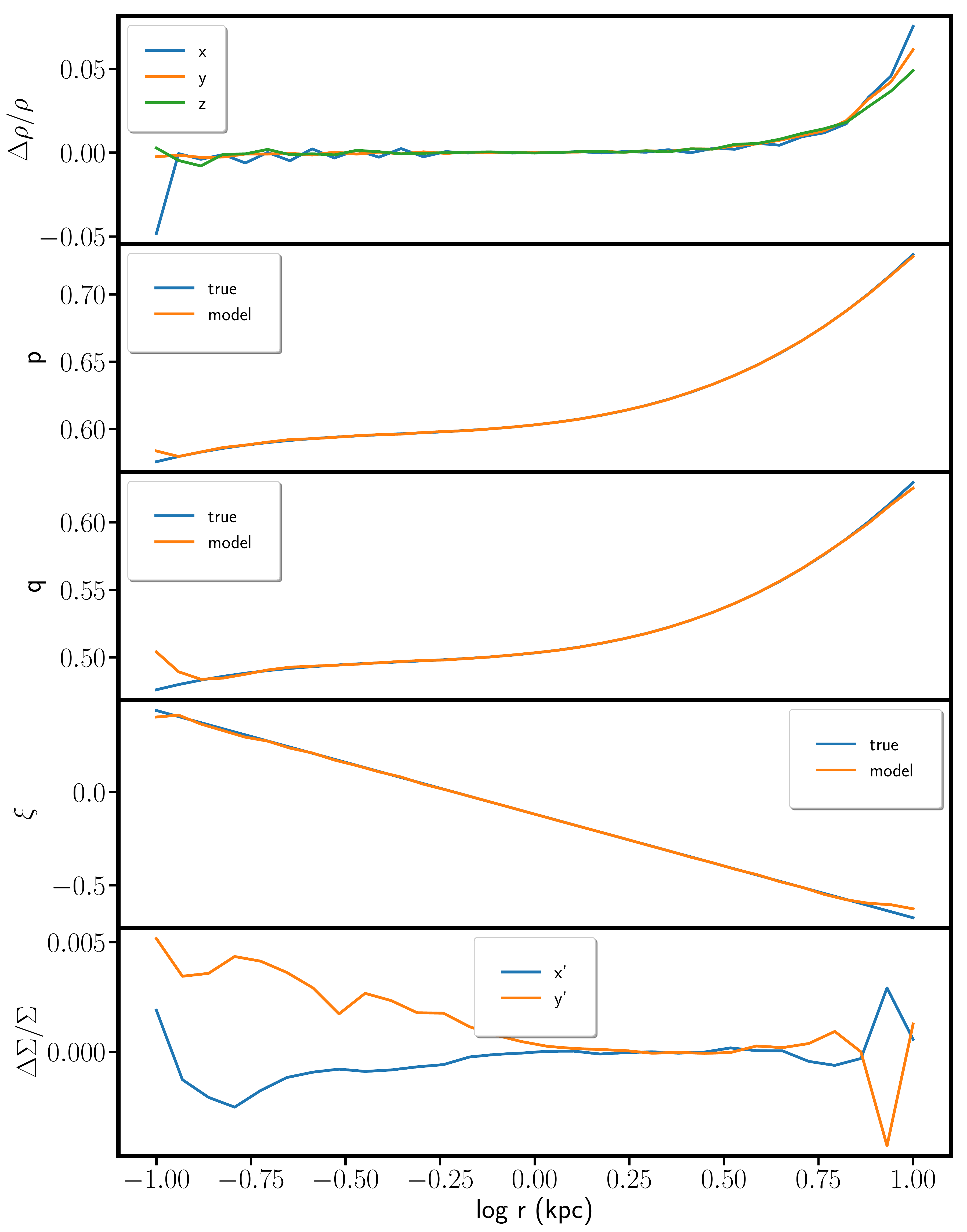}
	\vspace{-3mm}
	\caption{Same as Fig.~\ref{Fig.recovery} but using a Hernquist model which is less cuspy than the Jaffe model. Here too, the code fits both the intrinsic density and the projected SB very well.}
	\label{plot_Hernquist}
\end{figure}

We show in Fig.~\ref{Fig.recovery} the results of the deprojection for this last case, i.e. when all parameters need to be recovered. In the top panel, the three lines are the percentage differences between the true model and what the code reconstructs, computed along the principal axes. Also shown are the true $p$, $q$ and $\xi$ profiles superimposed with the reconstructed ones. Finally, in the bottom panel we show that the fit to observed SB is excellent. In all cases, the density is recovered well, within an accuracy of 1\%, out to the maximum radius sampled by the SB and down to a radius of the order of the resolution of the grid. For the very innermost and outermost points, all profiles start to deviate significantly from the true shapes.  This is mostly due to the extrapolation to large radii: we have repeated the test highlighted in Fig.~\ref{Fig.recovery} stopping the radial grid first at 20 kpc and then extending it out to 80 kpc. In Fig.~\ref{Fig.variable_grid_gammapq} we superimpose these results to those obtained for the grid extended out to 30 kpc, showing that the point at which the radial profiles start becoming unreliable also decreases. The last inner reliable point is set by a combination of radial extent and resolution of the grid. We will discuss it in detail in a future paper in combination with the PSF convolution.

Finally, provided that many ellipticals have cores, we also tested our algorithm with a \cite{Hernquist90} model (equation~\ref{eq:double-power-law} with $\alpha=1$ and $\beta=4$) using the same parameters as above but with $\alpha=1$ (see Section~\ref{Sec.proj}). We show in Fig.~\ref{plot_Hernquist} that we do not find significant differences with the results presented above.

These findings go even beyond what stated in Section~\ref{sec_deprojectellipsoids}, namely that the density can be uniquely recovered if it is stratified on perfect ellipsoids and we know the viewing angles. Here we achieve a very good recovery also when the density is stratified on \emph{deformed} ellipsoids \eqref{m}. This is not fully surprising, since the information available on the 4 planes in Fourier space discussed in Section~\ref{Sec.2} should be more than enough to constrain the four one-dimensional functions used in our procedure (assuming that the LOS is not parallel to a principal axes).

\section{Reconstruction of viewing angles}
\label{sec_viewingangles}

The projection geometry of a galaxy is, of course unknown. In the axisymmetric case, it is completely described by the inclination angle $i$ between the minor axis and the LOS, while in the triaxal case we need three angles, two ($\theta,\,\phi$) to specify the LOS direction and another one ($\psi$) to give a rotation around the LOS. However, it is unlikely that a given SB profile can be deprojected for every viewing geometry, which is something the fully non-parametric code is able to do by producing possibly unphysical densities. For example, only for a restricted set of viewing angle we can find an MGE \citep{Cappellari02} deprojection.

\begin{figure*}
    \subfloat[\label{Fig.pq_Nbody}]{\includegraphics[height=90mm]{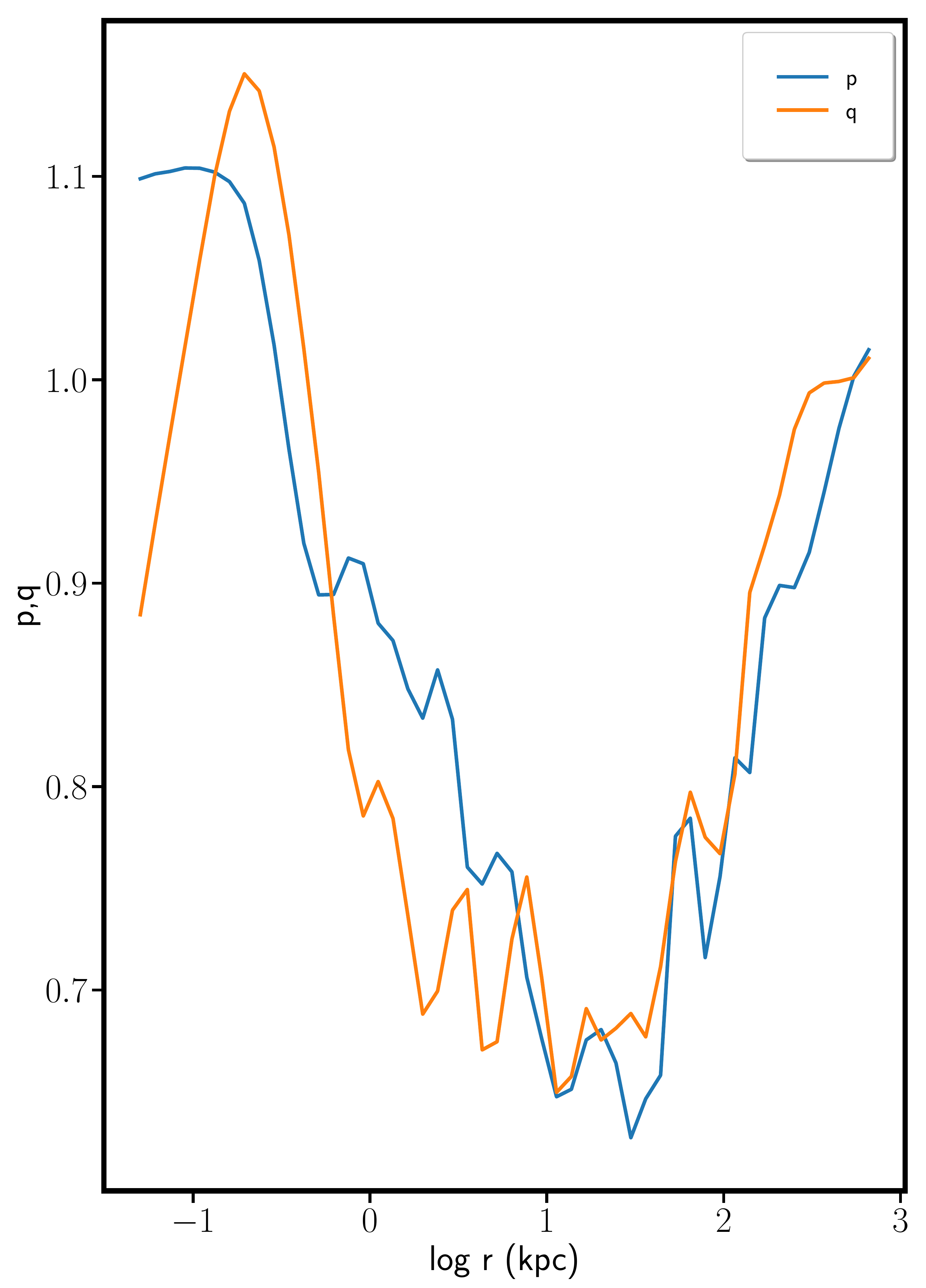}} 	\hfil
	\subfloat[\label{Fig.isofit_Nbody}]{\includegraphics[height=90mm]{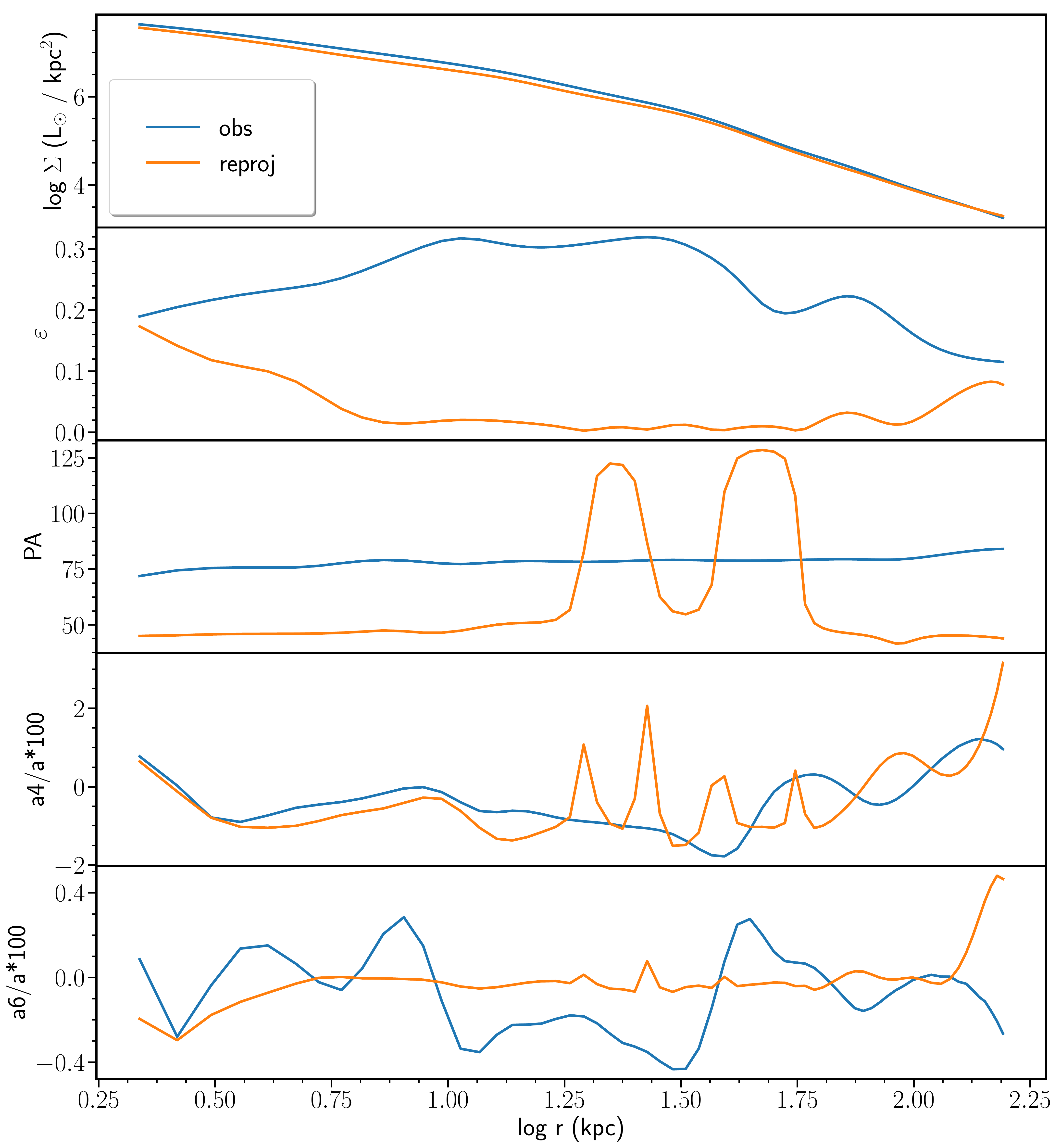}}

	\caption{Discarding \emph{NBODY} deprojections at wrong viewing angles.\emph{Left:} $p$ and $q$ profiles, obtained for viewing angles $\theta=40\degr$, $\phi=50\degr$, and $\psi=45\degr$. We formally discard this deprojection, because the profiles cross over. \emph{Right:} Isophote shape analysis as a function of semi-major axis for the re-projection along the $y$-axis (orange) of the density obtained by deprojecting at $\theta=30\degr$, $\phi=60\degr$, and $\psi=45\degr$. We discard this deprojection because of the unphysical jumps in the PA of the $y$-axis re-projection.}
    \label{Fig.bad_depros}
\end{figure*}

Here, we want to explore how significantly the range of allowed viewing angles can be narrowed using our constrained-shape deprojection approach.  To this end we deprojected three different models for several \emph{wrong} viewing angles and study the effects of such incorrect projection geometries. The models we consider are as follows.
\begin{itemize}
\item The ellipsoidal Jaffe model \emph{ELLIP} (Figs.~\ref{Fig.ell_contours} \& \ref{Fig.isoshapes});
\item The \emph{DISCYBOXY} Jaffe model (discy in the centre, boxy towards the outer regions, see Figs.~\ref{Fig.discy_contours}, \ref{Fig.boxy_contours} \& \ref{Fig.isoshapes}). The comparison between \emph{ELLIP} and \emph{DISCYBOXY} enables us to study the effects of deviations from perfect ellipticity;
\item Finally, we consider a more realistic case, allowing us to study the effects of noise. To this end, we use an $N$-body model drawn from simulations of \cite{Rantala18, Rantala19}, which has an effective radius of 7\,kpc and an assumed distance of 20 Mpc (model \emph{NBODY}). Its projections are described in Neureiter et al., submitted to MNRAS, where it is used to test our newly developed triaxial Schwarzschild code. The semi-major axis of the innermost isophote is 0.5'' ($\sim0.05\,$kpc), while the outermost radius is 100\,kpc. The grids onto which we place the SB and the intrinsic density have the same dimensions as those used for the Jaffe model. This gives a step of $\sim0.18 \log\,$kpc for the SB grid and of $\sim0.12\log\,$kpc for the $\rho$ grid. The SB grid has been chosen to be circular, while for the $\rho$ grid we have taken flattenings of $P=0.8$ \& $Q = 0.7$.
\end{itemize}

In all three cases, we project the true $\rho$ using $\theta=\phi=\psi=45\degr$. We first assume we knew the correct value of the angle $\psi$ and then consider two different wrong values of $\psi=30\degr$ and $\psi=60\degr$. For each one of these cases, we deproject the models on a grid of $\theta,\,\phi$ values linearly spaced from $0\degr$ to $90\degr$ with step of $5\degr$. The code is free to search for the best-fitting $p$, $q$ and $\xi$ profiles. The RMS we obtain for \emph{ELLIP} and \emph{DISCYBOXY} at the correct viewing angles is about 0.03\% in SB, 0.4\% (for \emph{ELLIP}) and 0.6\% (for \emph{DISCYBOXY}) in $\rho$.  For \emph{NBODY}, where noise is present, the RMS in SB is of the order of 0.8\% while for $\rho$ we get $\sim$14.4\%. In this case, the RMS value is mostly driven by the noise rather than by poor extrapolation.

\subsection{A recipe to compare deprojections obtained with different
  assumed viewing angles}
\label{sec_implausibledeproj}

Reconstructed densities that fit well the given SB for a given choice of viewing angles could generate unrealistic SB profiles when projected to different viewing angles. In the following we adopt some criteria to find and eliminate these cases. These criteria incorporate observations of massive elliptical galaxies as a class, e.g. their observed ellipticity distributions, frequency and strength of isophote twists etc. in a \emph{qualitative} way. We plan a more statistical analysis of this in a separate paper (de Nicola et al., in prep.). Depending on the class of galaxies considered, other criteria might be more useful. However, here we treat our mock SB data as if they were massive ellipticals to illustrate how well the viewing angles of these galaxies can be constrained photometrically.

\begin{figure*}
	%
	% to fill a page completely (no space left for text), use 
	% width=.47\linewidth
	%
	\subfloat[SB comparison, $\psi=30\degr$.\label{Fig.Jaffe_ell_sb30}]{\includegraphics[width=.35\linewidth]{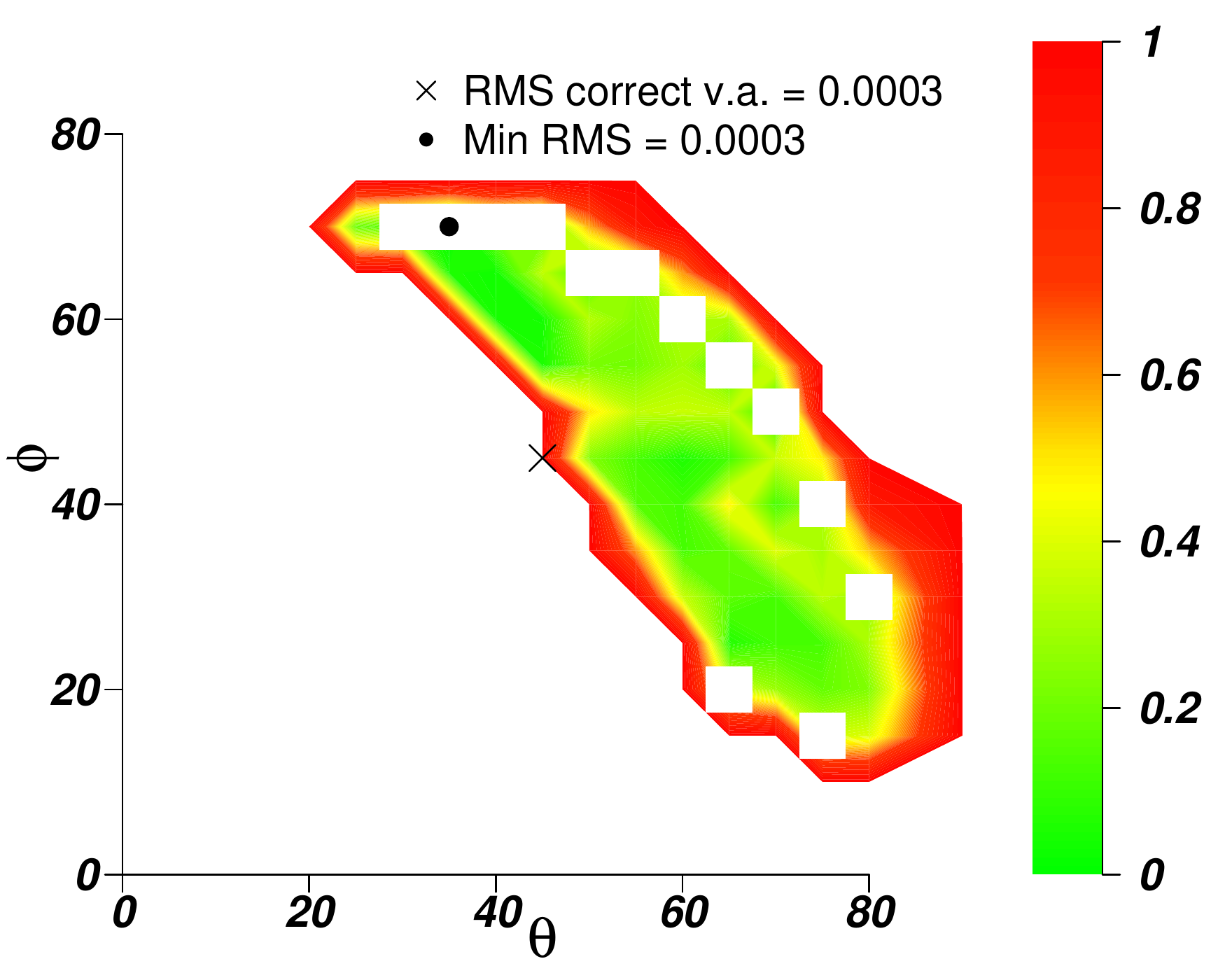}}
	\hfil
	\subfloat[$\rho$ comparison, $\psi=30\degr$.\label{Fig.Jaffe_ell_rho30}]{\includegraphics[width=.35\linewidth]{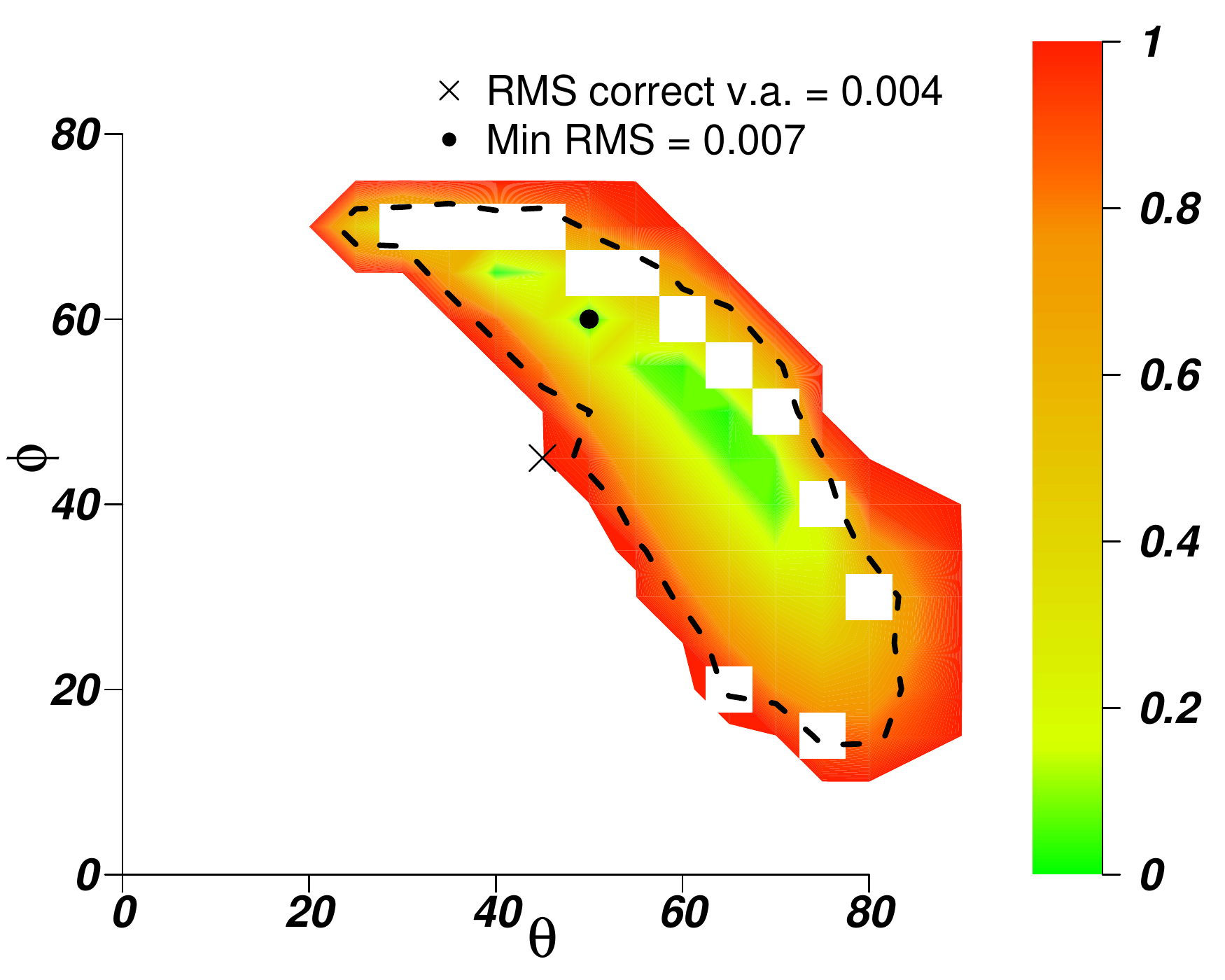}}
	\vspace{-2mm}

    \subfloat[SB comparison, $\psi=45\degr$. \label{Fig.Jaffe_ell_sb45}]{\includegraphics[width=.35\linewidth]{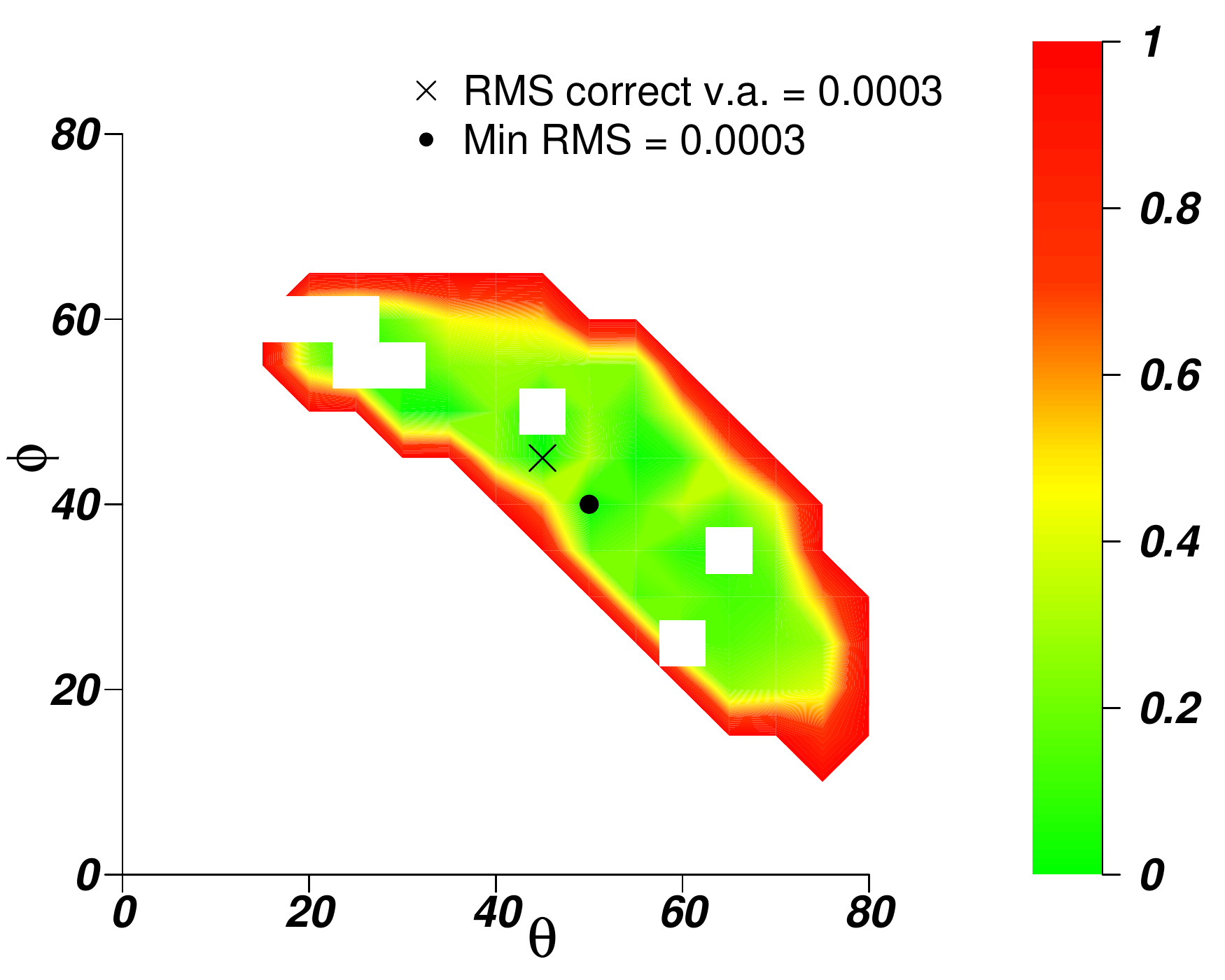}}
	\hfil
	\subfloat[$\rho$ comparison, $\psi=45\degr$.\label{Fig.Jaffe_ell_rho45}]{\includegraphics[width=.35\linewidth]{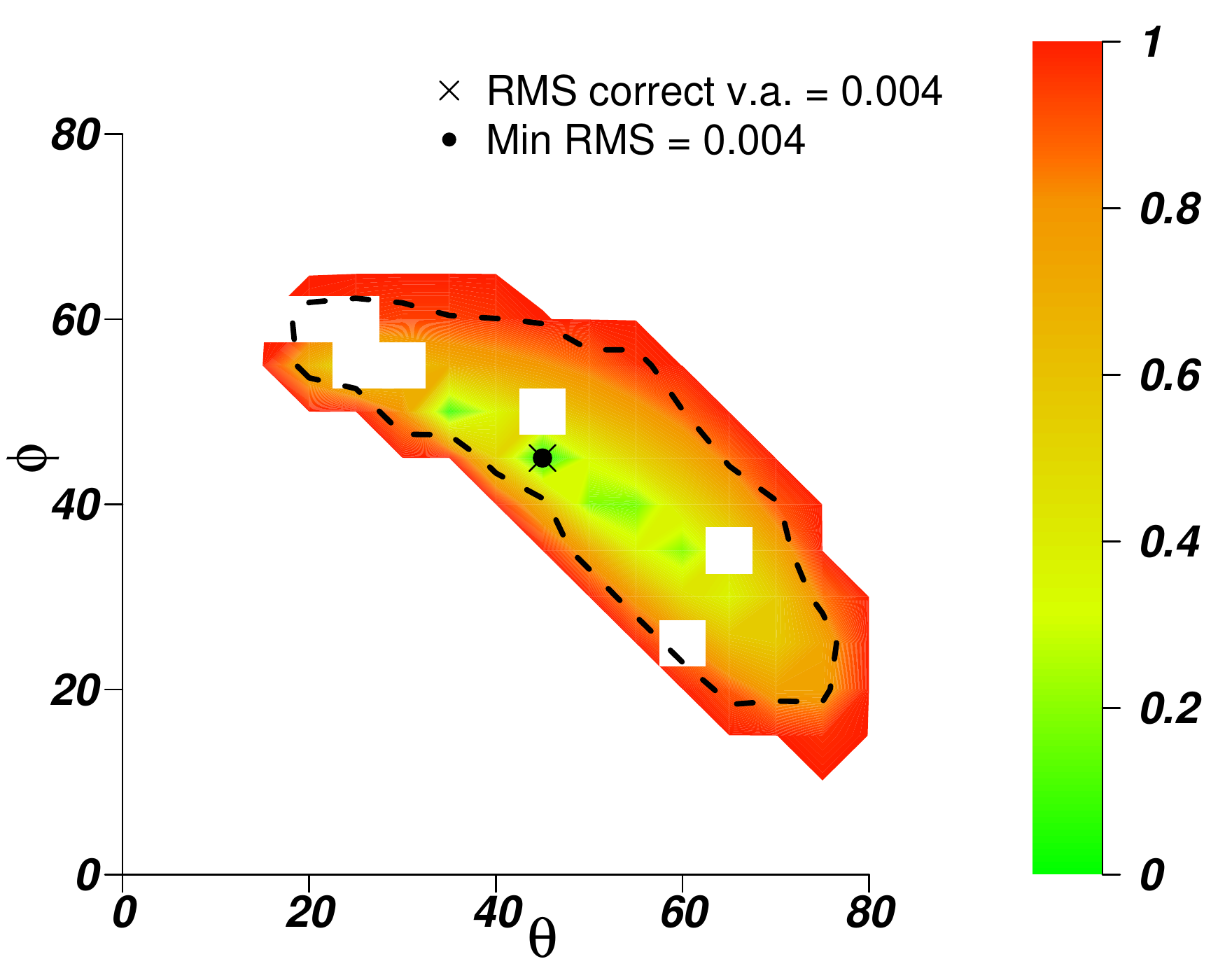}}
	\vspace{-2mm}

	\subfloat[SB comparison, $\psi=60\degr$. \label{Fig.Jaffe_ell_sb60}]{\includegraphics[width=.35\linewidth]{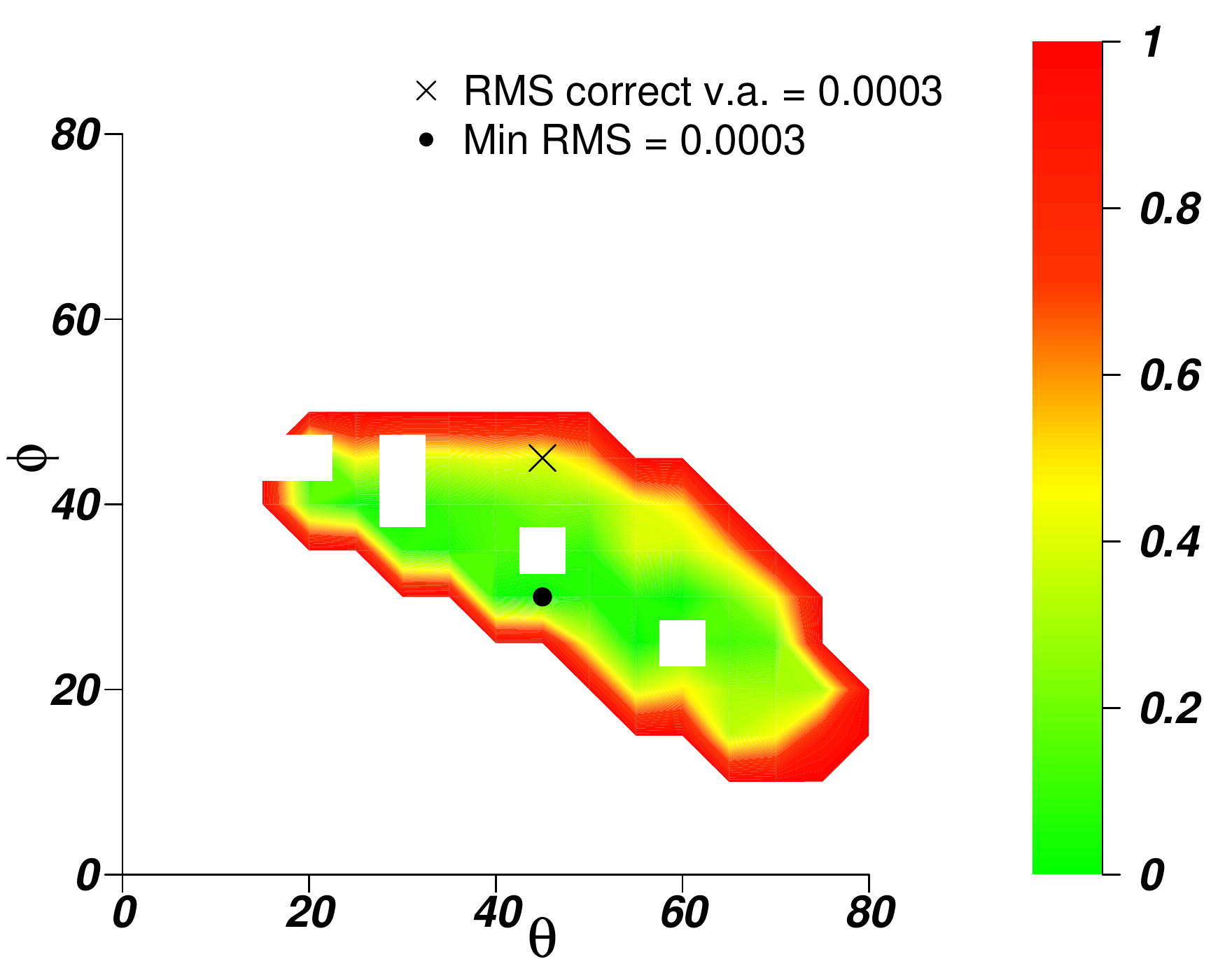}}
	\hfil
	\subfloat[$\rho$ comparison, $\psi=60\degr$.\label{Fig.Jaffe_ell_rho60}]{\includegraphics[width=.35\linewidth]{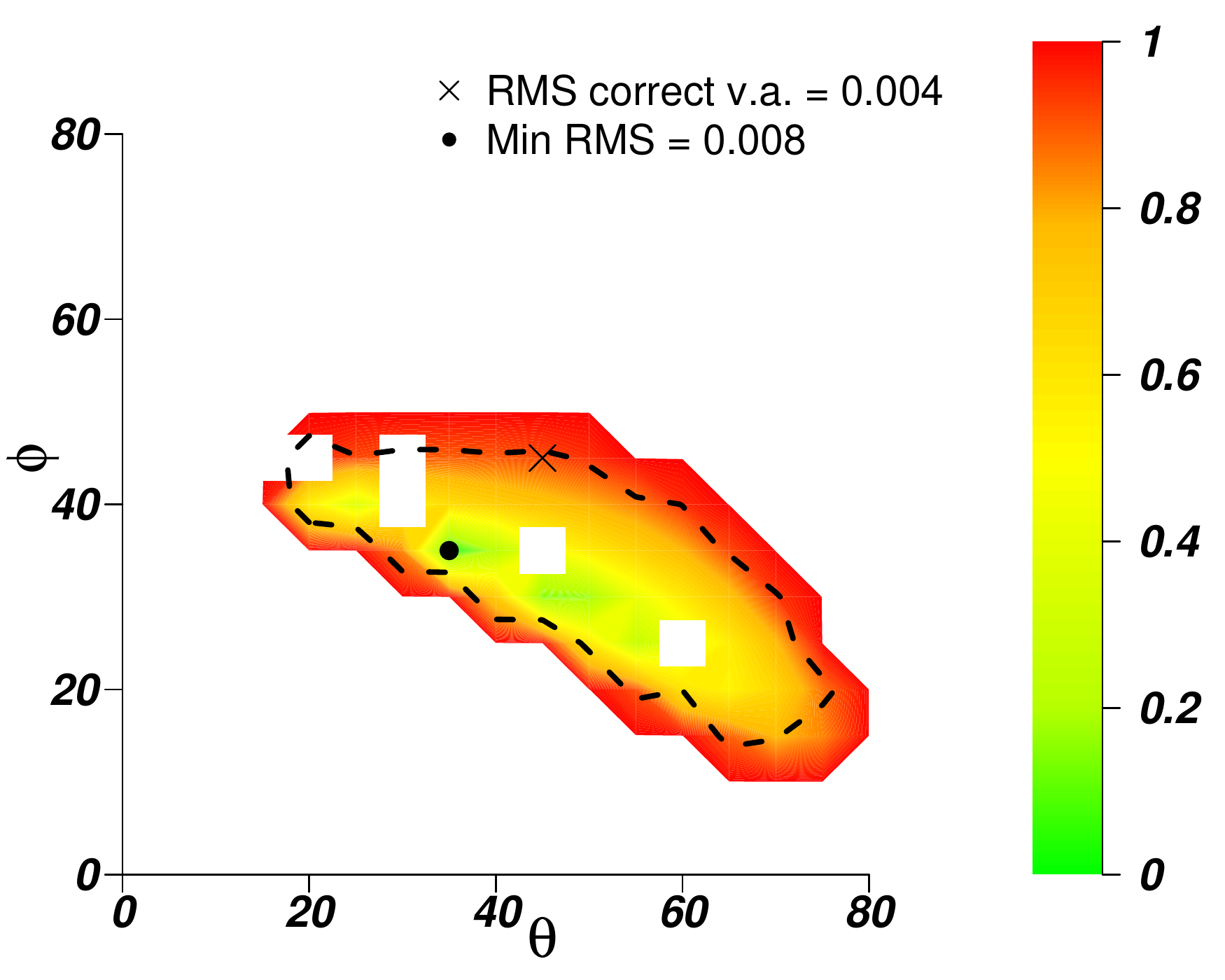}}

	\caption{Logarithmic RMS errors in SB (left) and $\rho$ (right) for model \emph{ELLIP} scaled to those obtained when deprojecting at the correct viewing angles ($\theta=\phi=\psi=45\degr$), obtained for constrained-shape deprojections at different assumed (wrong) viewing angles. The dashed contours on the right delimits the area inside which the RMS in SB is within twice the values for the correct viewing angles. The cross labels the correct $(\theta,\,\phi)$, while the black dot is at the minimum RMS. Empty (white) squares depict regions discarded because of crossing $p$ and $q$ profiles.}
    \label{Fig.Jaffe_ell_45}
\end{figure*}

Our main criterion to compare different deprojections is their relative likelihood, or the goodness of fit, respectively.  We discard all those deprojections which have an RMS (in SB) larger than 0.1\% for the Jaffe models \emph{ELLIP} and \emph{DISCYBOXY} and than 1\% for \emph{NBODY}. These values have been selected by checking the RMS (in SB) that we obtained when deprojecting the SB profiles for the true viewing angles ($\sim$0.03\% for \emph{ELLIP} and \emph{DISCYBOXY} and $\sim$0.8\% for \emph{NBODY}). The 1\% threshold we use for \emph{NBODY} is what we are likely to be using for real galaxies too. \\

Even those viewing directions that give an excellent fit to the observed SB can be ruled out if they happen to show $p$ and $q$ profiles which are not smooth, or have values which are either too low ($\leq 0.2$) or too high ($\geq 5$) with respect to the observed ellipticity distribution of elliptical galaxies. Finally, $p$ and $q$ profiles with interchanging principal axes are unlikely, since this would produce frequent and strong isophote twists (Fig.~\ref{Fig.twist_strong}), which are not observed often in massive ellipticals \citep{Goullaud18}. This means that we would accept a $p$ or $q$ profile which is always above unity but we would discard it in case it was above unity for some radii and below it for others\footnote{These conditions can either be verified \emph{a posteriori} or \emph{a priori} by imposing constraints on $p$ and $q$, both of which our code allows.}.

\begin{figure*}
	\subfloat[SB comparison, $\psi=30\degr$.\label{Fig.Jaffe_gamma_sb30}]
		{\includegraphics[width=.35\linewidth]{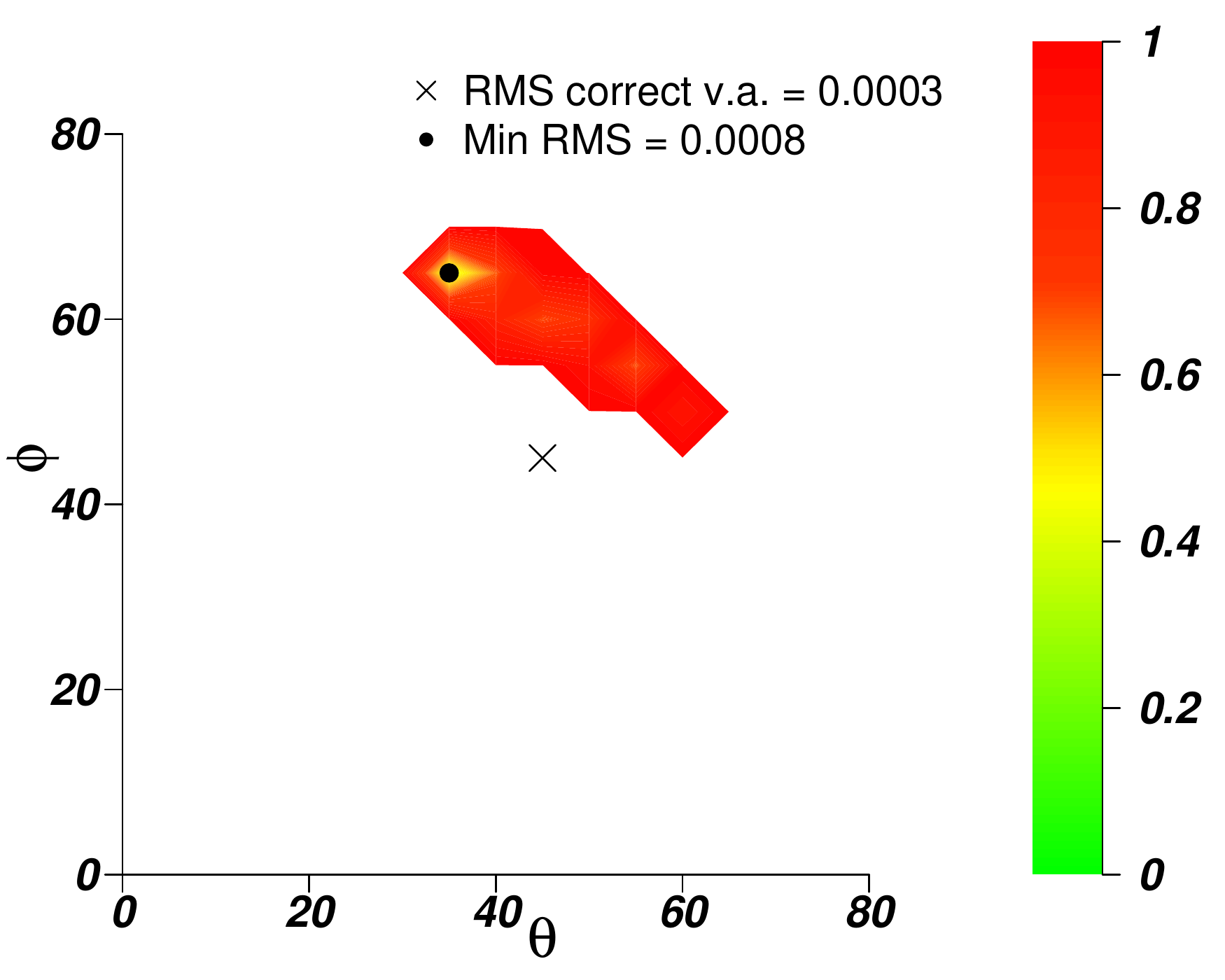}}
	\hfil
	\subfloat[$\rho$ comparison, $\psi=30\degr$.\label{Fig.Jaffe_gamma_rho30}]
		{\includegraphics[width=.35\linewidth]{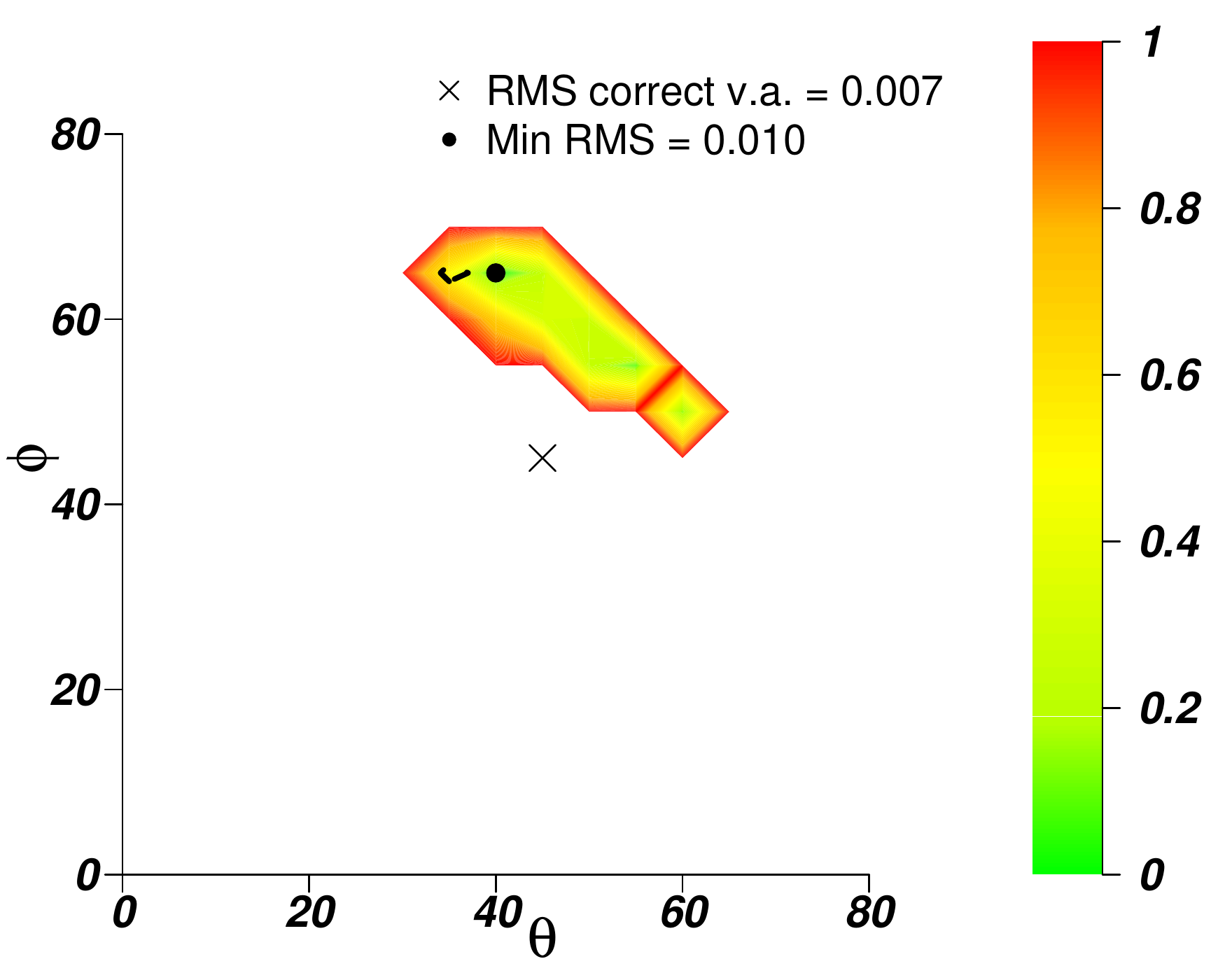}}
	\\
	\subfloat[SB comparison, $\psi=45\degr$. \label{Fig.Jaffe_gamma_sb45}]
		{\includegraphics[width=.35\linewidth]{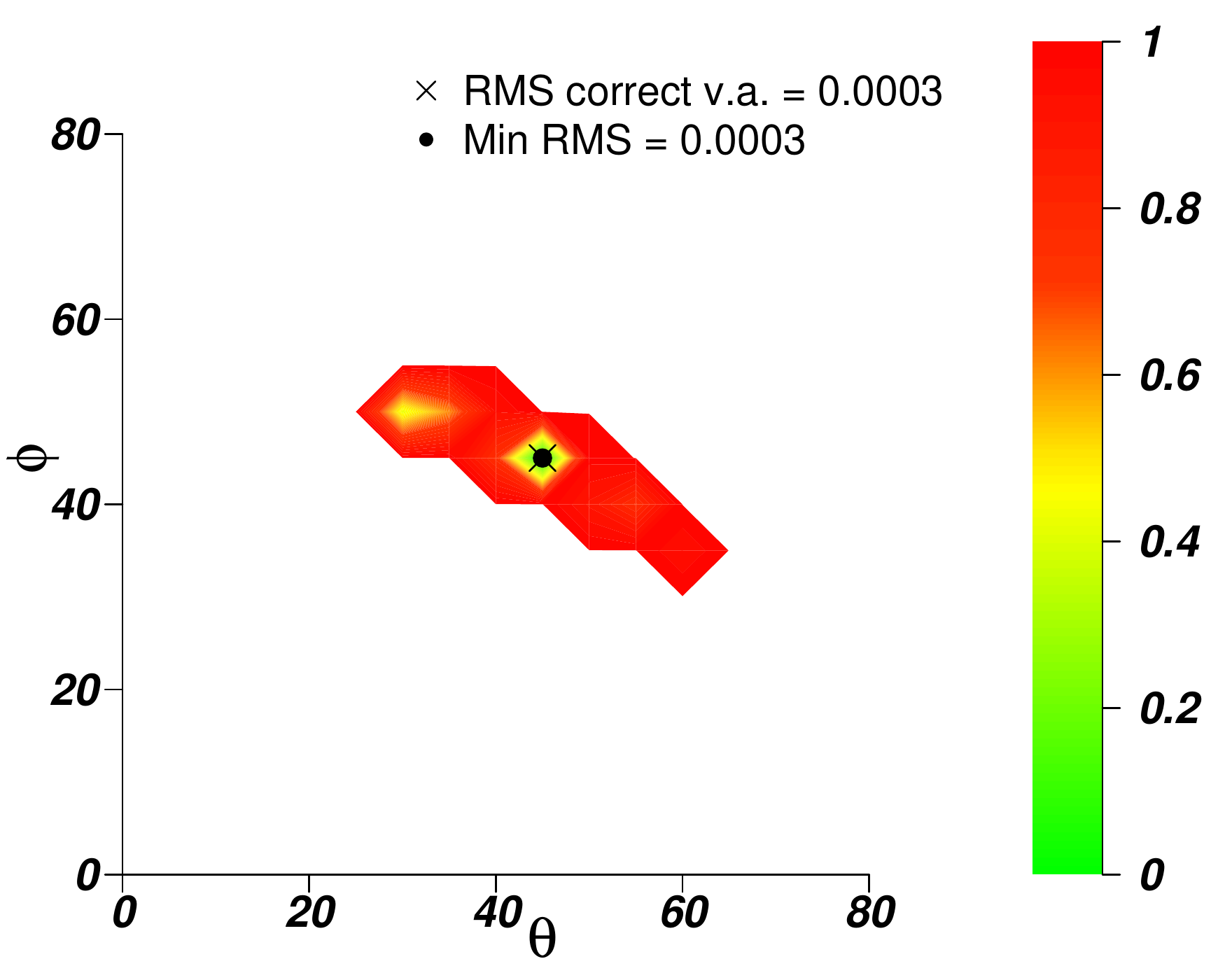}}
	\hfil
	\subfloat[$\rho$ comparison, $\psi=45\degr$.\label{Fig.Jaffe_gamma_rho45}]
		{\includegraphics[width=.35\linewidth]{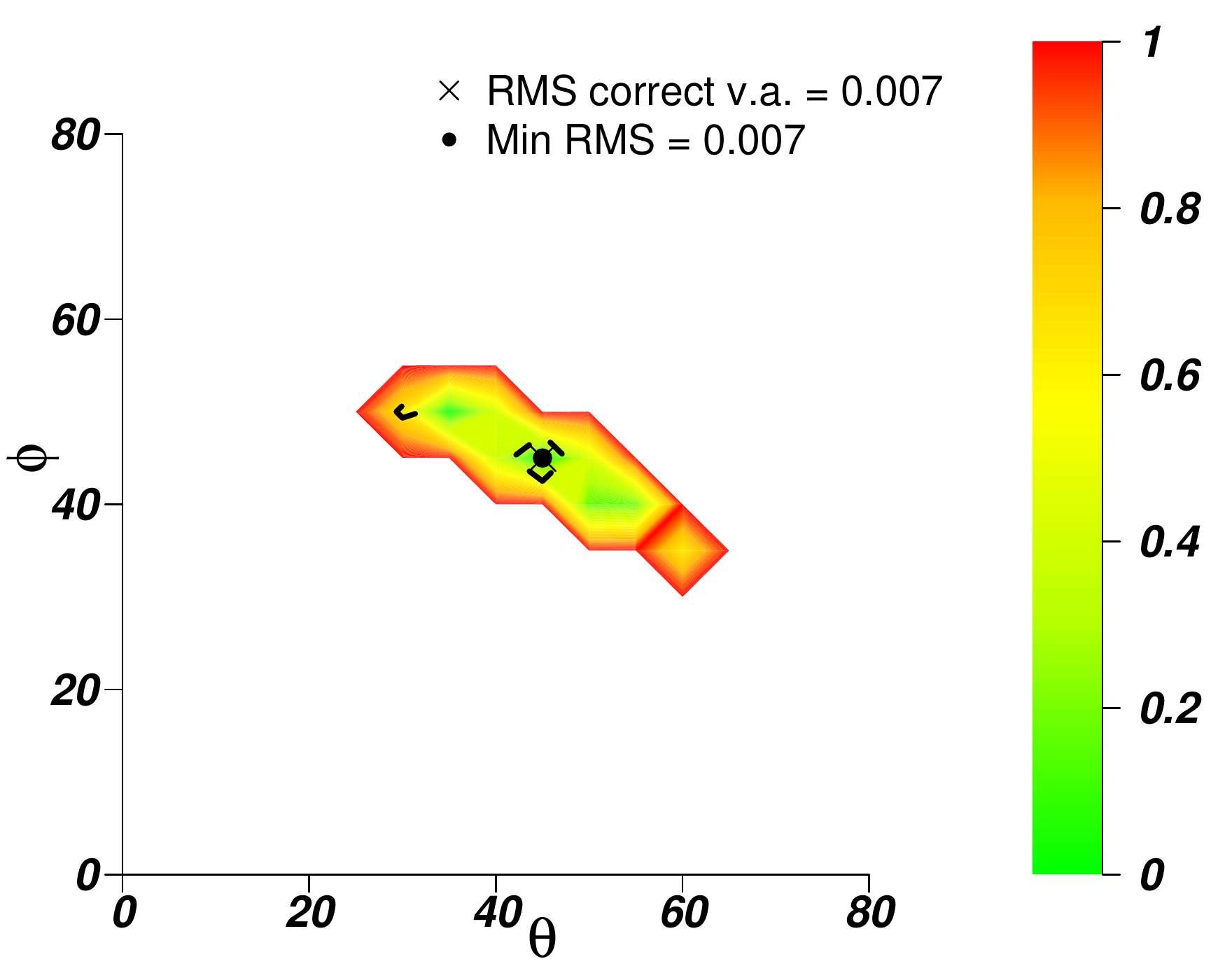}}
	\\
	\subfloat[SB comparison, $\psi=60\degr$. \label{Fig.Jaffe_gamma_sb60}]
		{\includegraphics[width=.35\linewidth]{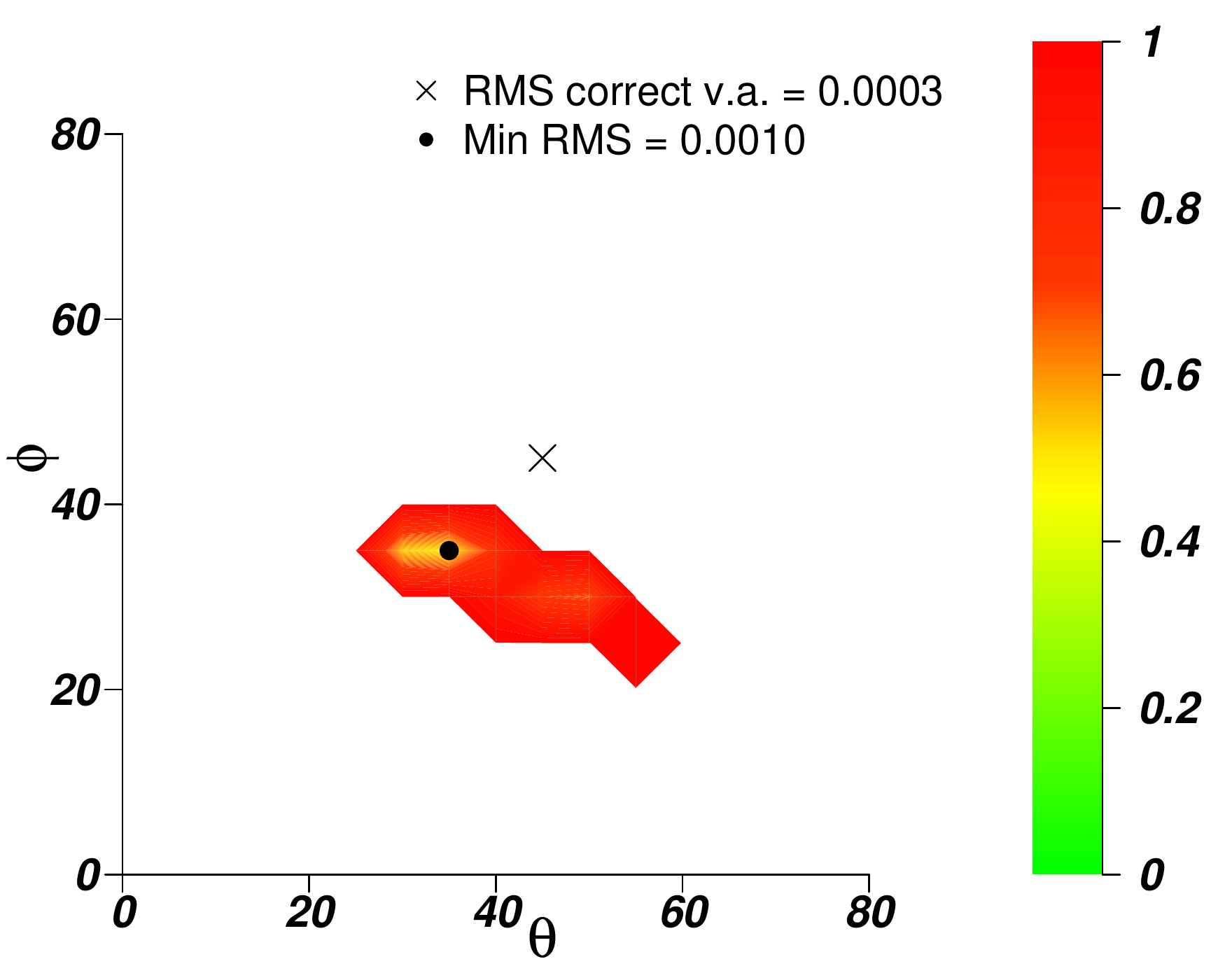}}
	\hfil
    \subfloat[$\rho$ comparison, $\psi=60\degr$.\label{Fig.Jaffe_gamma_rho60}]
    	{\includegraphics[width=.35\linewidth]{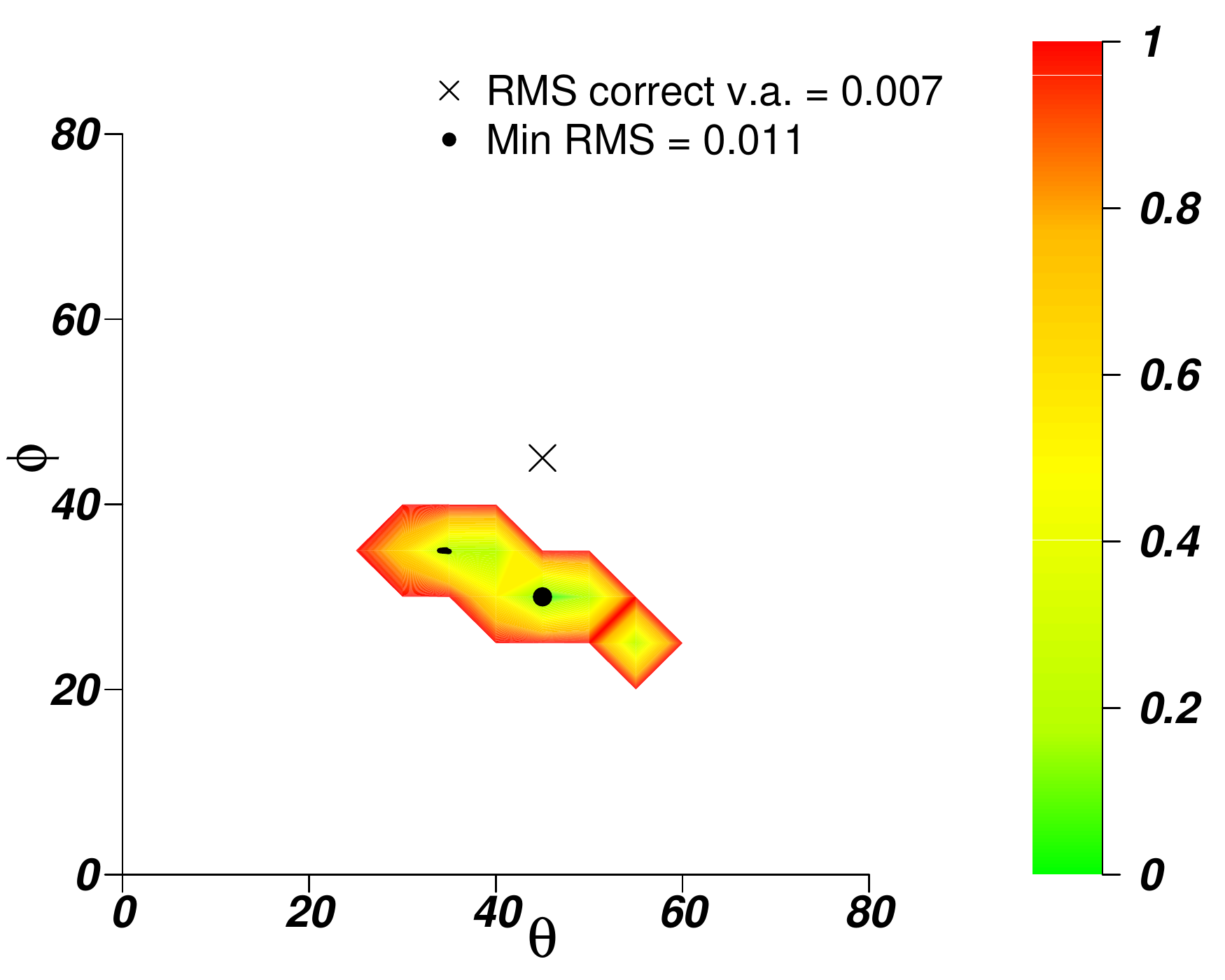}}
    \caption{Same as Fig.~\ref{Fig.Jaffe_ell_45} but for model \emph{DISCYBOXY}. We do not plot the white squares since in this case we do not need them to constrain the viewing direction.}
    \label{Fig.Jaffe_gamma_45}
\end{figure*}

Finally, we re-project the remaining densities along the principal axes and check the isophotal shapes, which is a technique already used in the axisymmetric case (e.g. \citealt{Jens05}). In fact, a plausible density for a giant elliptical galaxy is not expected to have too high (or too low) higher-order Fourier coefficients ($-5\le a_4\le 0.2$), too high ellipticity ($\geq 0.6$) or too severe twists ($\leq40\degr$). Examples of the second and third criteria are given in Fig.~\ref{Fig.bad_depros}.

\begin{figure*}
	\subfloat[SB comparison, $\psi=30\degr$.\label{Fig.Nbody_sb30}]
{\includegraphics[width=.35\linewidth]{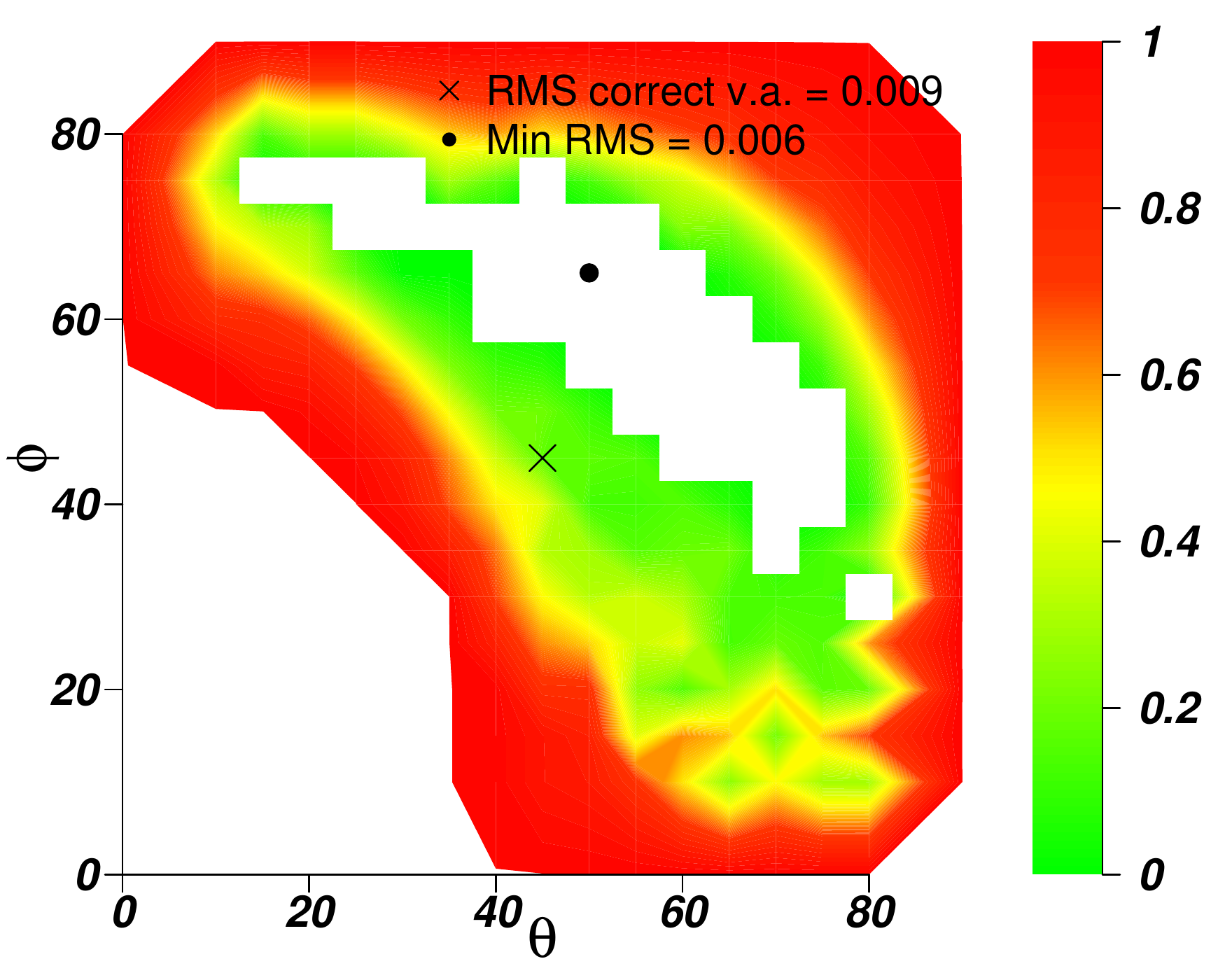}}
	\hfil
\subfloat[$\rho$ comparison, $\psi=30\degr$. \label{Fig.Nbody_rho30}]
{\includegraphics[width=.35\linewidth]{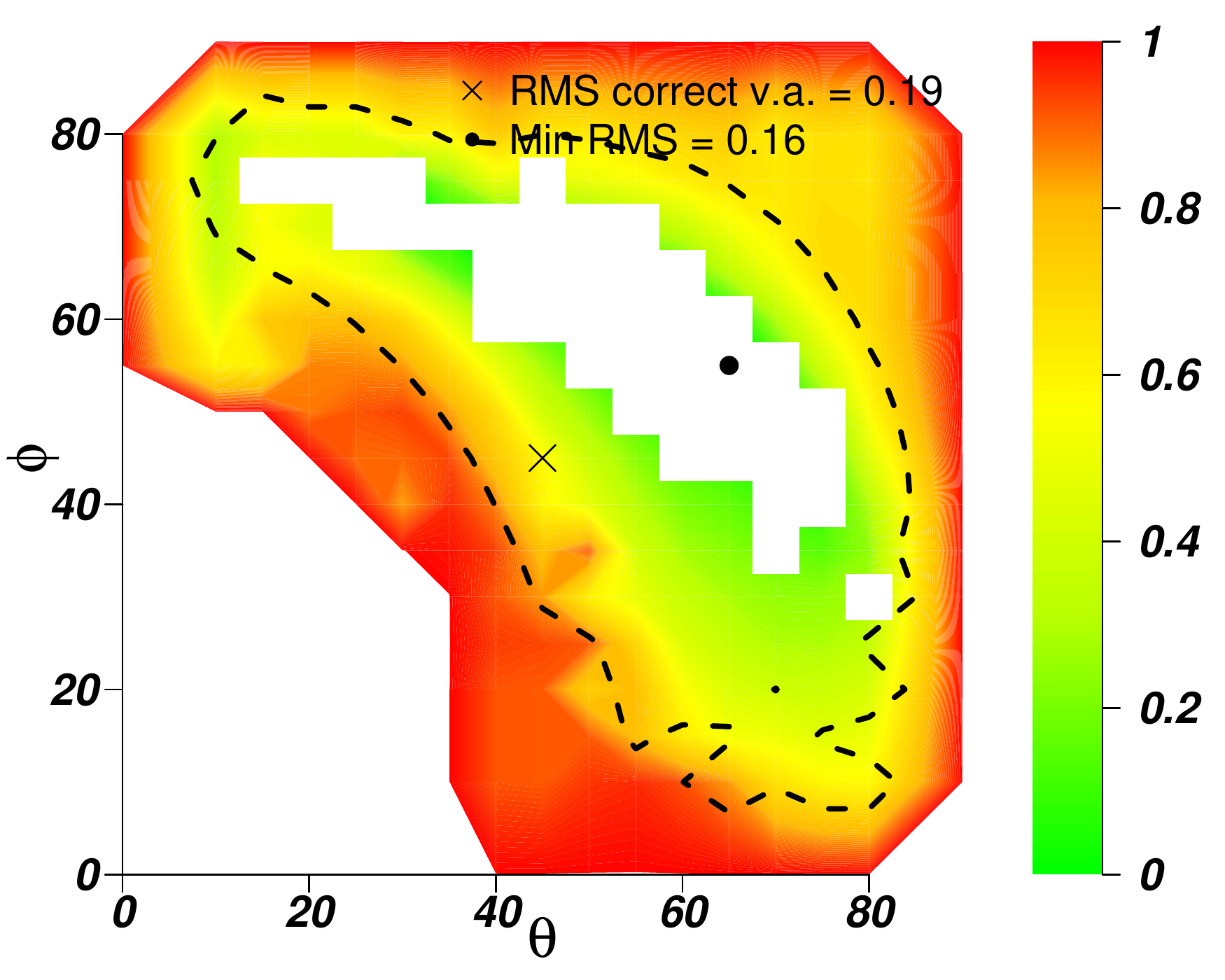}}
	\\
    \subfloat[SB comparison, $\psi=45\degr$. \label{Fig.Nbody_sb45}]
{\includegraphics[width=.35\linewidth]{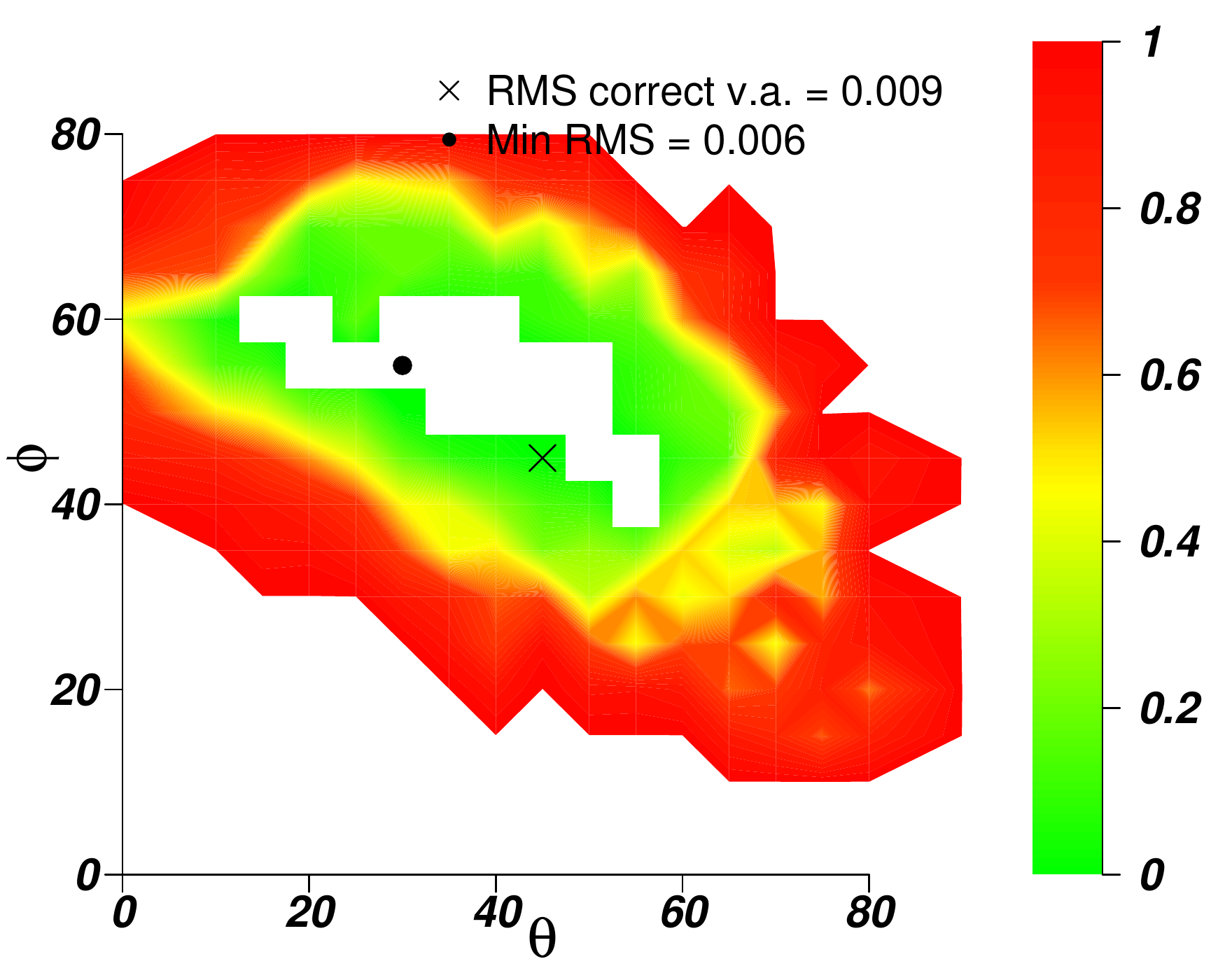}}
	\hfil
\subfloat[$\rho$ comparison, $\psi=45\degr$. \label{Fig.Nbody_rho45}]
{\includegraphics[width=.35\linewidth]{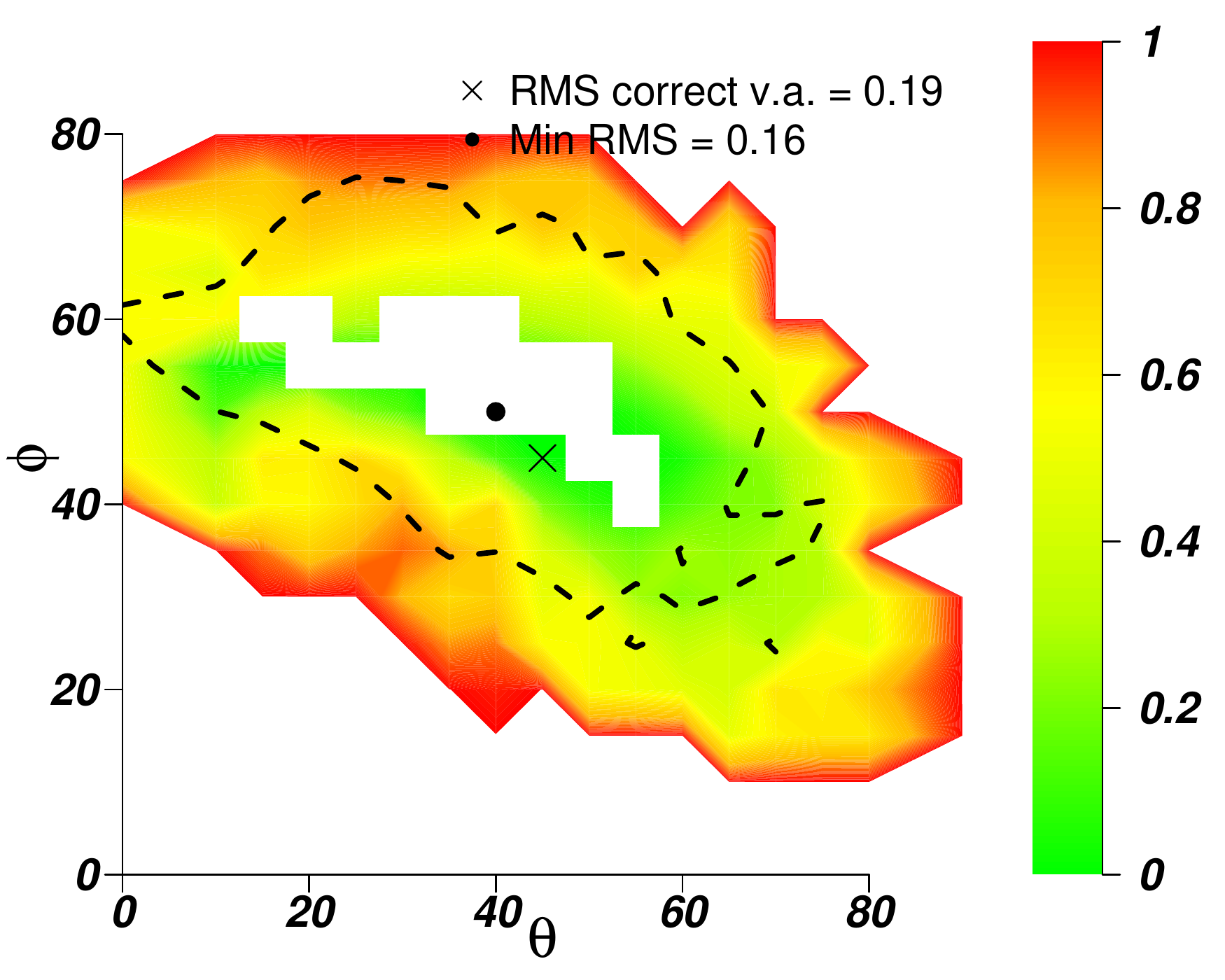}}
	\\
	\subfloat[SB comparison, $\psi=60\degr$. \label{Fig.Nbody_sb60}]
{\includegraphics[width=.35\linewidth]{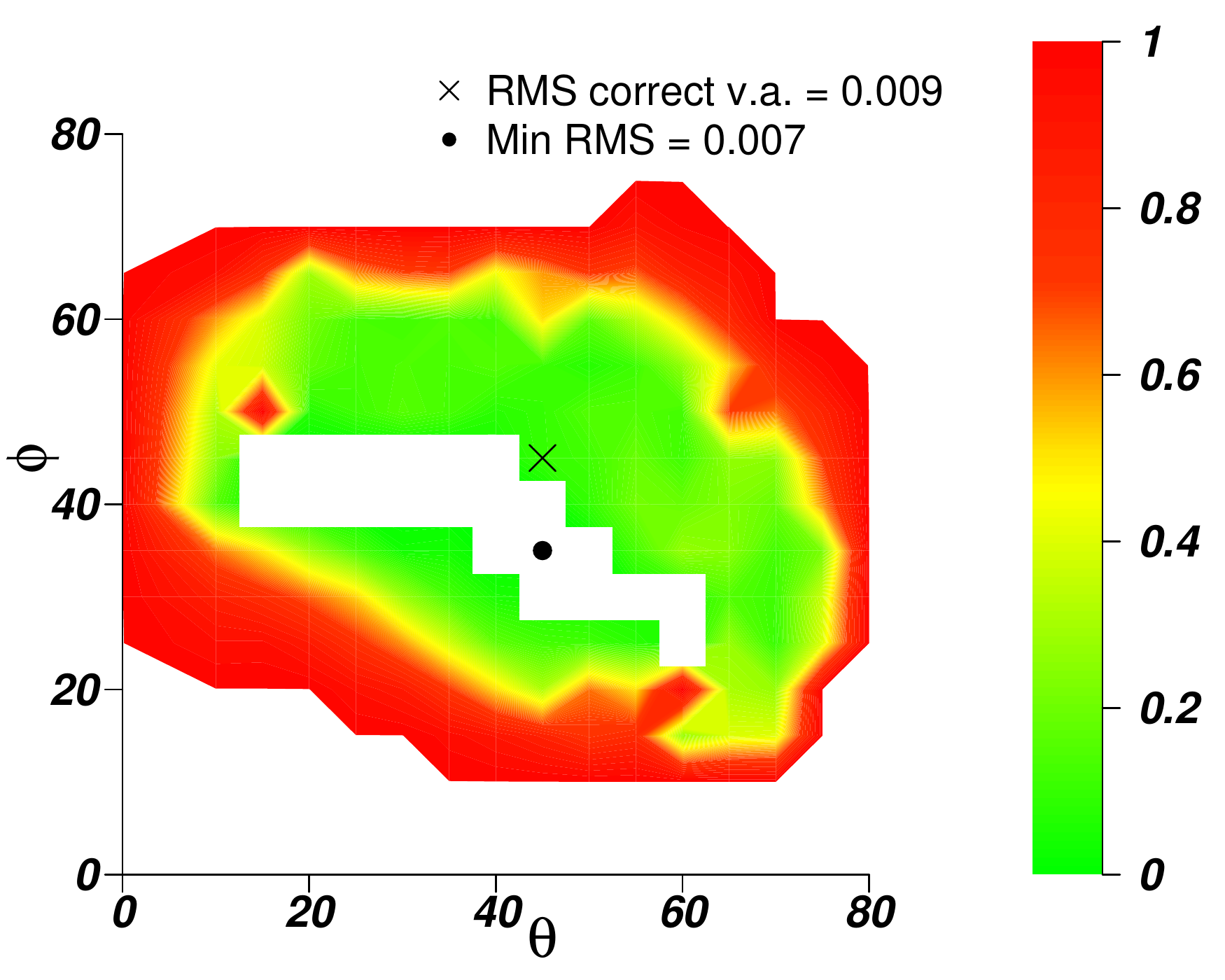}}
	\hfil
\subfloat[$\rho$ comparison, $\psi=60\degr$. \label{Fig.Nbody_rho60}]
{\includegraphics[width=.35\linewidth]{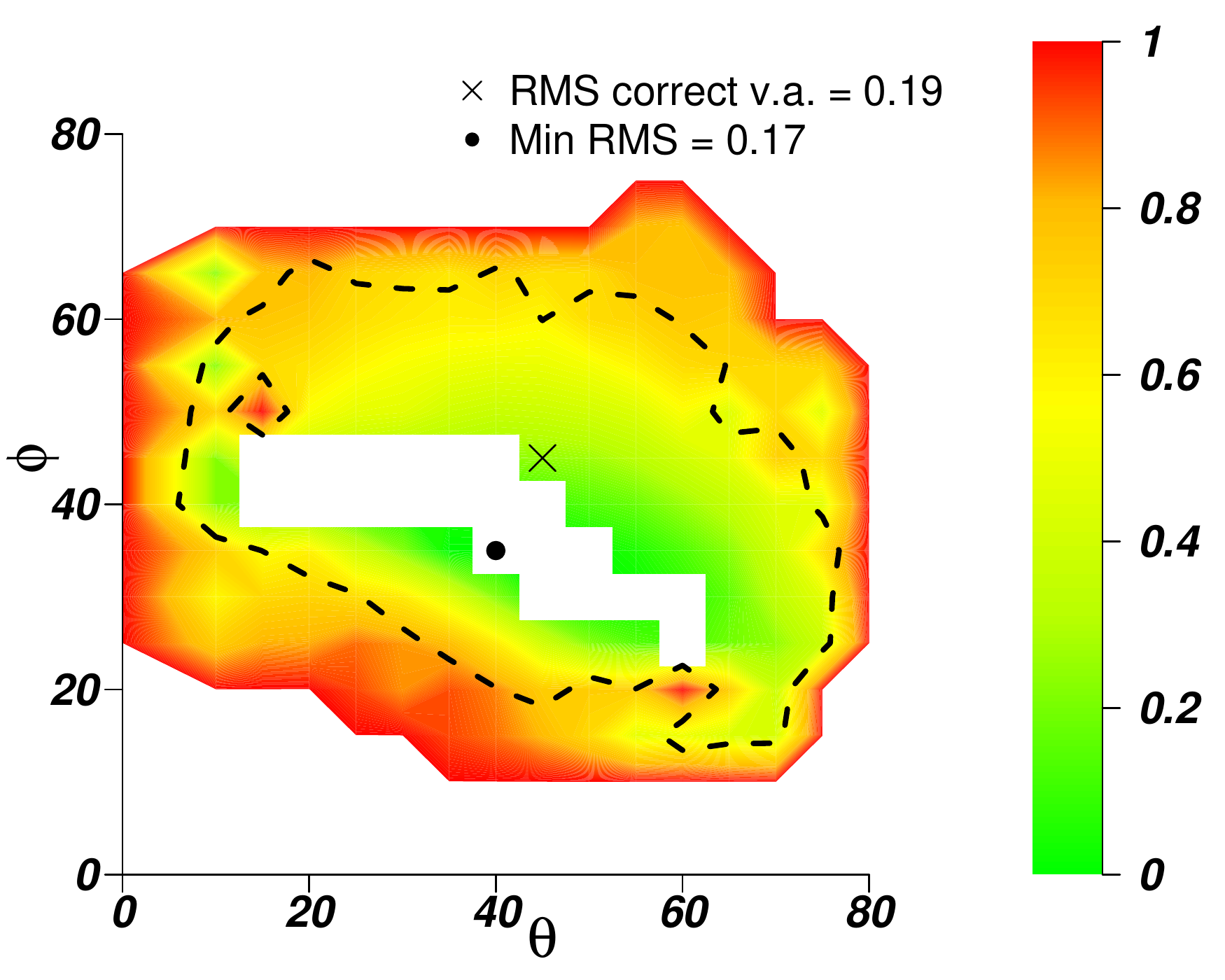}}

    \caption{Same as Figs.~\ref{Fig.Jaffe_ell_45} \& \ref{Fig.Jaffe_gamma_45} but for model \emph{NBODY}.}
    \label{Fig.Nbody_45}
\end{figure*}

\subsection{Results}
\label{sec_results}
In order to assess the results derived in the previous section, we plot in Figs.~\ref{Fig.Jaffe_ell_45}, \ref{Fig.Jaffe_gamma_45}, and \ref{Fig.Nbody_45} the RMS errors (both in SB and in $\rho$ and scaled to the RMS for the correct viewing angles) as a function of $(\theta,\,\phi)$ for the models \emph{ELLIP}, \emph{DISCYBOXY}, and \emph{NBODY}, respectively. In the top panels $\psi=45\degr$ (correct value), while the middle and the lower panels are for $\psi=30\degr$ and $\psi=60\degr$ respectively. In all these plots, a cross shows the correct $(\theta,\,\phi)$ and a block dot those corresponding to the minimum RMS. On the $\rho$ plots (right panels), we also show as dashed curve the contour delimiting the area inside which the RMS in SB is within twice the value for the correct viewing angles. White quadratic holes are regions omitted because of implausible $p$ or $q$ profiles or re-projections.

The main conclusions we can draw from these figures are as follows.
\begin{enumerate}
\item Regardless of the value of $\psi$, there are wrong viewing angles $(\theta,\,\phi)$ for which the code can find a fit to the SB even slightly better than that at the true viewing angles. This does not happen for model \emph{DISCYBOXY}, for which no solutions at wrong viewing angles are found, suggesting that the introduction of a variable $\xi$ profile shrinks the region of acceptable deprojections. Larger acceptable regions are found for model \emph{NBODY} (Fig.~\ref{Fig.Nbody_45}), where noise is present.
\item For \emph{ELLIP} and \emph{DISCYBOXY} the RMS differences for both SB and $\rho$ between the true and wrong viewing angles can be as high as three orders of magnitude (Figs.~\ref{Fig.Jaffe_ell_45} \& \ref{Fig.Jaffe_gamma_45}), but the noisy model \emph{NBODY} allows only for a one order of magnitude range in RMS (Fig.~\ref{Fig.Nbody_45}).  Furthermore, model \emph{NBODY}, because of its noise, is the only one for which the viewing angles that give the best RMS \emph{in $\rho$} are different from the true ones. However, these solutions have intersecting noisy $p$ and $q$ profiles, that according to our selection rules would be excluded (see Fig.~\ref{Fig.bad_depros}).
\item For \emph{ELLIP} and \emph{NBODY}, not only we find wrong intrinsic densities that project to a very good fit to the observed SB (as it happens in the non-parametric case), but also intrinsic densities with low RMS's which do not project to an acceptable fit to the observed SB profile.
\item If $\psi$ is wrong, then the $(\theta,\,\phi)$ pair that give the best RMS in SB is far off the true one. Moreover, the correct $(\theta,\,\phi)$ pair combined with the wrong $\psi$ can result in an RMS an order of magnitude larger than for the correct projection. Since the observed ellipticity and twist of a given model depend both on $(\theta,\,\phi)$ and the $p$, $q$ profiles, it is not immediately clear whether a set of wrong viewing angles cannot deliver a good solution, as the case discussed above for the two values of $\psi$ shows. In Fig.~\ref{Fig.twist_strong} we have already seen that intersecting $p$ and $q$ profiles help in generating large observed twists.
\item The conditions we apply to the $p$, $q$ profiles and the re-projections along the principal axes shrink the allowed range of viewing angles much more strongly for \emph{NBODY} than for models \emph{ELLIP} and \emph{DISCYBOXY}. The presence of noise allows to generate deprojected intrinsic densities that are 'stranger' and therefore more easily eliminated than in a noise-free case. Since the SB profile of an ordinary massive elliptical are not as noisy as our \emph{NBODY}, more deprojections are likely to survive these conditions when dealing with real galaxies.
\end{enumerate}

We have shown in a qualitative manner that the statistical photometric properties of massive ellipticals (observed ellipticities, isophotal distortions and isophote twists) can be modelled with deformed ellipsoidal intrinsic density distributions.  As long as the assumption of deformed-ellipsoidal density distributions holds, the range of possible deprojections shrinks considerably. In fact, since the deprojection becomes formally unique, comparing the fit quality of different deprojections at different assumed viewing angles can be used to narrow down the possible LOS of a massive galaxy just from photometric data. We plan to study in detail the intrinsic shape distribution of massive galaxies in a separate paper.

\section{Comparison with the MGE approach} 
\label{par.MGE}

\begin{table}
    \centering
    \caption{MGE fit to the Jaffe model detailed in this section. Luminosities are in counts/pixel/$10^8$, $\sigma$'s in pixels and position angles (PA) in degrees. From these values one can compute the SB profile using equation~(1) of \citet{Cappellari02}.}
    \begin{tabular}{c c c c}
    \hline
    $L$ & $\sigma$ & $q_s$ & PA \\ \hline
  1.80751  &    11.7388    &  0.72202  &   0.171955 \\
  1.11501   &   11.9416   &  0.485346  &   -1.85816 \\
  3.29186   &   23.4278   &  0.626633   &   -1.0591 \\
  2.82643   &   37.5102   &  0.630021   & -0.217961 \\
  3.43627   &   54.6153   &  0.629184   &  -1.75377 \\
  4.63525   &    90.624   &  0.632035   &  0.102047 \\
    2.874   &   147.446   &  0.705932   &        -5 \\
  3.04795   &   210.449   &   0.59828   &   2.22027 \\
  6.79399   &   404.492   &  0.817577   &  -0.29154 \\
         
    \hline
    \end{tabular}
    \label{Tab.MGE}
\end{table}

The Multi-Gaussian Expansion \citep{Cappellari02, VDB08} is a fast tool to deproject a SB profile, directly from a \texttt{FITS} file, assuming both the SB and the density profiles can be approximated as a sum of Gaussians. This analytic approach produces a fit
that can be reduced to a small set of numbers, yields smooth solutions, is fast, is bound to deliver reasonable re-projected SB whatever viewing angles are considered, and delivers a unique deprojection for a set of allowed viewing angles. However, this set can be empty if the gaussians required to get a good fit span a large range of twists or have very low flattenings. Therefore, here we first apply it to the models considered in the previous sections \emph{without} the additional flat components discussed in Appendix~\ref{sec_discs} to assess its performances in terms of quality of the reproduced density and set of allowed viewing angles. Then, we compare its results with those obtained by our code for a real galaxy.

\subsection{MGE performance on the Jaffe model}
\label{par.Jaffe_MGE}
We make sure that our coordinate system on the plane of the sky is consistent with what the MGE assumes, namely that the major axis of the innermost Gaussian component is aligned with the $x'$-axis. Since the twist of our Jaffe model is nearly zero in the innermost regions (see Fig.~\ref{Fig.twist_weak}), we can assume that aligning the innermost isophote with the $x'$-axis is to a very good approximation the same as aligning the innermost Gaussian of the MGE fit.  This is achieved by rotating our isophotes clockwise by the PA of the innermost isophote ($\sim 78\degr$) and adding this value to the $\psi = 45\degr$ we used above, giving a new $\psi_{\mathrm{MGE}} = 123\degr$.

\begin{figure*}
	\subfloat[$\psi = 108\degr$. \label{Fig.rms_MGE_2}]{\includegraphics[width=.32\linewidth]{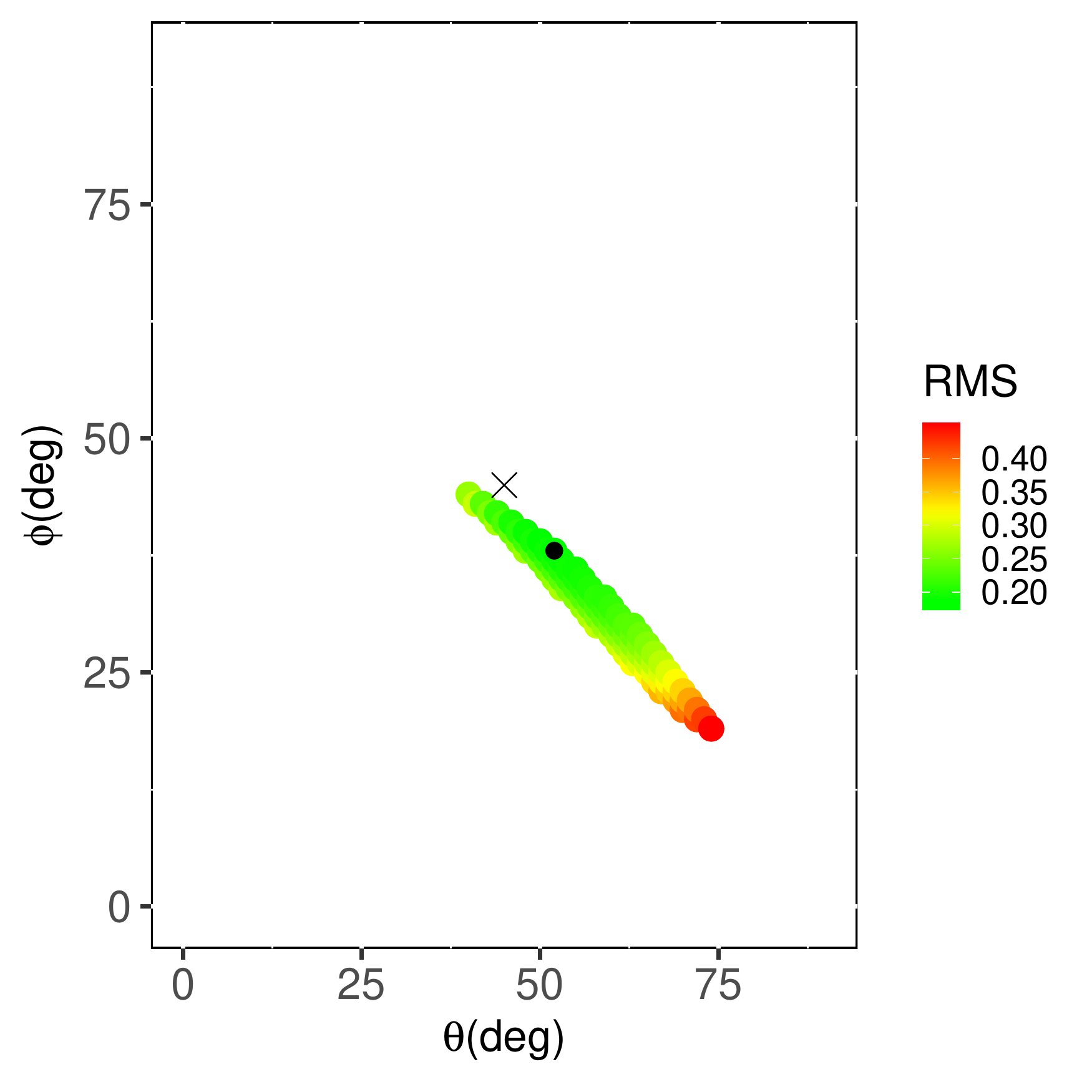}}
	\hfil
	\subfloat[$\psi = 123\degr$. \label{Fig.rms_MGE_1}]{\includegraphics[width=.32\linewidth]{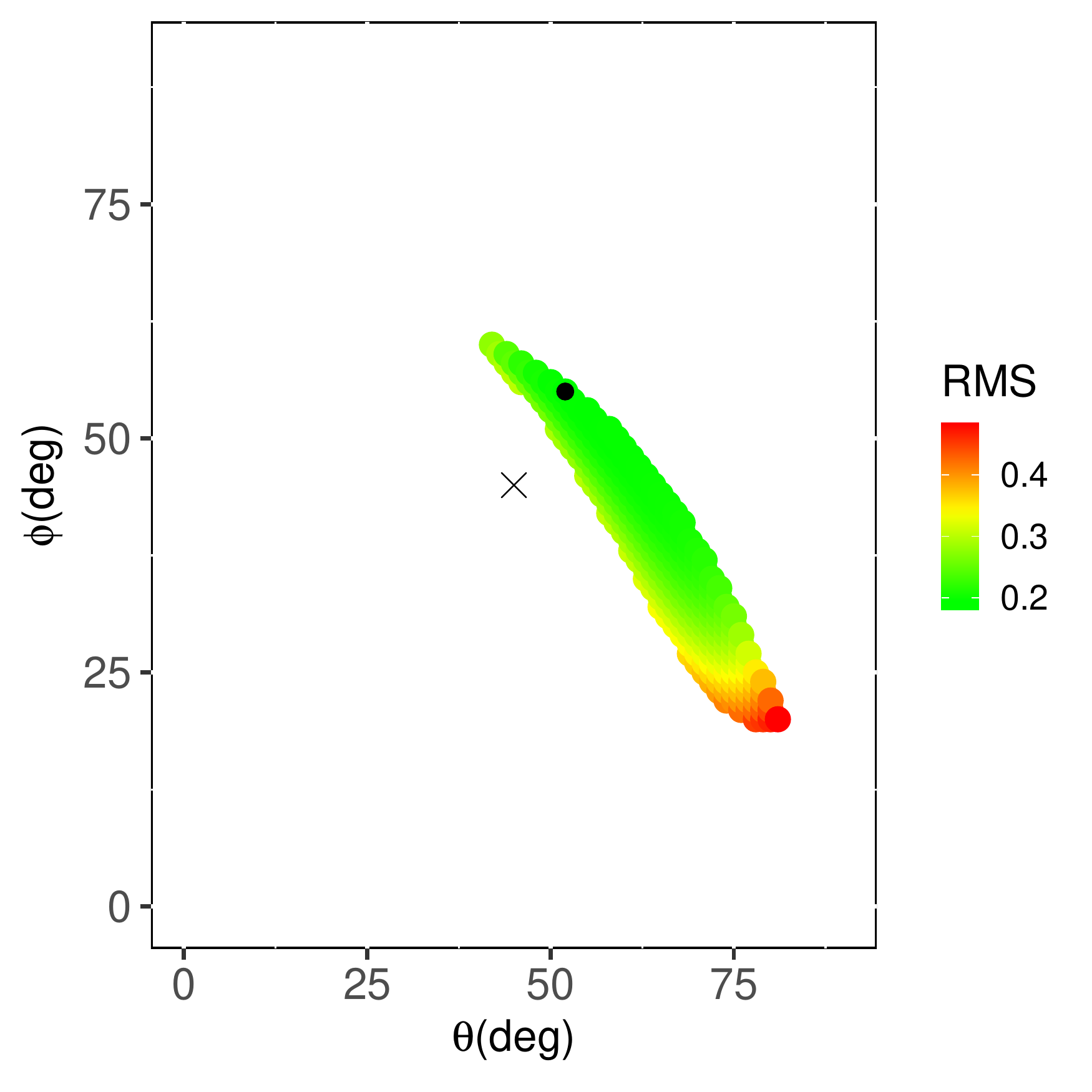}}
	\hfil
	\subfloat[$\psi = 138\degr$. \label{Fig.rms_MGE_3}]{\includegraphics[width=.32\linewidth]{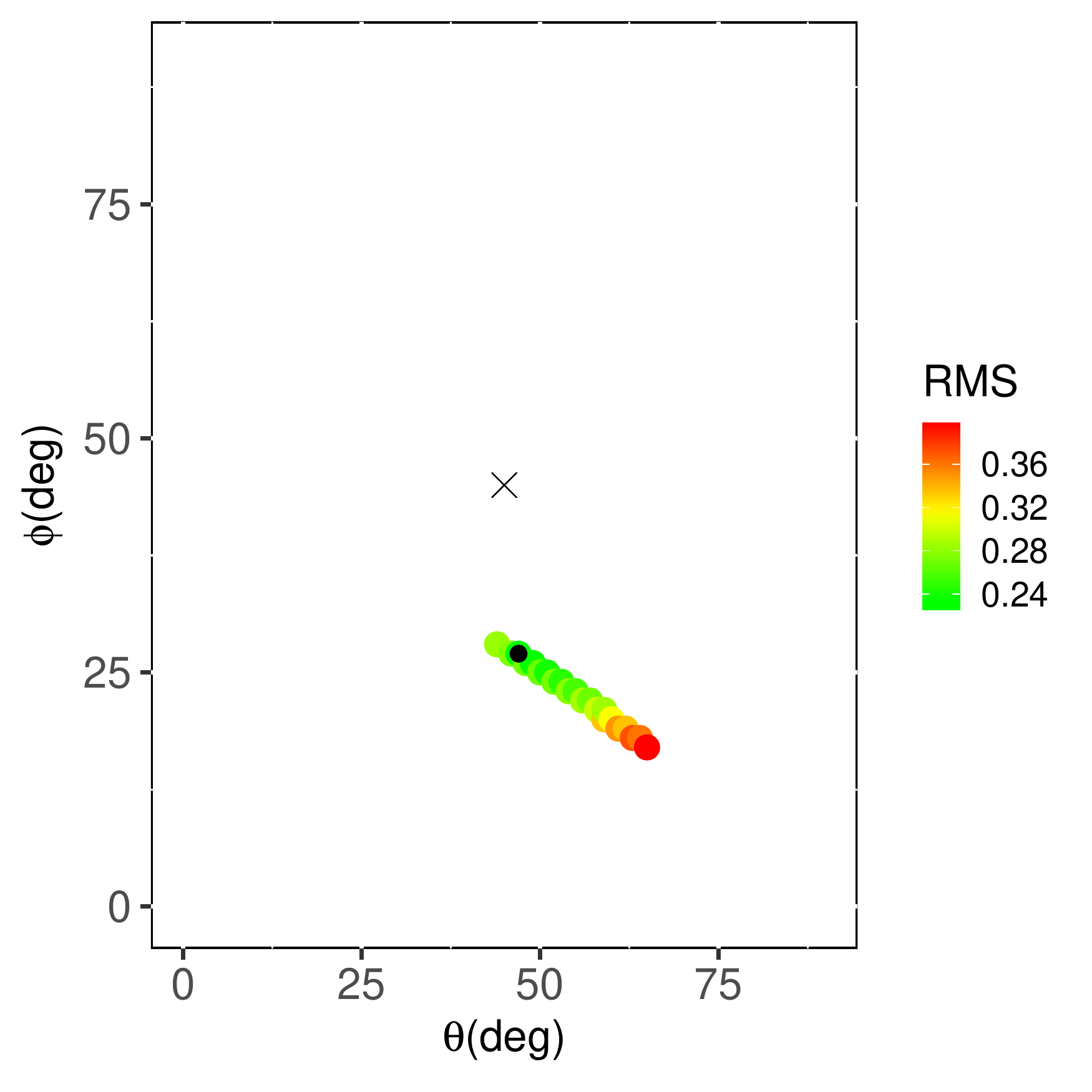}}
	\caption{RMS between the true intrinsic density and that recovered by the MGE for the true value of $\psi = 123\degr$ (middle) and for two wrong values of $\psi = 108\degr$ (left) and $\psi = 138\degr$ (right) as a function of all possible $\theta, \phi$ values compatible with this particular $\psi$. The cross and the black dot indicate the true ($\theta, \phi$) viewing angles and those at which the MGE deprojections give the least RMS, respectively.}
    \label{Fig.rms_MGE}
\end{figure*}

\begin{figure}
    \centering
    \includegraphics[width=\columnwidth,trim=0mm 1mm 2mm 2mm]{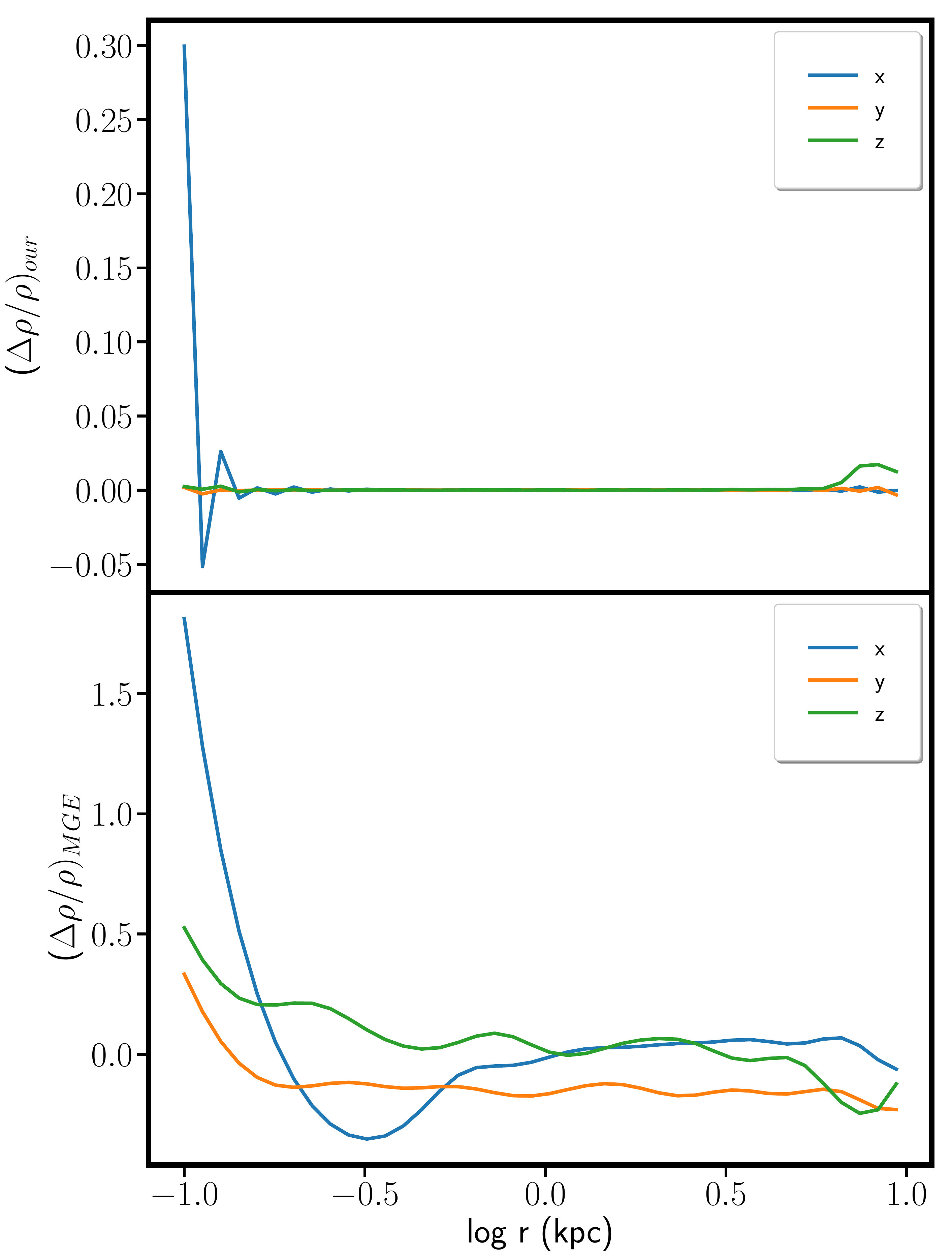}
    \caption{Comparison of the reconstructed intrinsic density of \emph{DISCYBOXY} between our code (top) and MGE (bottom) on the three principal axes. For both cases, we took the deprojection at the viewing angles giving the best RMS in $\rho$, even though these are not exactly the true ones.}
    \label{Fig.mine_mge}
\end{figure}

We project the density of the \emph{DISCYBOXY} model with viewing angles ($\theta=45\degr, \phi=45\degr, \psi_{\mathrm{MGE}}=123\degr$) and generate the galaxy image in \texttt{FITS} format as an input to the code of Michele Cappellari\footnote{\hyperlink{mcapp}{http://www-astro.physics.ox.ac.uk/$\sim$mxc/software/}} to produce the MGE fit. The procedure fits the image with a combination of $N=8$ Gaussians which we report in Table~\ref{Tab.MGE}. We tested several fits, each time by imposing different constraints on the flattenings $q_j'$ and the twist $\Delta\psi_j$, allowing for up to 30 Gaussians in the fit. The constraints are needed because by letting the code run unconstrained we obtained a solution for which no possible viewing angles are found; we ended up using $q_j' \in [0.2,1]$ and $\Delta\psi_j \in [-5,5]\degr$.  The RMS between the MGE SB and that we have on our grid \eqref{grid_sky} is 4.9\%. We then compute the intrinsic densities corresponding to the allowed viewing angles (see Section~\ref{sec:ellipsoidal} and equations (7-8) of \citealt{Cappellari02}). Clearly, while our code can produce a deprojection for each possible set of viewing angles, this is not possible for the MGE. Thus, to construct analogs of Fig.~\ref{Fig.Jaffe_ell_45} we have isolated all solutions which have the true $\psi = 123\degr$ (Fig.~\ref{Fig.rms_MGE_1}), then those at $\psi = 108\degr$ (Fig.~\ref{Fig.rms_MGE_2}) and finally those at $\psi = 138\degr$ (Fig.~\ref{Fig.rms_MGE_3}) and plot the RMS with respect to the true intrinsic density. The meaning of the black cross and dot are the same as in Fig.~\ref{Fig.Jaffe_ell_45}. The most significant findings are as follows.

\begin{itemize}
\item The quality of the MGE fit is poorer than the one achieved with the constrained-shape algorithm, delivering an RMS in SB of nearly 5\%. This is not surprising, since a superposition of a series of densities localized in shells provides much more flexibility than a set of Gaussians.
\item An MGE deprojection for the true viewing angles is possible.
\item Of all possible MGE deprojections, the one that gives the best RMS in $\rho$ ($\sim18\%$) is obtained for viewing angles of $\theta=49\degr,\,\phi=44\degr,\,\psi=119\degr$, different from the true values by a few degrees.
\item The RMS in $\rho$ yielded by the constrained-shape algorithm is significantly smaller (0.7\%) than the one achievable with the MGE approach, even omitting the last 10 radial points, where the Gaussians have a sharp cut-off (see also Fig.~\ref{Fig.mine_mge}).
\end{itemize} 

\subsection{Comparison using a real galaxy}
It is now interesting to compare our code with MGE for a real galaxy, which can neither be described exactly by a sum of Gaussians nor has the form of eq.~\ref{m}. We focus on the elliptical galaxy NGC5831, which has a $\sim35^\circ$ isophote twist and is also used by \citet{Cappellari02} as an example for the performance of MGE in presence of isophote twist. The released MGE Python code fits the photometry with a sum of 11 Gaussians yielding an RMS of 4.7\%; \citet{Cappellari02} quotes an even better RMS of 1.2\%, that we adopt as a benchmark.

\begin{figure*}
	\subfloat[$\psi = 30\degr$. \label{Fig.rms_N5831_30}]{\includegraphics[width=.32\linewidth]{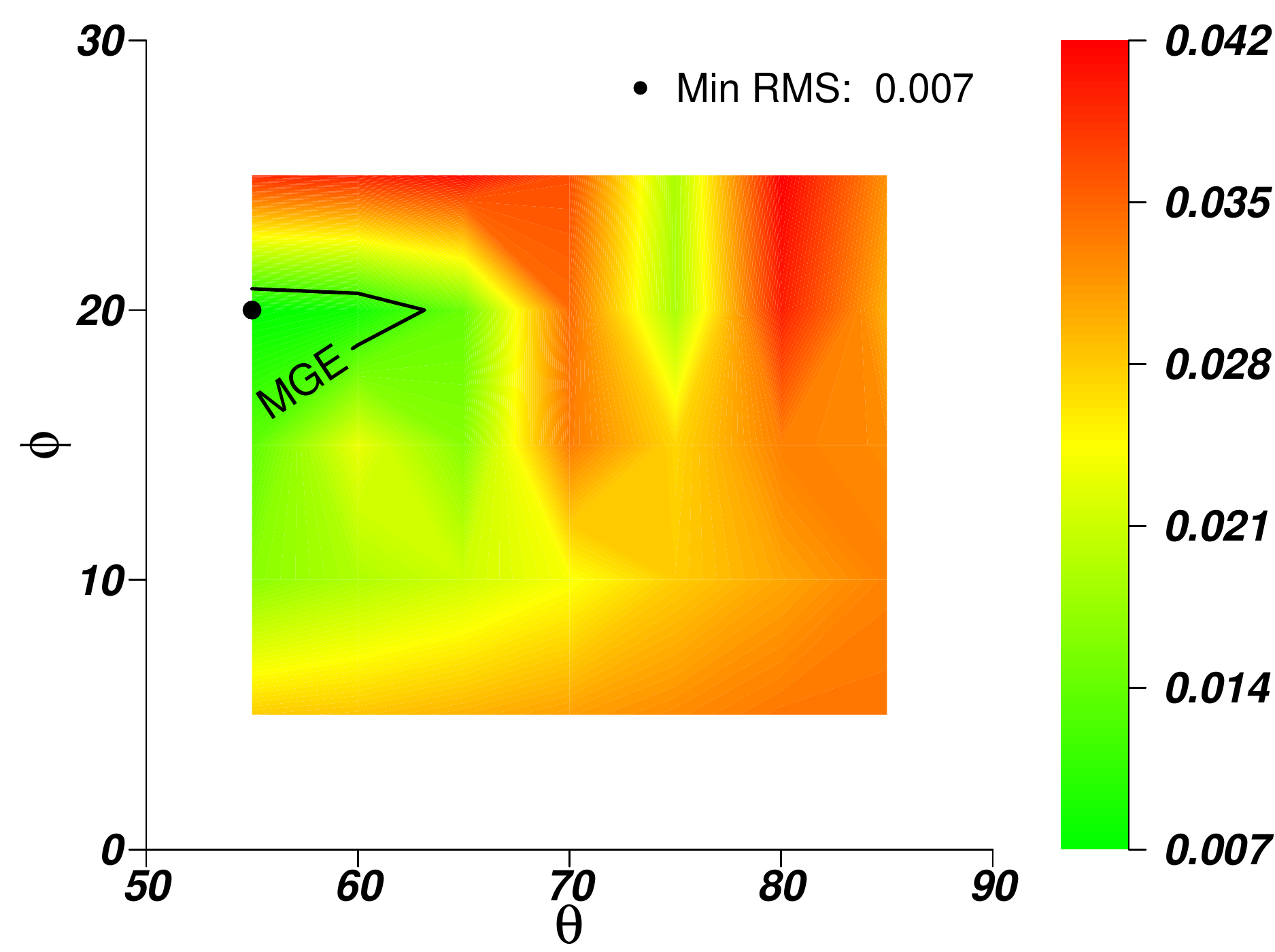}}
	\hfil
	\subfloat[$\psi = 40\degr$. \label{Fig.rms_N5831_40}]{\includegraphics[width=.32\linewidth]{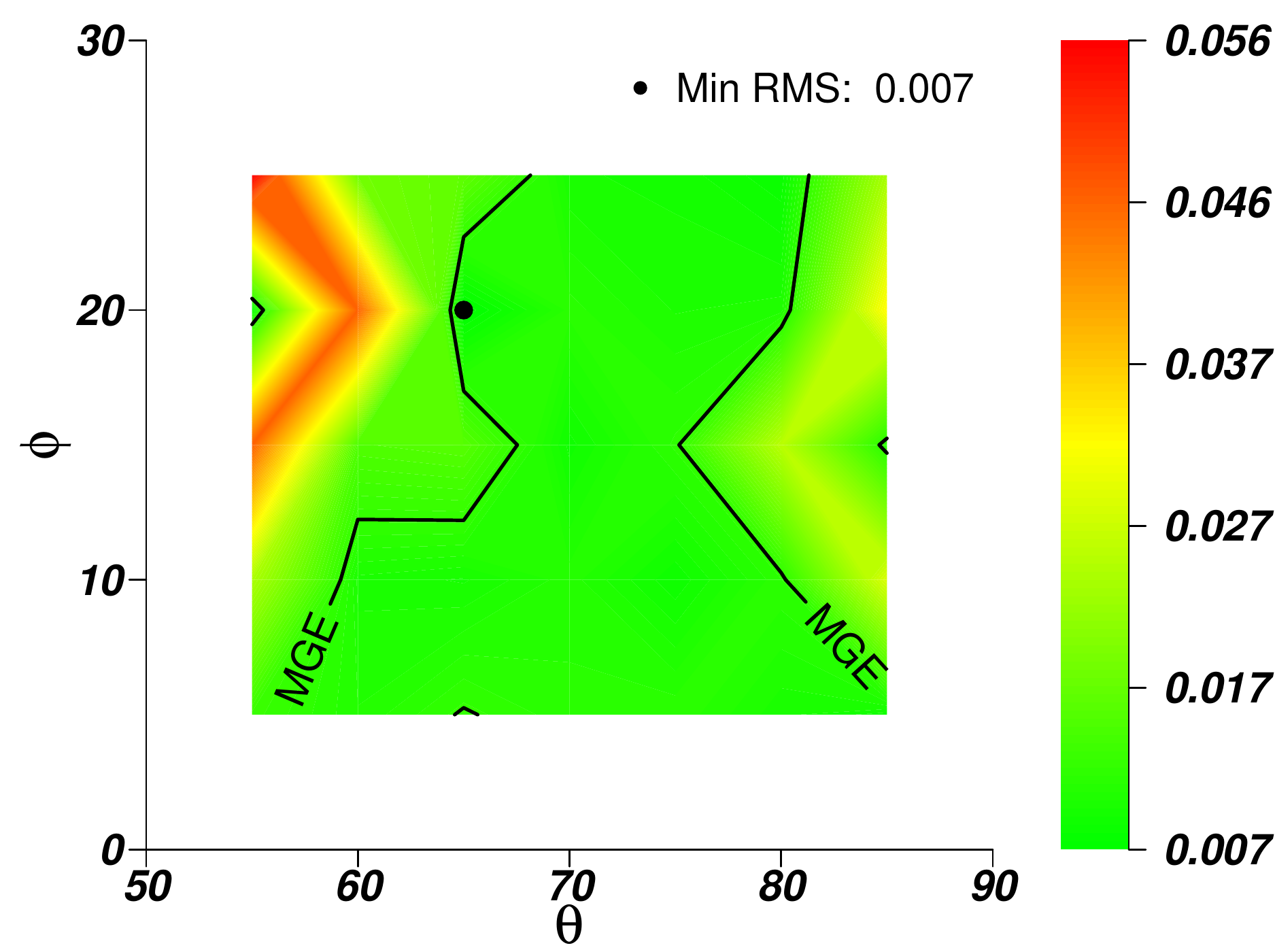}}
	\hfil
	\subfloat[$\psi = 50\degr$. \label{Fig.rms_N5831_50}]{\includegraphics[width=.32\linewidth]{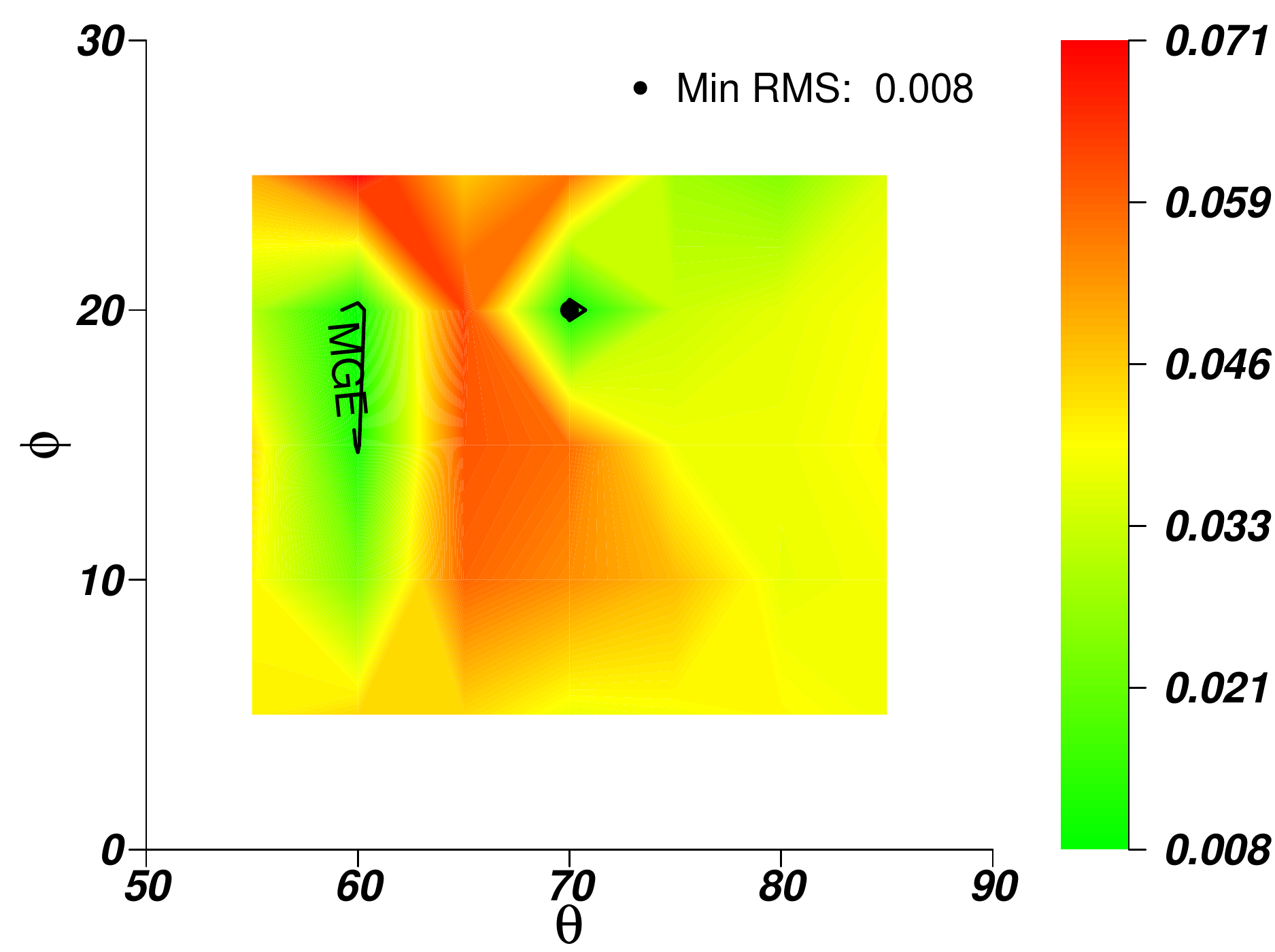}}
	\caption{RMS between the true SB and that reconstructed by our code for the elliptical galaxy NGC\,5831 for $\psi = 30\degr,\,40\degr$, and $50\degr$ as indicated. We sampled $\theta, \phi$ in the intervals where the MGE method allows for a solution. The black dots indicate the respective ($\theta, \phi$) with the smallest RMS. The black contours bracket the regions where RMS$\,\langle 1.2\%$.}
    \label{Fig.rms_N5831}
\end{figure*}

\begin{table}
\caption{Highest ellipticity $\varepsilon$ and ranges for $a_4$ and $a_6$ found when re-projecting the intrinsic densities obtained for NGC\,5831 along 60 random directions. No unphysical or unusual values for these coefficients are found.}
    \label{Tab.repro}
    \centering
    \begin{tabular}{c c c c}
    $\psi$ & $\varepsilon$ & $a_4$ & $a_6$ \\ \hline
  30$^\circ$  &    0.496    &  $\left[-0.51, 1.43\right]$  &   $\left[-0.09, 0.78\right]$ \\
  40$^\circ$    &   0.512   & $\left[-0.78, 2.06\right]$  &   $\left[-0.32, 1.12\right]$ \\
  50$^\circ$    &  0.478   &  $\left[-0.23, 1.65\right]$   &   $\left[-0.05, 0.91\right]$ \\ \hline
    \end{tabular}
\end{table}

We computed all the $(\theta,\,\phi,\,\psi)$ values compatible with the MGE fit, obtaining $\theta \in [55,85]^\circ,\, \phi \in [5,25]^\circ,\, \psi \in [106,114]^\circ$. Then, we used our near-ellipsoidal algorithm to deproject the surface density of the galaxy considering $\theta$ and $\psi$ in the interval allowed by the MGE fit with a step of $5^\circ$. Since the definition of $\psi$ adopted in the MGE formalism differs from ours (see Section~\ref{par.Jaffe_MGE}), we sampled $\psi \in [10,180]^\circ$ with a step of $10^\circ$.

%Similarly to Fig.~\ref{Fig.rms_MGE} \textbf{\B{WALTER: it's not similar to Fig.~\ref{Fig.rms_MGE}, which gives rms for the intrinsic density, not the surface density.}}, we plot in Fig.~\ref{Fig.rms_N5831} the results for the $\psi$ value with the largest number of good solutions ($40\degr$ in this case), then $\psi=50\degr$ and finally $\psi=30\degr$. 
In Fig.~\ref{Fig.rms_N5831}, we plot the RMS error of the surface density as function of $\theta$ and $\phi$ for $\psi=30\degr,\,40\degr$ and $50\degr.$ \textbf{In all three cases, we find viewing directions at which we can fit the surface brightness better than the MGE}. In particular, for all three $\psi$ values we find deprojections with RMS$\,<1\%$, with the $\psi=40\degr$ case having $\sim70$\% SBs below this threshold.

As outlined in Sec.~\ref{sec_implausibledeproj}, we reprojected the resulting intrinsic densities for 60 random viewing angles to ensure that the isophotes look reasonable. We show in Tab.~\ref{Tab.repro} the largest ellipticity $\varepsilon$ and the maximum/minimum a4/a6 values we found among all densities for a certain $\psi$. Here we do not apply any requirements on the twist angle $\tau$ since differently from most massive ellipticals this galaxy does have a strong twist in the outer regions. We see that the isophotes are never particularly flat and do not show anomalous $a_4$ or $a_6$ values. It is interesting to note that we do not find any re-projection yielding significantly boxy isophotes.\\

The main conclusion is that, although the MGE naturally rules out unsmooth densities and is fast, it might bias the region of allowed angles $\theta,\,\phi,\,\psi$ and deliver an SB fit and reconstructed density of relatively poor quality. We will investigate the impact of these shortcomings on dynamical modelling in a future paper.

\section{Conclusions}
\label{sec_conclusions}

We present two novel approaches to deproject elliptical galaxies under the assumption of triaxiality, the first fully non-parametric, the second stratified on deformed ellipsoidal shells. Both are able to deal with isophotal twist and can deproject systems that have the principal axes interchanging between them as a function of the distance from the centre.

The full non-parametric code can be used to explore a range of possible deprojections going beyond those allowed by current state-of-art algorithms, but at present does not allow for any control of the shape of the density. Our constrained-shape approach, on the other hand, allows for penalization towards discy/boxy shapes and controls the smoothness of the density along the major axis and of the density contours.  Tests performed with benchmark Jaffe and Hernquist models of varying axis ratios and shape biases show that the intrinsic density can be recovered very well when the viewing angles are known and far enough from the principal axes, much better than what can be achieved with a Multiple Gaussian Expansion. When dealing with a noisy system such as an $N$-body simulation, the SB can be fitted with an RMS $\sim1\%$, delivering a reconstructed density precise to 20\%, when the viewing angles are known.

We are able to constrain the possible range of viewing angles by mapping the RMS of the fitted SB as a function of $(\theta,\,\phi,\,\psi)$ and eliminating unphysical reconstructed densities, by examining their re-projected SB. This reduces the number of densities to be tested dynamically, which will be the subject of a forthcoming paper, towards the deprojections and dynamical modelling of real galaxies.

In this process we might discover that a number of galaxies appear similar to the \emph{LARGEDISC} discussed in Appendix~\ref{sec_discs}, where a massive disc component is present together with a triaxial bulge. For unfavourable viewing angles this component is invisible in projection. In Appendix \ref{sec_cloaked} we discuss a number of analytic descriptions of these possible \emph{cloaked densities}, the triaxial extension of axisymmetric \emph{conus densities}. In Appendix~\ref{sec_discs} we show how one can flag these cases. We explore how well we can deproject triaxial bodies where flattened axisymmetric components are present. We find that the constrained-shape approach performs well if these (disc like) components do not exceed 15\% of the total light. We develop a deformed-ellipsoidal shape plus axisymmetric component algorithm that is able to reconstruct well systems with nearly edge-on ($\approx 80\degr$) discs, or flag the possible presence of important (i.e.\ contributing $\approx 50$\% of the total light) disc components at unfavourable angles ($\le 45\degr$).

\section*{Acknowledgements}
We thank the anonymous referee for a constructive report that helped us improving the presentation of our results. We acknowledge the support by the DFG Cluster of Excellence ``Origin and Structure of the Universe''. The simulations have been carried out on the computing facilities of the Computational Center for Particle and Astrophysics (C2PAP). We thank B. Neureiter, A. Rantala, T. Naab and M. Frigo for providing us with the density of the $N$-body model.

\section*{Data Availability}
The data underlying this article will be shared on reasonable request to the corresponding author.
	
\bibliographystyle{mnras}
\bibliography{bibl}

\appendix

%%%%%%%%%%%%%%%%%%%%
\section{Constructing cloaked densities}
\label{sec_cloaked}
We are now considering various ways to construct analytical models that project to nothing. Such \emph{cloaked densities} can be added or subtracted to any model without changing its projection but potentially with drastic changes to its spatial shape.

%%%%%%%%%%%%%%%%%%
\begin{table*}
    \caption{\textbf{Functions used by ellipsoidal cloaked densities.}  Analytical 3D spherical distributions $\varphi(r)$, normalised to $\varphi(0)=1$, with Fourier transform $\hat{\varphi}(k)$ that vanishes for $k\ge1$. Functions with $\forall r: \varphi(r)>0$ have 3D \citet{Wendland1995,Wendland2005} functions as Fourier transform. $\sigma_k^2$ is the 1D variance of $\hat{\varphi}(k)$ and determines the width of $\varphi$ since $\varphi(r\to0)=1-\tfrac12\sigma_k^2r^2+O(r^4)$.
	\label{tab:func:ell}
}

\medskip
\begin{tabular}{lll@{$\;\approx\;$}ll}
	$\hat{\varphi}(k<1)/\hat{\varphi}(0) $ & $\varphi(r)$
	& %$\varphi(r\to0)$ or
	\multicolumn{2}{c}{$\sigma_k^2$} & $\varphi>0$
	\\[1ex]
	\hline
	\\[-2ex]
	$1$	& $3r^{-2}\left[\sinc r-\cos r\right]$
	& $\tfrac15$ & 0.2 %$1-\frac1{10}r^2+\frac1{280}r^4\dots$	
	& no	\\[2ex]
	$1-k$
	& $12r^{-4}\left[2-2\cos r-r\sin r\right]$
	& $\tfrac2{15}$ & 0.1333 %$1-\frac1{15}r^2+\frac1{560}r^4\dots$
	& no 	\\[2ex]
	$(1-k)^2$ & $60r^{-4}\left[2+\cos r-3\sinc r\right]$
	& $\frac2{21}$ & 0.0952 %$1-\frac1{21}r^2+\frac1{1008}r^4\dots$
	& yes	\\[2ex]
	$(1-k)^3(1+3k)$
	& $630r^{-7}\left[r(8+7\cos r)+(r^2-15)\sin r\right]$
	& $\frac1{12}$ & 0.0833 %$1-\frac1{24}r^2+\frac1{1320}r^4\dots$
	& no	\\[2ex]
	$(1-k)^4(1+4k)$
	& $5040r^{-8}\left[9r\sin r+(24-r^2)\cos r+4r^2-24\right]$
	& $\frac1{15}$ & 0.0667 %$1-\frac1{30} r^2 + \frac1{1980} r^4 \dots $
	& yes	\\[2ex]
	$\tfrac12(1+\cos k\pi)$
	& $\frac{3\pi^2}{\pi^2-6}\left[r^{-2}(\sinc r-\cos r)-\frac{1}{\pi^2-r^2}\left(\frac{\pi^2+r^2}{\pi^2-r^2}\sinc r + \cos r\right)\right]$
	& $\frac{\pi^4-20\pi^2+120}{5\pi^2(\pi^2-6)}$ & 0.1048
	& no	\\[2ex]
	\hline
\end{tabular}
\end{table*}
%%%%%%%%%%%%%%%%%%

%%%%%%%%%%%%%%%%%%%%
\subsection{Cloaked densities via differentiation}
We are after triaxial functions whose Fourier transform vanishes on the four planes $\vec{\ell}_i\cdot\vec{k}=0$. A simple such function is 
\begin{multline}
    	\label{eq:cloak:op}
(\vec{\ell}_1\cdot\vec{k})(\vec{\ell}_2\cdot\vec{k})(\vec{\ell}_3\cdot\vec{k})(\vec{\ell}_4\cdot\vec{k}) \\ %\nonumber
= (\ell_xk_x)^4 + (\ell_yk_y)^4 + (\ell_zk_z)^4
- 2 (\ell_xk_x)^2 (\ell_yk_y)^2 \\
- 2 (\ell_xk_x)^2 (\ell_zk_z)^2 
- 2 (\ell_yk_y)^2 (\ell_zk_z)^2,
\end{multline}
which is positive in each of the funnels around one fundamental axis and negative in the three-sided funnels in the middle of each octant (see Fig.~\ref{fig:proj}). If we multiply the Fourier transform $\hat{f}(\vec{k})$ of some triaxial function $f(\vec{r})$ with \eqref{eq:cloak:op}, the corresponding density 
\begin{align}
	\label{eq:invisible:diff}
	\rho_0(\vec{r}) &\equiv
	(\vec{\ell}_1\cdot\vec{\nabla})
	(\vec{\ell}_2\cdot\vec{\nabla})
	(\vec{\ell}_3\cdot\vec{\nabla})
	(\vec{\ell}_4\cdot\vec{\nabla}) f(\vec{r})
\end{align}
is invisible when seen along any of the four LOS $\vec{\ell}_i$ -- this is also obvious by doing the projection via integration by parts. Defining $\vec{R}=(X,Y,Z)^t\equiv(x/\ell_x,y/\ell_y,z/\ell_z)^t$, this can be expressed as
\begin{align}
	\label{eq:invisible:diff2}
	\rho_0(\vec{r}) &=
	\left[\partial_X^4 + \partial_Y^4 + \partial_Z^4 - 2\partial_X^2\partial_Y^2 - 2\partial_X^2\partial_Z^2 - 2\partial_Y^2\partial_Z^2
	\right]f(\vec{r}).	
\end{align}
Applying this procedure to a spherical Gaussian $f(\vec{r})=G(\vec{r})\equiv\exp(-\tfrac12\vec{r}^2)$, we find $\rho_{0}=h(\vec{r})G(\vec{r})$ with
\begin{eqnarray}
	h(\vec{r}) &=& \ell_x^4 (x^4-6x^2+3) + \ell_y^4 (y^4-6y^2+3) + \ell_z^4 (z^4-6z^2+3)
	\nonumber \\ &&
			- 2 \ell_x^2\ell_y^2(x^2-1)(y^2-1)
			- 2 \ell_x^2\ell_z^2(x^2-1)(z^2-1)
	\nonumber \\ &&
			- 2 \ell_y^2\ell_z^2(y^2-1)(z^2-1).
	\label{eq:cloak:diff:gauss}
\end{eqnarray}
This function itself is not bounded from below, i.e.\ approaches $-\infty$ in certain directions. This means that $[1+h(\vec{r})]G(\vec{r})$ is not a physical model, but $h(\vec{r})G(\vec{r})$ is, of course, bounded and can be added to another model such that the total is still non-negative.

When applying the recipe to a triaxial Gaussian $G(\vec{r})=\exp(-\frac12\vec{r}^t\cdot\mat{C}^{-1}\cdot\vec{r})$, then we again obtain $\rho_{0}=h(\vec{r})G(\vec{r})$ with $h(\vec{r})$ as given in equation~\eqref{eq:cloak:diff:gauss} after the replacements $\vec{\ell}\to\mat{C}^{-1/2}\cdot\vec{\ell}$ and $\vec{r}\to\mat{C}^{-1/2}\cdot\vec{r}$.

%%%%%%%%%%%%%%%%%%
\begin{figure*}
    \centering
    \includegraphics[width=100mm,angle=270]{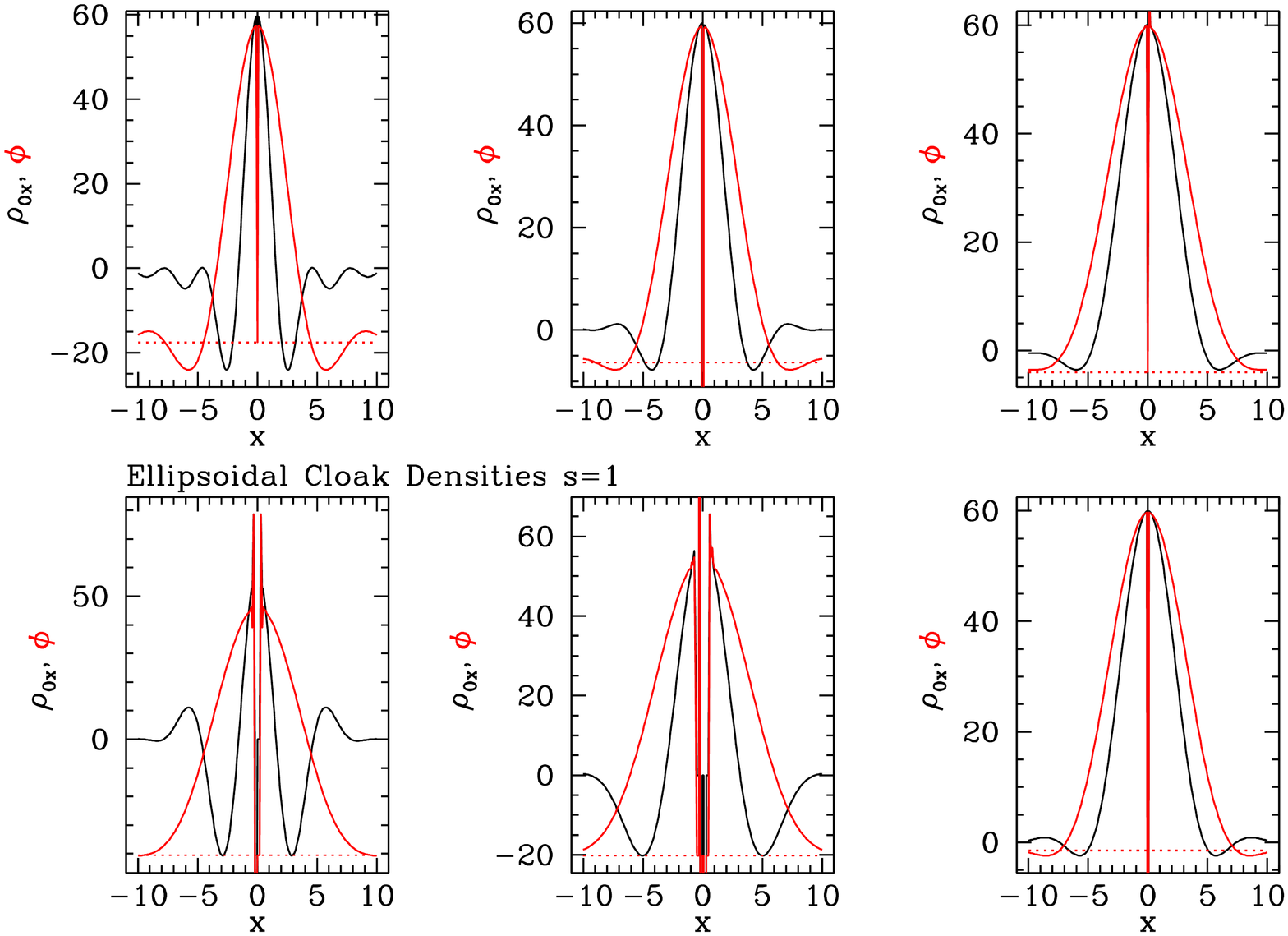}
    \caption{Plots of the six functions $\varphi$ of Table \ref{tab:func:ell} (red); the horizontal line shows the zero level) and the corresponding $\rho_{0x}$
    for $s=1$ (black, see equation~\ref{eq:rho0i}).}
    \label{fig:func:ell}
\end{figure*}
%%%%%%%%%%%%%%%%%%

Another option is to pick $f(\vec{r})=F(|\vec{R}|)$. Then
\begin{align}
	\rho_0(\vec{R}) &= 3\frac{F'}{R^3}-3\frac{F''}{R^2}+2\frac{F'''}{R}
	\nonumber \\ &+ 
	Q(\vec{R})
	\left(-15\frac{F'}{R^3}+15\frac{F''}{R^2}-6\frac{F'''}{R}+F''''\right),
\end{align}
where
\begin{equation}
	Q (\vec{R})
	\equiv R^{-4}\left[X^4+Y^4+Z^4 - 2X^2Y^2 - 2X^2Z^2 - 2Y^2Z^2\right],
\end{equation}
which is maximal at $Q=1$ on the axes, minimal at $Q=-1/3$ at $X^2=Y^2=Z^2$, and vanishes for $X\pm Y\pm Z=0$ (with both signs independent), corresponding to $x/\ell_x \pm y/\ell_y \pm z/\ell_z = 0$, which holds on four planes, which in a sense are the reciprocal planes to those in $\vec{k}$ space where $\hat{\rho}=0$.

\begin{figure*}
    \centering
    \includegraphics[width=0.4\linewidth]{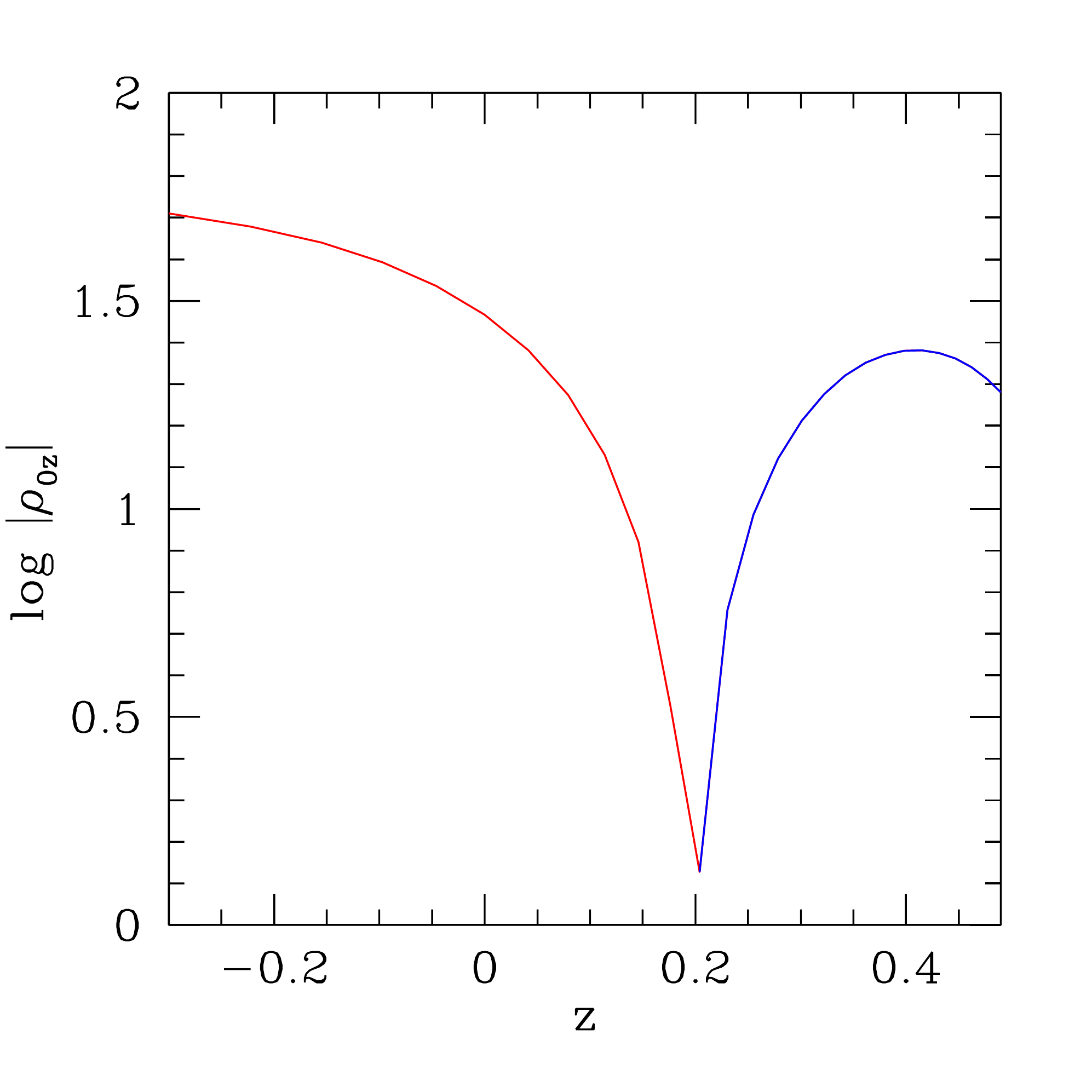}
    \includegraphics[width=0.4\linewidth]{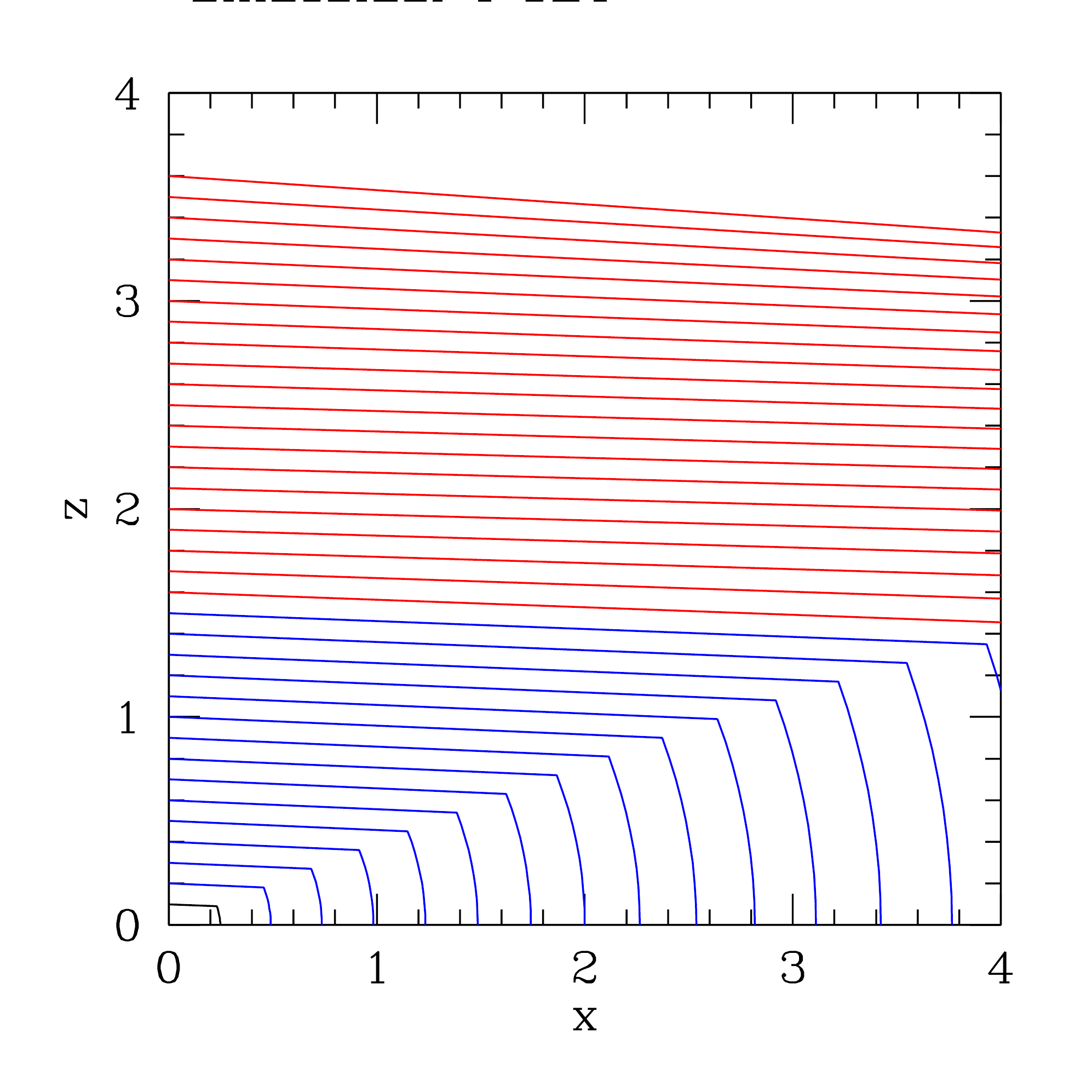}
    \caption{The equivalent of the bottom plot of Fig. \ref{Fig.konus} (left) and the middle plot of Fig. \ref{Fig.konus_contour} for the density $\rho_{0z}$
    with $s=1$ (see equation~\ref{eq:rho0i}) and inverted colours (red for positive densities and blue for negative values).}
    \label{fig:cloackexample}
\end{figure*}

\subsection{Cloaked densities via Fourier transform of compact functions}
\label{sec:cloak:FT}
When constructing a cloaked density via differentiation as in the previous sub-section, one has little control over the resulting shape. Here, we consider methods to construct cloaked densities with certain properties as Fourier transform of functions $\hat{f}(\vec{k})$ that vanish everywhere except for a finite triaxial region. The simplest such functions are obtained by shifting a symmetric function of compact support by an amount $1/s$ along $\hat{\vec{x}}_i$ and superimpose it with a version shifted in the other direction:
\begin{equation}
	\hat{\rho}(\vec{k}) = \textstyle \frac12\sum_{\pm}
	\hat{f}(\vec{k}\pm\uvec{x}_i/s)
\end{equation}
with density
\begin{equation}
	\label{eq:cloak}
	\rho(\vec{r}) = \cos(\vec{r}\cdot\uvec{x}_i/s) f(\vec{r}).
\end{equation}
This is a way to hide a disc perpendicular to the $x_i$ axis, with typical scale height $s$, and with extent given by the typical scale of $f$.

Another possibility is to shift $\hat{f}$ along a direction in the three-sided funnels in the centre of each octant (see Fig.~\ref{fig:proj}).

\subsubsection{Ellipsoidal cloaked densities}
One option is to take $\hat{f}(\vec{k})$ to be ellipsoidal. Let $\varphi(r)$ be a spherical function whose 3D Fourier transform $\hat{\varphi}(k)$ vanishes for $k>1$. From such a function, we may construct an invisible model via the above recipe as
\begin{equation}
	\rho_{0i} = \cos\pfrac{\vec{r}\cdot\uvec{x}_i}{s}\,\frac{\varphi(m)}{abc}
\label{eq:rho0i}
\end{equation}
For this to be invisible its Fourier transform must not intersect the plane $\vec{\ell}\cdot\vec{k}=0$ which requires
\begin{align}
	\label{eq:suppress}
	s < \ell_i \sigma_\ell,\qquad
	\sigma_\ell^{-2} = \vec{\ell}\cdot\mat{C}^{-1}\cdot \vec{\ell}.
\end{align}
So, not surprisingly it is easier to hide a disc that is near-perpendicular to the LOS (large $\ell_i$) than other discs.  Possible functions $\varphi(r)$ are listed in Table~\ref{tab:func:ell} and shown in Fig.~\ref{fig:func:ell}. Fig.~\ref{fig:cloackexample} shows the qualitative equivalent of the bottom plot of Fig. \ref{Fig.konus} (left) and the middle plot of Fig. \ref{Fig.konus_contour} for the density $\rho_{0z}$ with $s=1$ (see equation~\ref{eq:rho0i}).

%%%%%%%%%%%%%%%%%%%%%
\subsubsection{Cuboidal cloaked densities}
\label{ref:cloak:cuboidal}
Instead of shifting ellipsoidal Fourier distributions, to generate cloaked densities, one may also use cuboidal distributions of the form 
\begin{equation}
	\hat{f}(\vec{k}) =
	\hat{h}_x(ak_x)\,\hat{h}_y(bk_y)\,\hat{h}_z(ck_z)
\end{equation}
with $\hat{h}_i(k)\neq0$ only for $|k|<1$. For example the top-hat function and its $n$-fold self-convolution\footnote{These functions, also known as \cite{Schoenberg1946} B-splines, are (modulo a scaling) identical to the \cite{Irwin1927}-\cite{Hall1927} probability density for the sum $k$ of $n$ independent variables, each drawn form a uniform distribution between $-1/n$ and $1/n$. The only difference to the common use of these functions is that we revert the role of the function and its Fourier transform so that the latter has compact support.}, which correspond to
\begin{equation}
	\label{eq:wn}
	h(x) = b_n(x) \equiv \sinc^n(x/n)
\end{equation}
with
\begin{equation}
	b_1(x) = \sinc x \equiv \frac{\sin x}{x},
\end{equation}
which has as Fourier transform the top-hat function $\hat{b}_1=\frac12$ for $|x|<1$ and 0 otherwise. The scaling of the argument by $1/n$ in~\eqref{eq:wn} ensures that $\hat{h}_n(k)=0$ for $|k|>1$. Possible 3D densities are then
\begin{align}
	\rho_{i,\mat{n}}(\vec{r}) = \frac{1}{abc}
	\cos\pfrac{r_i}{s}\, b_{\s{n}_x}\pfrac{x}{a}\, b_{\s{n}_y}\pfrac{y}{b}\,
						 b_{\s{n}_z}\pfrac{z}{c}
\end{align}
with parameters $\mat{n}=(\s{n}_x,\s{n}_y,\s{n}_z)$, $\vec{a}=(a,b,c)$, $s$ and $i$. At large distances, these functions decay as $1/\vec{x}^{\mat{n}}\equiv 1/x^{\s{n}_x}y^{\s{n}_y}z^{\s{n}_z}$. In order for this density to be invisible, its Fourier transform must not intersect the plane $\vec{\ell}\cdot\vec{k}=0$, which requires that
\begin{equation}
	\frac{\ell_i}{s} > \frac{\ell_x}{a} + \frac{\ell_y}{b} + \frac{\ell_z}{c}.
\end{equation}

%%%%%%%%%%%%%%%%%%%%
\subsection{Cloaked conus densities}
\label{sec:conus}
The method of the previous sub-section cannot create centrally diverging cloaked densities, because such distributions have power on all scales and their Fourier transform is not confined to a compact region. This is, however, only a shortcoming of this particular method and not inherent to cloaked densities: one may superpose many such models with ever smaller $\mat{C}$ and $s$ to create a cuspy cloaked density.

%%%%%%%%%%%%%%%%%%
%%%%%%%%%%%%%%%%%%
\begin{table*}
	\caption{\textbf{Functions for elliptic conus densities.} $\hat{\varphi}_1(x)$ is obtained from $\hat{\varphi}(k)$ via equation~\eqref{eq:phi1:phi2} (modulo a constant factor) and $\hat{\varphi}^{(2n)}_1(x)$ serves as vertical density profile for the conus density disc.
	\label{tab:func:conus:ell}}
\medskip
\begin{tabular}{llll}
	$n$ & $\hat{\varphi}^{(2n)}_1(x<1)$ & $\hat{\varphi}(k<1)$ & comments
	\\[0.5ex]
	\hline
	\\[-2ex]
	1	& $1-6x^2+5x^4$ & $(1-k^2)^{5/2}$ & $\hat{\varphi}^{(2n)}_1$ discontinuous at $x=1$
	\\[1ex]
	1	& $(1-7x^2)(1-x^2)$ & $(1-k^2)^{7/2}$ &
	\\[1ex]
	2	& $3-30x^2+35x^4$ & $(1-k^2)^{7/2}$ & $\hat{\varphi}^{(2n)}_1$ discontinuous at $x=1$
	\\[1ex]
	2	& $(3-42x^2+63x^4)(1-x^2)$ & $(1-k^2)^{9/2}$ & 
	\\[1ex]
	\hline
\end{tabular}
\end{table*}

Alternatively, we may construct a cloaked density from a Fourier transform that is defined everywhere inside a cone around one of the fundamental axes. Without loss of generality, we take this to be the $z$ axis. Taking the cone to be elliptic, this gives the ansatz
\begin{equation}
	\hat{\rho}(\vec{k}) = \hat{\varphi}\left(\sqrt{a^2k_x^2+b^2k_y^2}\Big/k_z\right)
		\hat{f}(|k_z|),
\end{equation}
where as before $\hat{\varphi}(k)$ vanishes for $k>1$, while $\hat{f}$ is as of yet unspecified. For this to be invisible
\begin{equation}
	\ell_x^2/a^2 + \ell_y^2/b^2 > \ell_z^2.
\end{equation}
Fourier transforming $\hat{\rho}(\vec{k})$ first in $x$ and $y$ and then in $z$ gives
\begin{align}
    	\rho(\vec{r})
	&= \frac{1}{ab} \int_{-\infty}^{+\infty}
	\Exp{\I k_z z}\;k_z^2\hat{f}(k_z)\;\varphi(k_z\mu)\, \diff k_z 
	\nonumber \\
	&= \frac{1}{ab\mu} \int_{-\infty}^{+\infty}
	\Exp{\I\kappa z/\mu}\;\pfrac{\kappa}{\mu}^2\hat{f}\pfrac{\kappa}{\mu}\;\varphi(\kappa)\,
	\diff\kappa
\end{align}
with $\mu^2\equiv x^2/a^2+y^2/b^2$. For this to result in a closed functional form, the freedom for the function $\hat{f}(k)$ must be exploited. If, for example, one takes $k^2\hat{f}(|k|)=(-\I k)^{2n}$ with\footnote{Or $n>1$ if $\hat{f}(0)=0$ is required.} $n>0$, then
\begin{equation}
	\rho(\vec{r})
	= \frac{1}{ab\mu^{2n+1}} \hat{\varphi}_1^{(2n)}\pfrac{z}{\mu},
\end{equation}
where $\hat{\varphi}_1$ is the one-dimensional Fourier transform of $\varphi(r)$, which in turn was the two-dimensional Fourier transform of $\hat{\varphi}$ that vanishes at $k>1$. By comparing their respective inverse Fourier transforms, one finds
\begin{equation}
	\label{eq:phi1:phi2}
	\hat{\varphi}_1(x)
	= 2\int_0^{\sqrt{1-x^2}} \hat{\varphi}\left(\sqrt{x^2+k^2}\right) \diff k
	= 2\int_x^1 \frac{\hat{\varphi}(r)\,r\,\diff r}{\sqrt{r^2-x^2}}
.
\end{equation}
It follows that $\hat{\varphi}_1(x)$ also vanishes at $x>1$, which implies that the density vanishes for $|z|>\mu$, i.e.\ $\rho(\vec{r})$ describes a flaring elliptic disc with vanishing column density and power-law mid-plane profile.  Possible functions $\hat{\varphi}_1(r)$ are listed in Table~\ref{tab:func:conus:ell}.

\subsection{Near-invisible densities}
We now consider simple analytic density distributions with projections that do not vanish exactly, but are potentially very small. These may be useful in numerical work, for example as a perturbation to be added to another model as input for an iterative deprojection algorithm, or as a component of a superposition-based deprojection.

\subsubsection{Near-invisible ellipsoidal models}
The idea here is to replace the functions of compact support used in the previous sub-section with more general ellipsoidal models, i.e. use the recipe~\eqref{eq:cloak} with some model $f(\vec{r})$ whose Fourier transform $\hat{f}(\vec{k})$ may not vanish anywhere. Then, of course, the resulting $\hat{\rho}(\vec{k})$ will not vanish on the four planes $\vec{\ell}_i\cdot\vec{k}=0$, but can be small on these planes if $\hat{f}(\vec{k})$ decays sufficiently fast and the scale $s$ is sufficiently small, such that the projection $\Sigma$, though not vanishing, is hardly visible.

\begin{figure*}
	\subfloat[$\theta=80\degr$ \label{Fig.ell_profile_disc}]{\includegraphics[height=115mm]{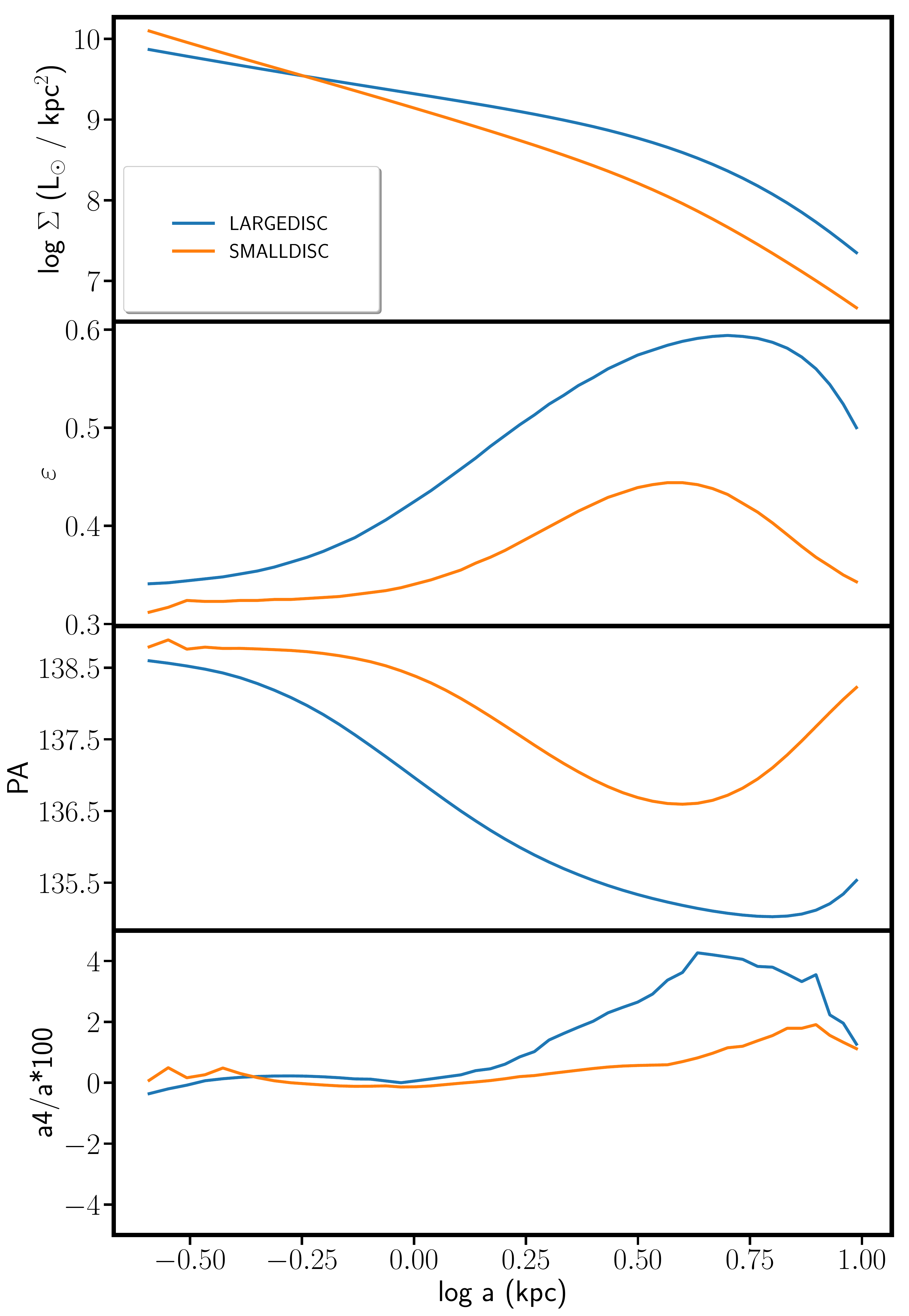}}
	\hfil
	\subfloat[$\theta=45\degr$ \label{Fig.PA_profile_disc}]{\includegraphics[height=115mm]{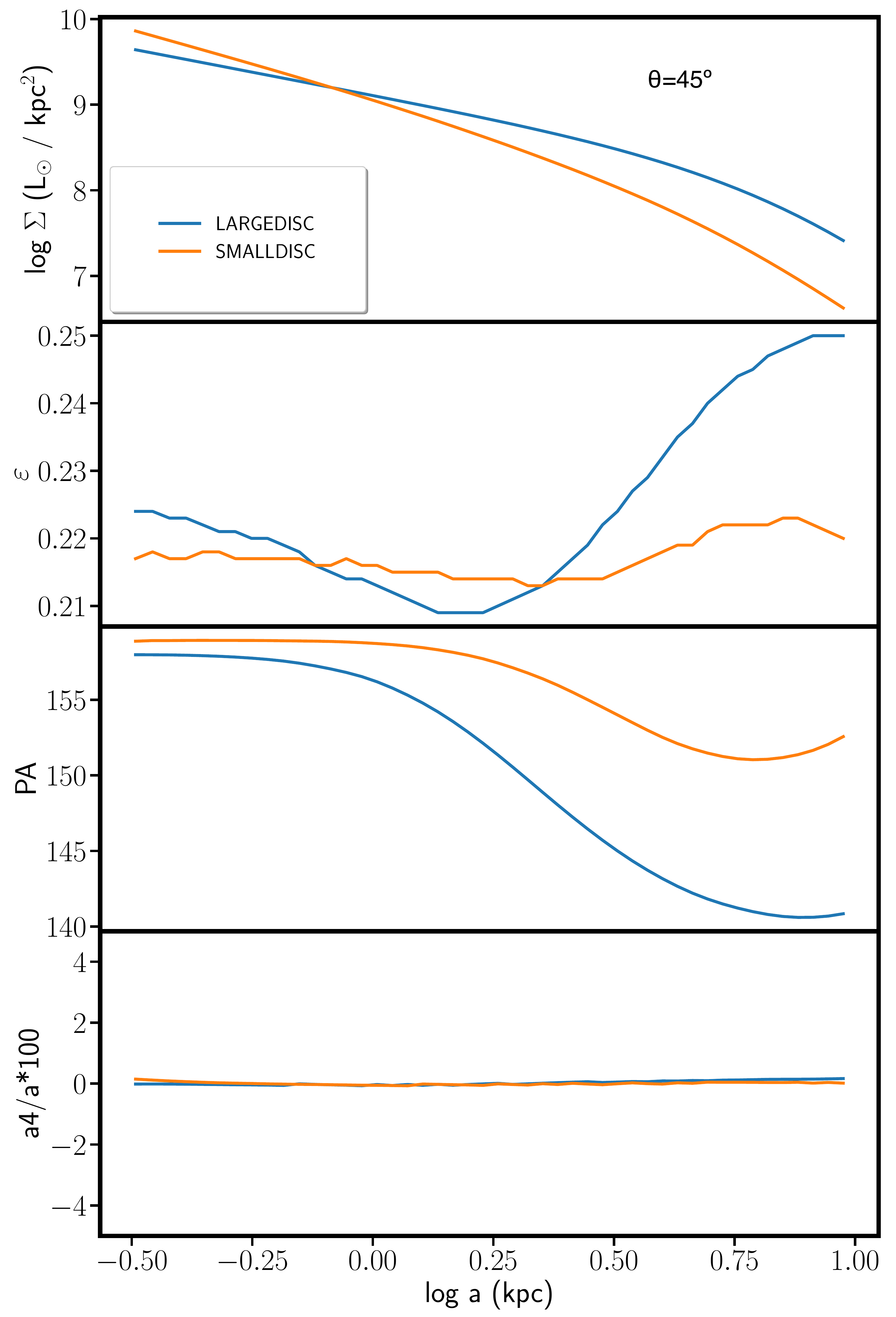}}
 
	\caption{Surface brightess ellipticity $\varepsilon$, $a4$ and twist profiles as a function of semi-major axis when we project the \emph{LARGEDISC} or the \emph{SMALLDISC} model at $\theta = 80\degr$ (left) or $\theta=45\degr$ (right), in both cases with $\phi=\psi=45\degr$. High ellipticity and the positive $a4$, which are clear markers of a disc-like component, are only present for $\theta=80\degr$ but not $45\degr$.}
	\label{Fig.isophotes_disc} 
\end{figure*}

The simplest case is a near-invisible ellipsoidal Gaussian, when this recipe gives density
\begin{equation}
	\rho(\vec{r}) = \cos\pfrac{\vec{r}\cdot\uvec{x}_i}{s}
		\frac{\exp\left(-\frac12
		\vec{r}^t\cdot\mat{C}^{-1}\vec{r}\right)}{\sqrt{(2\pi)^3|\mat{C}|}} .
\end{equation}
Adding such a model generates a disc in the plane perpendicular to $\uvec{x}_i$ with projected surface density
\begin{equation}
	\label{eq:proj:Gauss:mod}
	\Sigma(\xi,\eta) = 
	\exp\left(-\tfrac12\ell_i^2\sigma^2_\ell/s^2\right)\;
	\cos\left[\frac{1}{s}
	\begin{pmatrix}\xi\\\eta\end{pmatrix}\cdot
	\begin{pmatrix}\xi_i+\ell_i\sigma_\ell^2\sigma_{\xi\ell}^{-2} \\
				  \eta_i+\ell_i\sigma_\ell^2\sigma_{\eta\ell}^{-2}
	\end{pmatrix}\right]\;
	\Sigma_0(\xi,\eta),
\end{equation}
where $\sigma_\ell$ was given in equation~\eqref{eq:suppress},
\begin{align}
	\sigma_{\xi\ell}^{-2} = \vec{\xi}^t\cdot\mat{C}^{-1}\cdot\vec{\ell},
	\qquad
	\sigma_{\eta\ell}^{-2} = \vec{\eta}^t\cdot\mat{C}^{-1}\cdot\vec{\ell},
\end{align}
while
\begin{equation}
	\Sigma_0(\xi,\eta) = \frac{\exp\left(-\tfrac12(\xi,\eta)^t\cdot\bar{\mat{C}}^{-1}\cdot(\xi,\eta)\right)}{\sqrt{(2\pi)^2|\bar{\mat{C}}|}}
\end{equation}
is the the projected density of an ellipsoidal Gaussian. Thus, $\Sigma$ differs from that of an ellipsoidal Gaussian by both a cosine modulation and suppression factor. For a substantial suppression $s\ll\ell_i\sigma_\ell$, which favours discs near-perpendicular to the LOS so that $\ell_i$ is large.

%%%%%%%%%%%%%
\subsubsection{Near-invisible elliptical discs}
We can also use a Gaussian for $\hat{\varphi}$ in the recipe of \S\ref{sec:conus}, i.e.
\begin{equation}
	\hat{\rho}(\vec{k}) = k_z^2 \Exp{-\frac12(a^2k_x^2+b^2k_y^2)/k_z^2}.
\end{equation}
\begin{equation}
	\rho(\vec{r}) = \frac{1}{(2\pi)^{3/2} ab}\frac1{\mu^5}
	\left(3-6\frac{z^2}{\mu^2}+\frac{z^4}{\mu^4}\right)
	\Exp{-\tfrac12z^2/\mu^2}.
\end{equation}

\section{Probing the effects of hidden discs}
\label{sec_discs}

Massive elliptical galaxies have nearly elliptical isophotes and this justifies the assumption of the deformed ellipsoidal deprojection algorithm discussed in the previous sections. However, even these objects could harbour (possibly faint) disc components, possibly nearly invisible in projection (see discussion in Appendix \ref{sec_cloaked}).  Here we explore the effects of hidden discs by considering a flat component whose intrinsic light density $\rho_{D}$ is described by a double exponential profile, reminiscent of those observed for spiral galaxies:
\begin{equation}
    \rho_D = N \Exp{-\frac{\sqrt{x^2 + y^2}}{h}} \Exp{-\frac{z}{h_z}}.
    \label{eq.rho_disc}
\end{equation}
We choose the scale length and height to be $h=0.5$ and $h_z=0.1$, respectively, such that the half-light radius is similar to the one of the Jaffe model used above and the structure is flatter than the most flatten elliptical galaxies known. $N$ is a normalization factor used to vary the disc mass. The density contours in the meridional plane are rhombi, i.e. quite different from the deformed ellipses of equation \eqref{m}. We deproject the projection of $\rho_D$ using our implementation of M99's code, finding, as expected, that the deprojection is unique for $\theta=90\degr$ and it can be tuned towards the true density by using the $d_4$ parameter of the code to obtain discy isophotes at lower $\theta$ angles.

\begin{figure*}
	\subfloat[SB comparison, $\theta = 15\degr$.\label{Fig.disc_sb_theta15}]
		{\includegraphics[width=.35\linewidth]
			{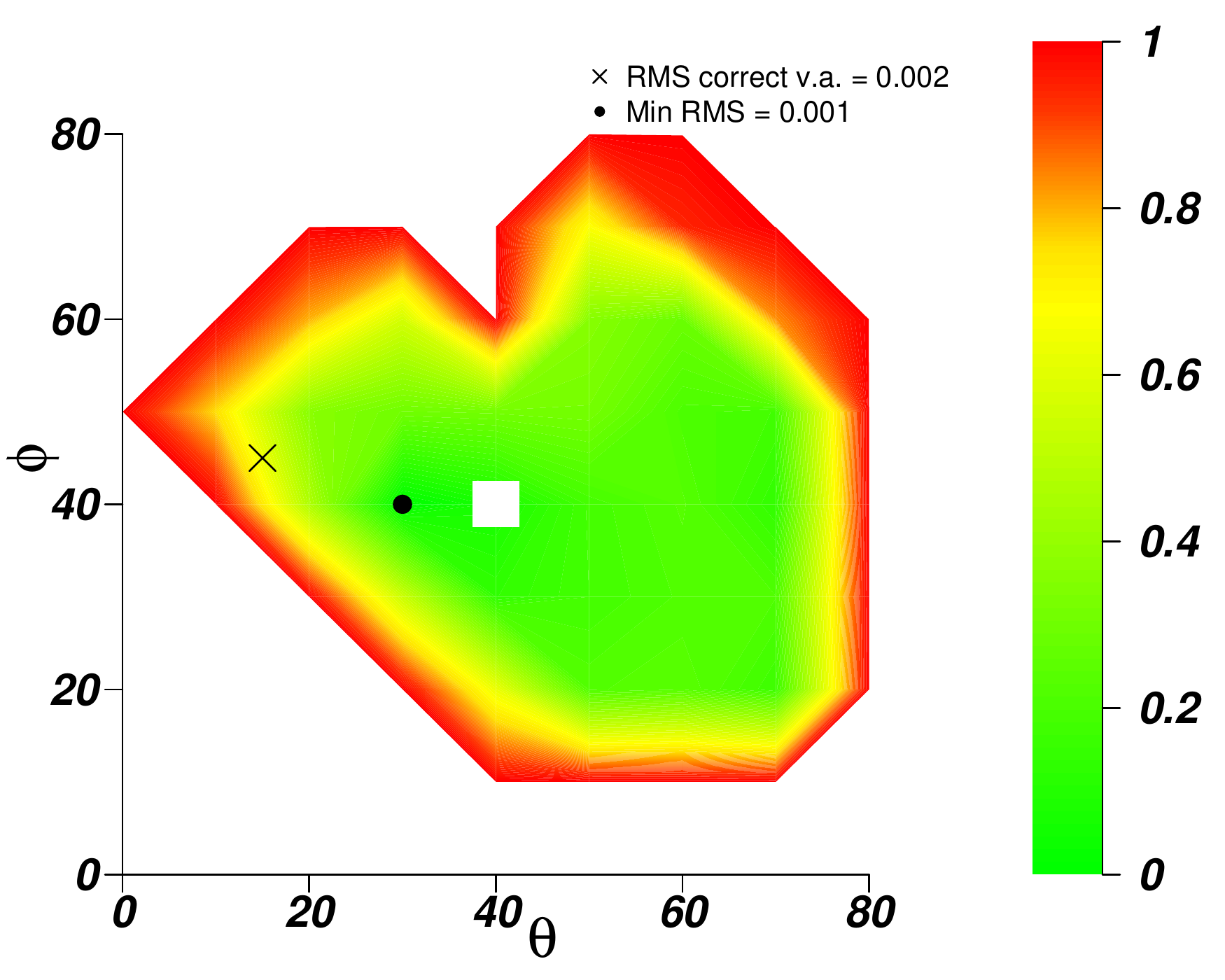}}
	\hfil
	\subfloat[$\rho$ comparison, $\theta = 15\degr$.\label{Fig.disc_rho_theta15}]
		{\includegraphics[width=.35\linewidth]
			{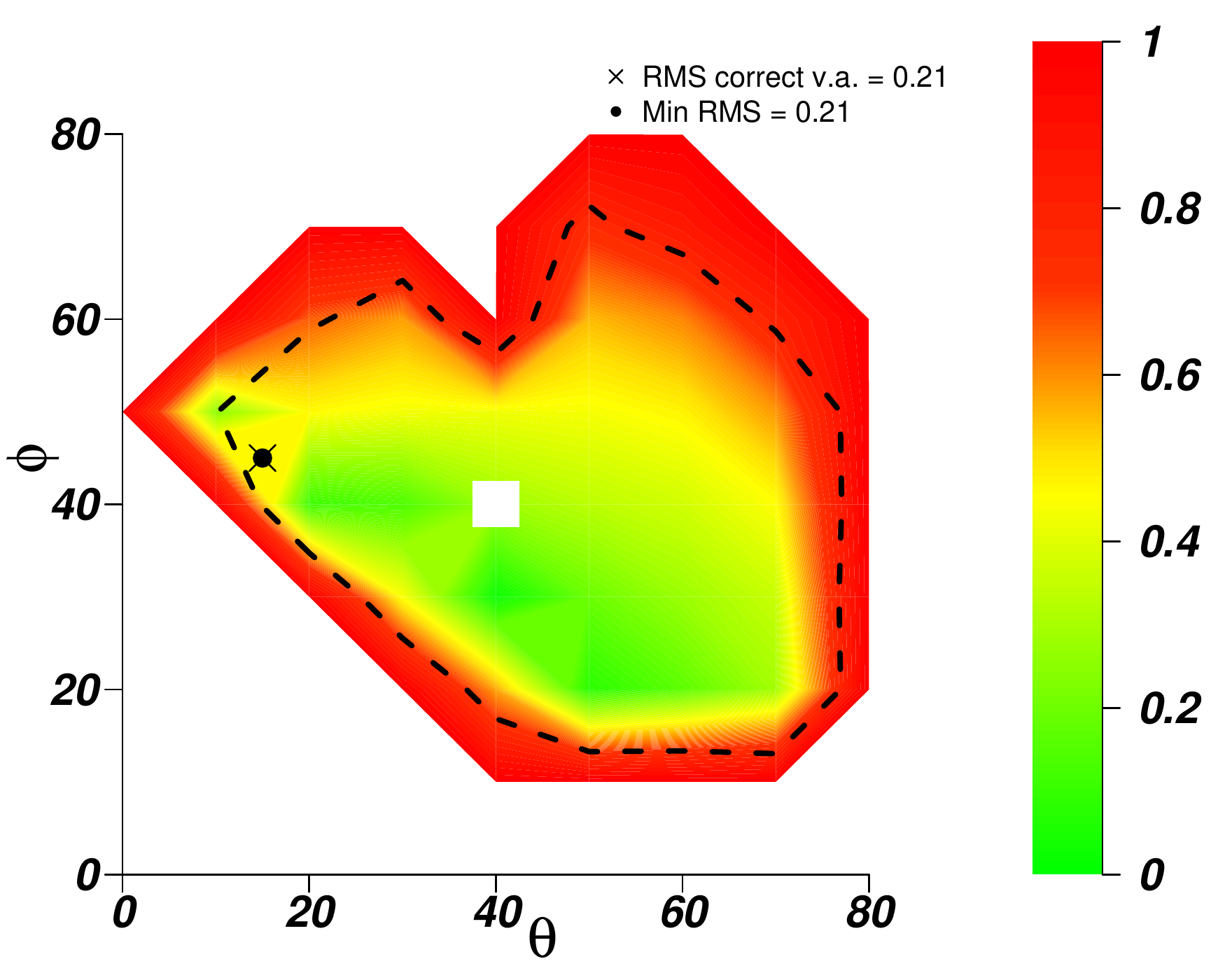}}
	\\
	\subfloat[SB comparison, $\theta = 45\degr$.\label{Fig.disc_sb_theta45}]
		{\includegraphics[width=.35\linewidth]
			{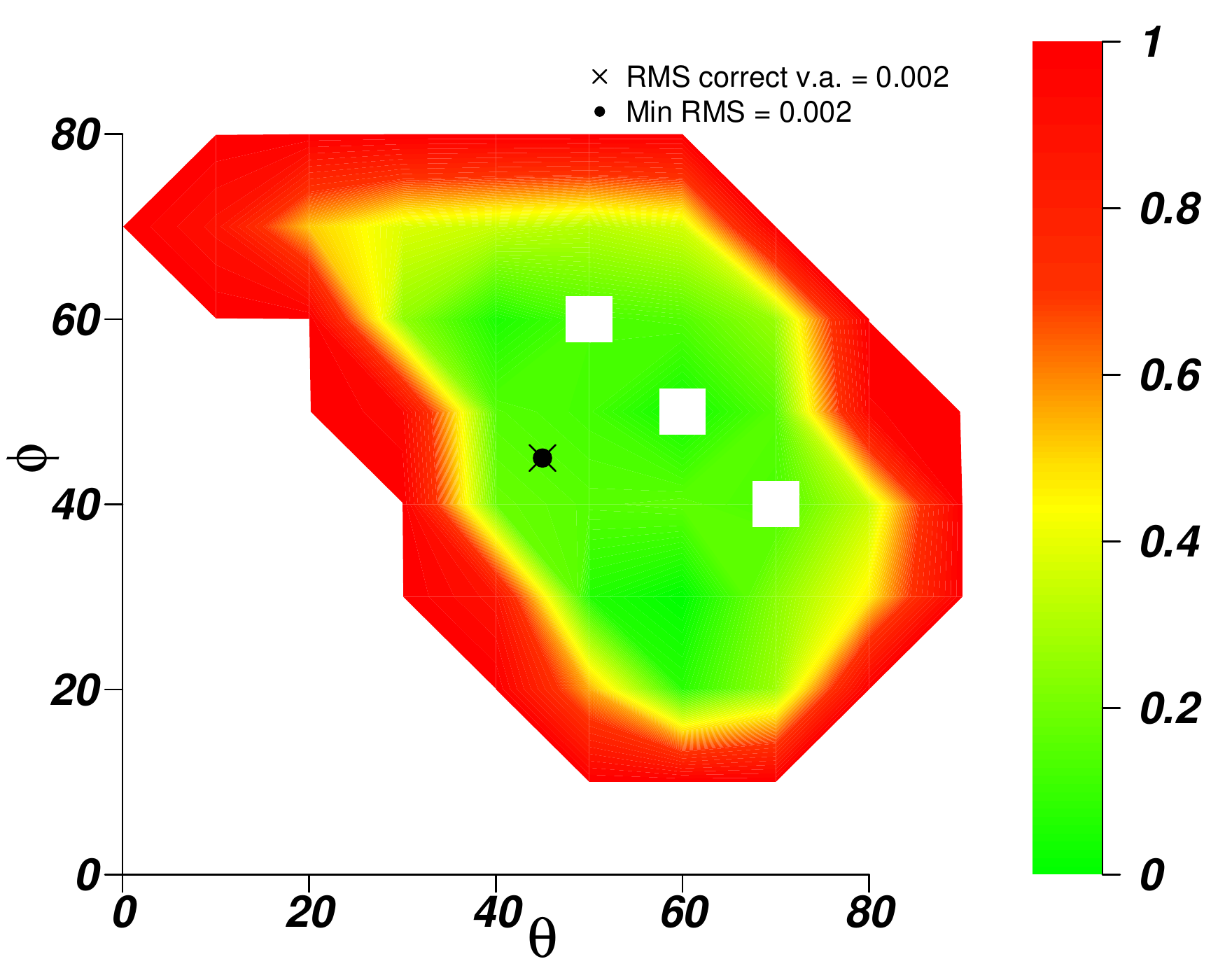}}
	\hfil
	\subfloat[$\rho$ comparison, $\theta = 45\degr$.\label{Fig.disc_rho_theta45}]
		{\includegraphics[width=.35\linewidth]
			{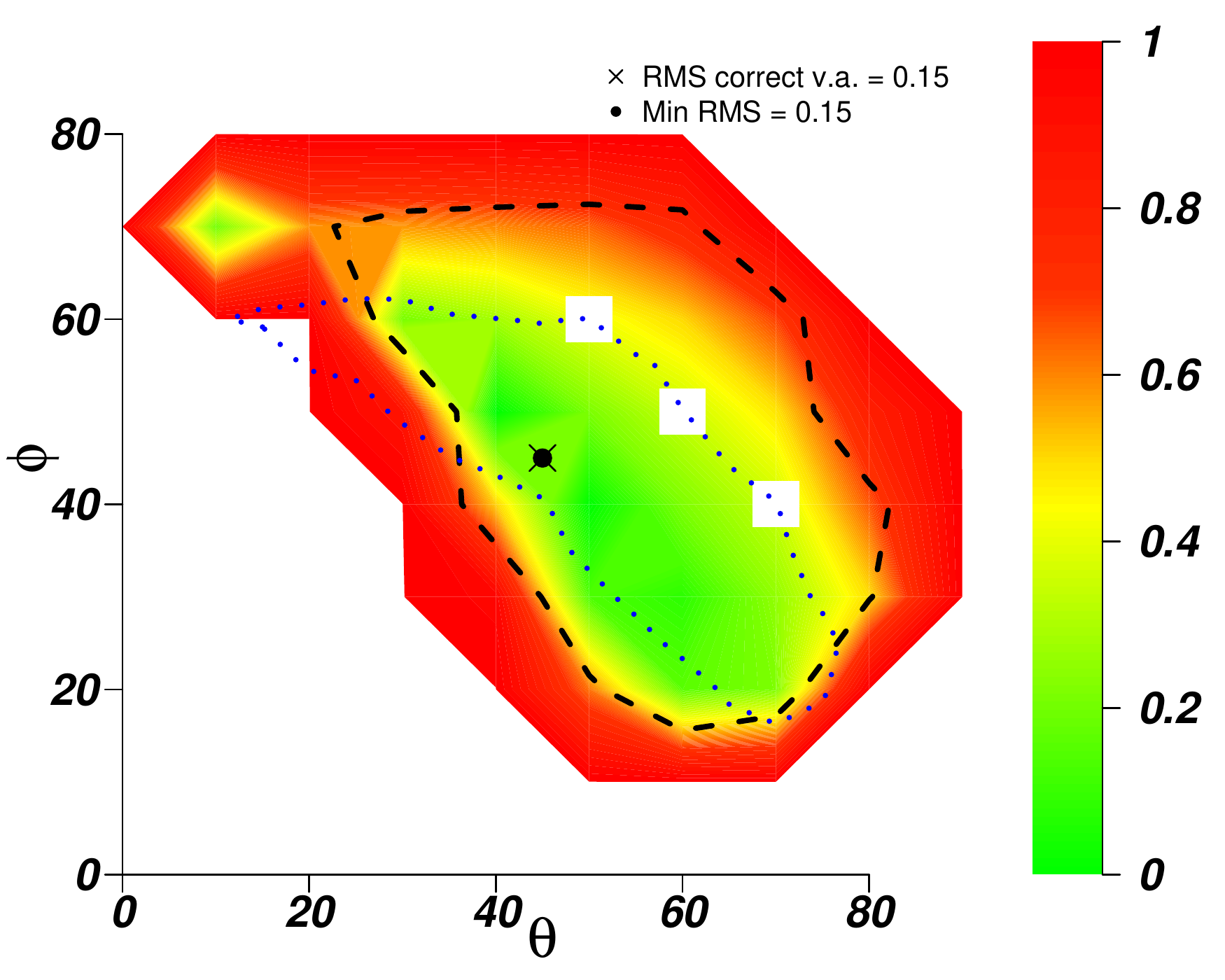}}
	\\
	\subfloat[SB comparison, $\theta = 80\degr$.\label{Fig.disc_sb_theta80}]
		{\includegraphics[width=.35\linewidth]
			{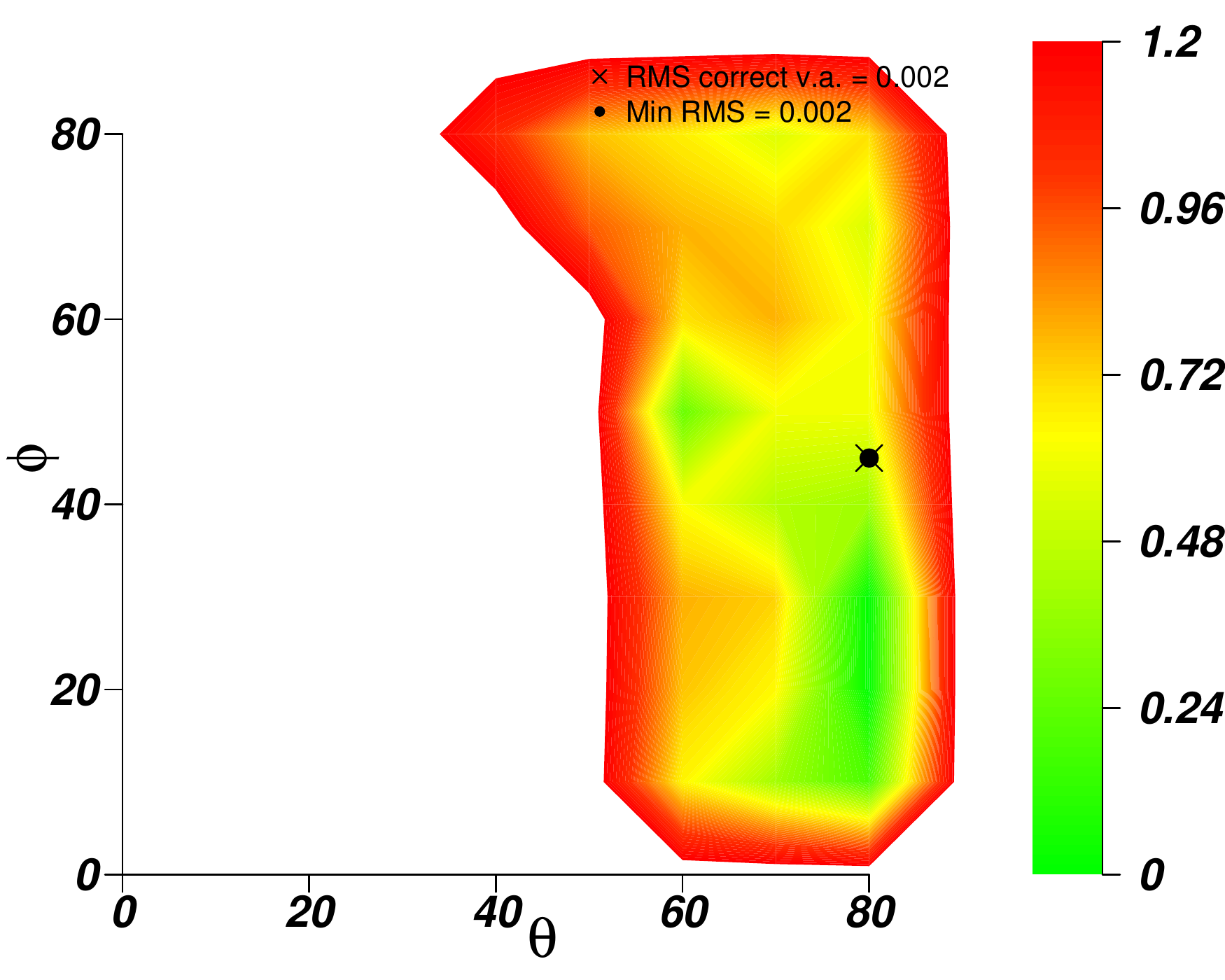}}
	\hfil
    \subfloat[$\rho$ comparison, $\theta = 80\degr$.\label{Fig.disc_rho_theta80}]
		{\includegraphics[width=.35\linewidth]
			{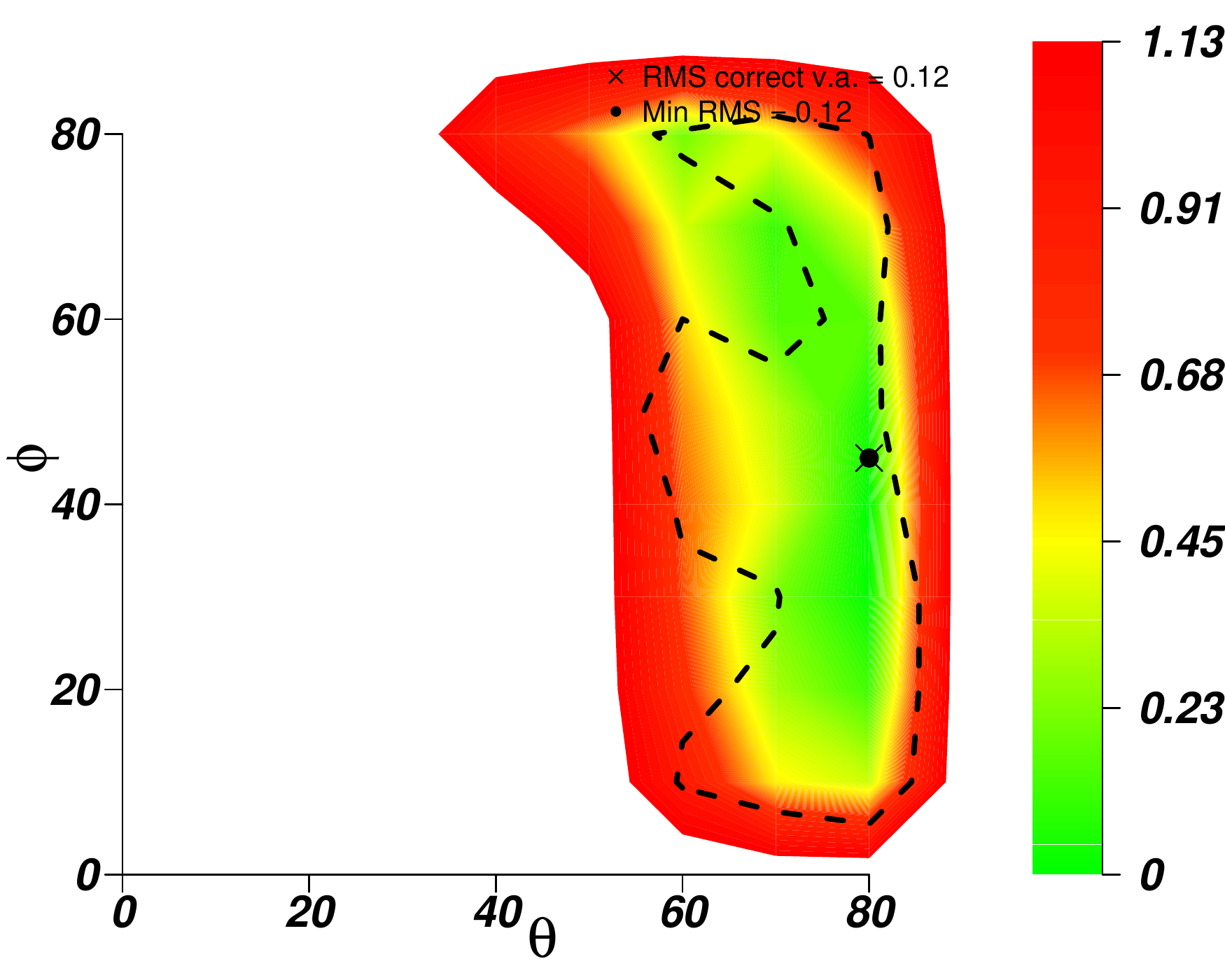}}

	\caption{Same as Figs.~\ref{Fig.Jaffe_ell_45}-\ref{Fig.Nbody_45} for model \emph{SMALLDISC}. As in Fig.~\ref{Fig.Jaffe_ell_45} (reproduced here by the blue dashed contour in the middle left panel), the area of good fits overlaps well with that where the intrinsic density matches the true one, but is larger. The true viewing angles are recovered well except for the small discrepancy at $\theta=15\degr$.}
    \label{Fig.disc_maps}
\end{figure*}

As a second step, we sum to the density of \emph{ELLIP} the density $\rho_D$ with normalisation $N$ chosen such that the two components have mass ratios of 1 (\emph{LARGEDISC}) or 5.67 (\emph{SMALLDISC}, where the flattened component has 15\% of the total mass). We project these densities for $\theta=80\degr$, $45\degr$, and $15\degr$ with $\phi=\psi=45\degr$.  Decreasing $\theta$ makes it easier to hide the flattened component in projection. For \emph{SMALLDISC} (and even more for \emph{LARGEDISC}), the isophotes of the projected density at $\theta=80\degr$ show a clear signature (high ellipticity and $a_4$ values, see Fig.~\ref{Fig.isophotes_disc}, left). At $\theta = 45\degr$ the only possible signature for \emph{LARGEDISC} is a $\sim 20\degr$ twist (Fig.~\ref{Fig.isophotes_disc}, right), which lies just on the threshold of what we can observe in massive ellipticals (see Fig.~\ref{Fig.twist_weak}).

\begin{figure}
	\hfil
	\includegraphics[width=78mm]{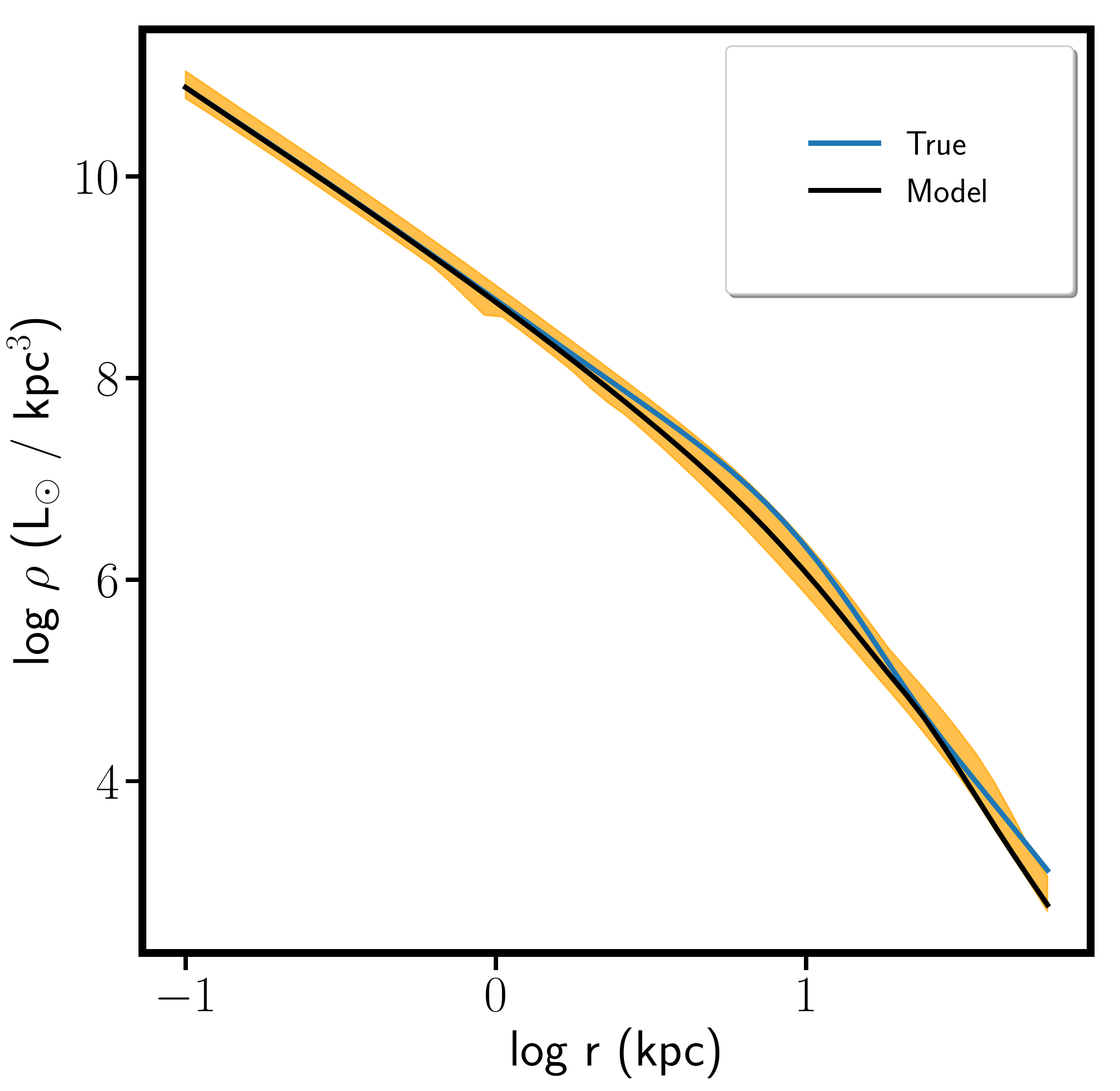}
	\hfil

	\caption{The intrinsic density along the major axis of model \emph{SMALLDISC} (blue), and the range of densities recovered with the constrained-shape method for viewing angles compatible with the surface brightness obtained projecting at $\theta=\phi=\psi=45\degr$. The black line shows the deprojection assuming these angles.}
	\label{Fig.radprofs_disc15}
\end{figure}

We are always able to deproject \emph{SMALLDISC} using the constrained-shape method, matching well the projected surface brightness and with resonably good precision the intrinsic density, getting RMS in $\rho$ of 12\%, 15\%, 20\% at $\theta=80\degr,\,45\degr,\,15\degr$, respectively. This corresponds to the range in density errors found when reconstructing the viewing angles for the Jaffe-only density (see Figs.~\ref{Fig.Jaffe_ell_45} and \ref{Fig.radprofs_disc15}).  However, the region of allowed viewing angles in these cases is larger (Fig.~\ref{Fig.disc_maps}).

For \emph{LARGEDISC} the situation is more difficult.  Given the strongly non-elliptical isophotes of the $\theta=80\degr$ projection, the constrained-shape algorithm is unable to deliver projected densities matching the true ones. We cure this problem by modifying the deprojection algorithm: we add a non-parametric, axisymmetric, flattened component, that is added to the one with deformed ellipsoidal shape, and optimize it subject to regularization constraints together with the first component through the Metropolis procedure. With this code we are able to reproduce well the SB profile, recovering the intrinsic density with an RMS of less than 9\%. Of course, since in this case the disc's signature can be seen in the photometry (Fig.~\ref{Fig.ell_profile_disc}), we may also directly subtract it from the galaxy image as done by \cite{Scorza90}.

\begin{figure}
	\hfil
	\includegraphics[width=78mm]{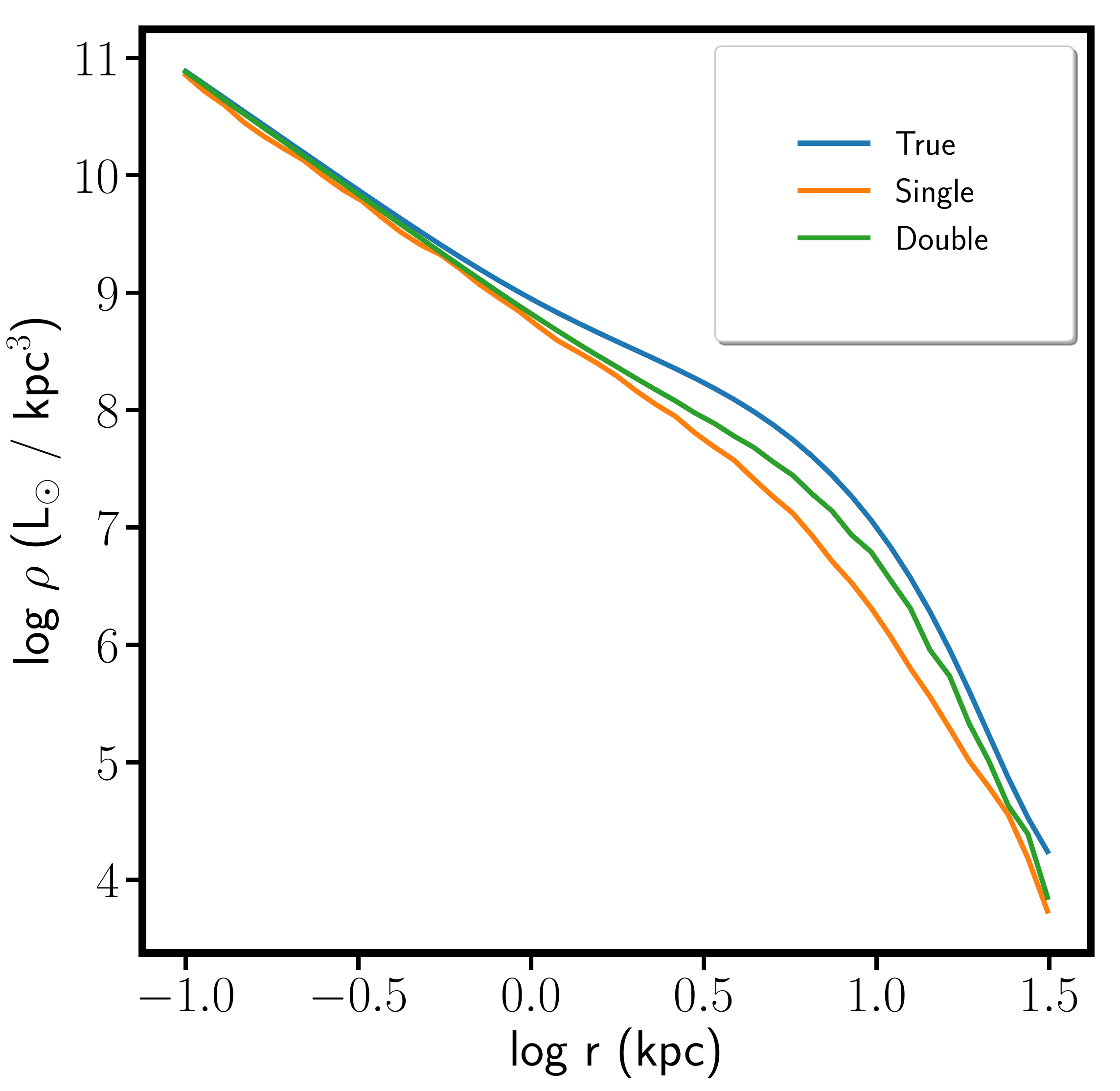}
	\hfil
	\caption{The intrinsic density along the major axis of \emph{LARGEDISC} (blue), the density  recovered with the constrained-shape method without (orange, `single') and with a complementary axisymmetric model (green, `double') assuming the true viewing angles $\theta=\phi=\psi=45\degr$.}
	\label{Fig.radprofs_disc50}
\end{figure}

When we project \emph{LARGEDISC} at $\theta=45\degr$ or $15\degr$, the disc becomes impossible to spot from a photometric analysis alone (Fig.~\ref{Fig.PA_profile_disc}) and the constrained-shape algorithm is able to reproduce the observed surface brightness very well. However, the intrinsic density can only be recovered up to an RMS of $\sim 36$\% (or even worse when $\theta=15\degr$). Using the modified, constrained-shape-plus-axisymmetric-component algorithm we are able to reproduce the observed surface brightness to the same precision and the intrinsic density with an RMS of $\sim 22$\% (see Fig.~\ref{Fig.radprofs_disc50}). We \emph{do not} see such a strong difference between the densities reconstructed with or without complementing the constrained-shape method with an axisymmetric component for models \emph{SMALLDISC}, \emph{ELLIP}, or \emph{DISCYBOXY}.

This exploration can guide us when deprojecting the surface photometry of real elliptical galaxies that do not have clear signs for the presence of a disc component. If a disc component is present, we expect that the differences between intrinsic densities recovered with and without a complementary axisymmetric component exceed the variations observed as function of assumed viewing angles.

% DON'T TOUCH THESE LINES!
\bsp	% typesetting comment
\label{lastpage}
\end{document}